\documentclass{emulateapj}
\usepackage{color}
\usepackage{amsmath}
\usepackage{threeparttable}
\usepackage{natbib}
\usepackage{graphicx}
\usepackage{color}
\usepackage{xspace}
\usepackage{mathrsfs}
\usepackage{rotating}
\usepackage{ulem}

\DeclareGraphicsExtensions{.jpg,.pdf,.png,.ps}

\newcommand{\yes}{\textcolor{blue}{\bf Yes}}
\newcommand{\no}{\textcolor{red}{\bf No}}
\newcommand{\marginal}{\textcolor{blue}{\it Marginal}}

\newcommand{\LCDM}{\ensuremath{\Lambda{\rm CDM}}}

\newcommand{\lcdm}{\LCDM}
\newcommand{\ns}{\ensuremath{n_{s}}\xspace}
\newcommand{\nrun}{\ensuremath{dn_s/d\ln k}\xspace}
\newcommand{\wmap}{\textit{WMAP}\xspace}
\def\cl{$C_{\ell}$}

\def\bea{\begin{eqnarray}}
\def\eea{\end{eqnarray}}

\newcommand{\zeq}{\ensuremath{z_{\rm eq}}}

\newcommand{\omegae}{\ensuremath{\Omega_{\rm e}}}
\newcommand{\od}{\ensuremath{\Omega_{\rm de}}}
\newcommand{\odo}{\ensuremath{\Omega_{\rm de}^0}}
\newcommand{\omo}{\ensuremath{\Omega_{\rm m}^0}}
\newcommand{\oeqs}{\ensuremath{w_0}}
\newcommand{\beq}{\begin{equation}}
\newcommand{\eeq}{\end{equation}}
\newcommand{\beqa}{\begin{eqnarray}}
\newcommand{\eeqa}{\end{eqnarray}}

\newcommand{\sumnu}{\ensuremath{\sum m_\nu}}
\newcommand{\neff}{\ensuremath{N_{\rm eff}}}
\newcommand{\thetas}{\ensuremath{\theta_{s}}}
\newcommand{\yp}{\ensuremath{Y_p}}

\newcommand{\degs}{deg$^2$}

\def\microKsq{\mu{\mbox{K}}^2}

\def\be{\begin{equation}}
\def\ee{\end{equation}}

\def\ell{l}

\newcommand{\WMAP}{{\it WMAP}}
\newcommand{\wseven}{\textit{WMAP}7} 
\newcommand{\comment}[1]{{}}

\newcommand{\omk}{\mbox{$\Omega_k$}}

 \newcommand{\ltsima}{$\; \buildrel < \over \sim \;$}
 \newcommand{\ltsim}{\lower.5ex\hbox{\ltsima}}

  \newcommand{\sptcl}{\ensuremath{{\rm SPT_{\rm CL}}}}

 \newcommand{\chisq}{\ensuremath{\chi^2}}
 \newcommand{\ho}{\ensuremath{H_{0}}}
 
 \newcommand{\boss}{\ensuremath{{\rm BAO}_{\rm BOSS}}}

\def\Davis{1}
\def\Berkeley{2}
\def\KICPChicago{3}
\def\PhysicsUChicago{4}
\def\UChicago{5}
\def\EFIChicago{6}
\def\AAUChicago{7}
\def\Argonne{8}
\def\NIST{9}
\def\McGill{10}
\def\FermiLab{11}
\def\JPL{12}
\def\Caltech{13}
\def\Colorado{14}
\def\NASA{15}
\def\LBNL{16}
\def\Michigan{17}
\def\Munich{18}
\def\ExcellenceCluster{19}
\def\MPE{20}
\def\CaseWestern{21}
\def\Minnesota{22}
\def\ArtInstChicago{23}
\def\CfA{24}
\def\BCCP{25}

\begin{document}

\title{Constraints on Cosmology from the
Cosmic Microwave Background Power Spectrum of the 2500 $\deg^2$ SPT-SZ Survey}

\slugcomment{Submitted to \apj}


\author{
  Z.~Hou,\altaffilmark{\Davis}
  C.~L.~Reichardt,\altaffilmark{\Berkeley}
  K.~T.~Story,\altaffilmark{\KICPChicago,\PhysicsUChicago}
  B.~Follin,\altaffilmark{\Davis}
  R.~Keisler,\altaffilmark{\KICPChicago,\PhysicsUChicago}
  K.~A.~Aird,\altaffilmark{\UChicago}
  B.~A.~Benson,\altaffilmark{\KICPChicago,\EFIChicago}
  L.~E.~Bleem,\altaffilmark{\KICPChicago,\PhysicsUChicago}
  J.~E.~Carlstrom,\altaffilmark{\KICPChicago,\PhysicsUChicago,\EFIChicago, \AAUChicago,\Argonne}
  C.~L.~Chang,\altaffilmark{\KICPChicago,\EFIChicago,\Argonne}
  H-M. Cho, \altaffilmark{\NIST}
  T.~M.~Crawford,\altaffilmark{\KICPChicago,\AAUChicago}
  A.~T.~Crites,\altaffilmark{\KICPChicago,\AAUChicago}
  T.~de~Haan,\altaffilmark{\McGill}
  R.~de~Putter,\altaffilmark{\JPL,\Caltech}
  M.~A.~Dobbs,\altaffilmark{\McGill}
  S.~Dodelson,\altaffilmark{\KICPChicago,\AAUChicago,\FermiLab}
  J.~Dudley,\altaffilmark{\McGill}
  E.~M.~George,\altaffilmark{\Berkeley}
  N.~W.~Halverson,\altaffilmark{\Colorado}
  G.~P.~Holder,\altaffilmark{\McGill}
  W.~L.~Holzapfel,\altaffilmark{\Berkeley}
  S.~Hoover,\altaffilmark{\KICPChicago,\PhysicsUChicago}
  J.~D.~Hrubes,\altaffilmark{\UChicago}
  M.~Joy,\altaffilmark{\NASA}
  L.~Knox,\altaffilmark{\Davis}
  A.~T.~Lee,\altaffilmark{\Berkeley,\LBNL}
  E.~M.~Leitch,\altaffilmark{\KICPChicago,\AAUChicago}
  M.~Lueker,\altaffilmark{\Caltech}
  D.~Luong-Van,\altaffilmark{\UChicago}
  J.~J.~McMahon,\altaffilmark{\Michigan}
  J.~Mehl,\altaffilmark{\Argonne,\KICPChicago}
  S.~S.~Meyer,\altaffilmark{\KICPChicago,\PhysicsUChicago,\EFIChicago,\AAUChicago}
  M.~Millea,\altaffilmark{\Davis}
  J.~J.~Mohr,\altaffilmark{\Munich,\ExcellenceCluster,\MPE}
  T.~E.~Montroy,\altaffilmark{\CaseWestern}
  S.~Padin,\altaffilmark{\KICPChicago,\AAUChicago,\Caltech}
  T.~Plagge,\altaffilmark{\KICPChicago,\AAUChicago}
  C.~Pryke,\altaffilmark{\KICPChicago,\EFIChicago,\AAUChicago,\Minnesota}
  J.~E.~Ruhl,\altaffilmark{\CaseWestern}
  J.T.~Sayre,\altaffilmark{\CaseWestern}
  K.~K.~Schaffer,\altaffilmark{\KICPChicago,\EFIChicago,\ArtInstChicago}
  L.~Shaw,\altaffilmark{\McGill}
  E.~Shirokoff,\altaffilmark{\Berkeley}
  H.~G.~Spieler,\altaffilmark{\LBNL}
  Z.~Staniszewski,\altaffilmark{\CaseWestern}
  A.~A.~Stark,\altaffilmark{\CfA}
  A.~van~Engelen,\altaffilmark{\McGill}
  K.~Vanderlinde,\altaffilmark{\McGill}
  J.~D.~Vieira,\altaffilmark{\Caltech}
  R.~Williamson,\altaffilmark{\KICPChicago,\AAUChicago} and
  O.~Zahn\altaffilmark{\BCCP}
 }

\altaffiltext{\Davis}{Department of Physics, University of California, One
Shields Avenue, Davis, CA, USA 95616}
\altaffiltext{\Berkeley}{Department of Physics, University of California,
Berkeley, CA, USA 94720}
\altaffiltext{\KICPChicago}{Kavli Institute for Cosmological Physics, University
of Chicago, 5640 South Ellis Avenue, Chicago, IL, USA 60637}
\altaffiltext{\PhysicsUChicago}{Department of Physics, University of Chicago,
5640 South Ellis Avenue, Chicago, IL, USA 60637}
\altaffiltext{\UChicago}{University of Chicago, 5640 South Ellis Avenue,
Chicago, IL, USA 60637}
\altaffiltext{\EFIChicago}{Enrico Fermi Institute, University of Chicago, 5640
South Ellis Avenue, Chicago, IL, USA 60637}
\altaffiltext{\AAUChicago}{Department of Astronomy and Astrophysics, University
of Chicago, 5640 South Ellis Avenue, Chicago, IL, USA 60637}
\altaffiltext{\Argonne}{Argonne National Laboratory, 9700 S. Cass Avenue,
Argonne, IL, USA 60439}
\altaffiltext{\NIST}{NIST Quantum Devices Group, 325 Broadway Mailcode 817.03,
Boulder, CO, USA 80305}
\altaffiltext{\McGill}{Department of Physics, McGill University, 3600 Rue
University, Montreal, Quebec H3A 2T8, Canada}
\altaffiltext{\FermiLab}{Center for Particle Astrophysics, Fermi National 
Accelerator Laboratory, Batavia, IL, USA 60510}
\altaffiltext{\JPL}{Jet Propulsion Laboratory, California Institute of Technology, 
4800 Oak Grove Drive, Pasadena, CA, USA 91109}
\altaffiltext{\Caltech}{California Institute of Technology, MS 249-17, 1216 E.
California Blvd., Pasadena, CA, USA 91125}
\altaffiltext{\Colorado}{Department of Astrophysical and Planetary Sciences and
Department of Physics, University of Colorado, Boulder, CO, USA 80309}
\altaffiltext{\NASA}{Department of Space Science, VP62, NASA Marshall Space
Flight Center,Huntsville, AL, USA 35812}
\altaffiltext{\LBNL}{Physics Division, Lawrence Berkeley National Laboratory,
Berkeley, CA, USA 94720}
\altaffiltext{\Michigan}{Department of Physics, University of Michigan, 450
Church Street, Ann  Arbor, MI, USA 48109}
\altaffiltext{\Munich}{Department of Physics,
Ludwig-Maximilians-Universit\"{a}t, Scheinerstr.\ 1, 81679 M\"{u}nchen, Germany}
\altaffiltext{\ExcellenceCluster}{Excellence Cluster Universe, Boltzmannstr.\ 2,
85748 Garching, Germany}
\altaffiltext{\MPE}{Max-Planck-Institut f\"{u}r extraterrestrische
Physik, Giessenbachstr.\ 85748 Garching, Germany}
\altaffiltext{\CaseWestern}{Physics Department, Center for Education and
Research in Cosmology and Astrophysics, Case Western Reserve
University,Cleveland, OH, USA 44106}
\altaffiltext{\Minnesota}{Department of Physics, University of Minnesota, 116
Church Street S.E. Minneapolis, MN, USA 55455}
\altaffiltext{\ArtInstChicago}{Liberal Arts Department, School of the Art
Institute of Chicago, 112 S Michigan Ave, Chicago, IL, USA 60603}
\altaffiltext{\CfA}{Harvard-Smithsonian Center for Astrophysics, 60 Garden
Street, Cambridge, MA, USA 02138}
\altaffiltext{\BCCP}{Berkeley Center for Cosmological Physics, Department of
Physics, University of California, and Lawrence Berkeley National Laboratory,
Berkeley, CA, USA 94720}

\begin{abstract}
We explore extensions to the \lcdm\ cosmology using
measurements of the cosmic microwave background (CMB) from the
recent SPT-SZ survey, along with data from \wseven{} and
measurements of \ho{} and BAO.  We check for consistency 
within \lcdm{} between these datasets, and find some tension.  The CMB alone gives weak support to physics beyond \lcdm, due to a slight trend relative to \lcdm{} of decreasing power towards smaller angular scales.
While it may be due to statistical fluctuation, this trend could
also be explained by several extensions.  We consider running of the
primordial spectral index ($\nrun$), as well as two
extensions that modify the damping tail power (the primordial
helium abundance \yp{} and the effective number of neutrino species
\neff{}) and one that modifies the large-scale power due to the
integrated Sachs-Wolfe effect (the sum of neutrino masses \sumnu).
These extensions have similar observational consequences and are
partially degenerate when considered simultaneously. 
Of the 6 one-parameter extensions considered, we find
CMB to have the largest preference for \nrun with $-0.046<\nrun<-0.003$ at 95\% confidence, which strengthens to a
$2.7\,\sigma$ indication of $\nrun<0$ from CMB+BAO+\ho{}.
Detectable $\nrun\ne0$ is difficult to explain in the context of
single-field, slow-roll inflation models.  We find $\neff=3.62\pm0.48$ for the CMB, which tightens to $\neff=3.71\pm0.35$ from CMB+BAO+\ho{}.
Larger values of \neff{} relieve the mild tension between CMB, BAO and \ho{}.  
When the Sunyaev-Zel’dovich selected galaxy cluster abundances (\sptcl) data are also included, we obtain $\neff=3.29\pm0.31$.  Allowing for \sumnu gives a $3.0\,\sigma$
detection of $\sumnu>0$ from CMB+BAO+\ho+\sptcl.  The median
value is $(0.32\pm0.11)$\,eV, a factor of six above the lower
bound set by neutrino oscillation observations.  All datasets except
\ho{} show some preference for massive neutrinos; data combinations
including \ho{} favor nonzero masses only if BAO data are also
included.  We also constrain the two-parameter extensions $\neff+\sumnu$ and $\neff+\yp$ to explore constraints on additional light species and big bang nucleosynthesis respectively.
\end{abstract}


\section{Introduction}
\label{sec:intro}
Measurements of the damping tail of the cosmic microwave background (CMB) power spectrum 
provide new insights into the spectrum of the primordial density
fluctuations, the contents of the pre-recombination plasma, and, through the
effect of gravitational lensing, the low-redshift properties of the universe. 
 In this paper, we study the constraints that can be placed on extensions to the standard
 \LCDM{} cosmological model by recent South Pole Telescope (SPT) CMB power spectrum measurements presented in a companion paper
\citep[hereafter S12]{story13} and complementary lower-redshift probes.
For each extension, we discuss both the cosmological constraints and the
physical origin of these constraints.  

As shown by S12, the SPT bandpowers lead to constraints on the six standard \LCDM{} parameters that are consistent with and comparable to those from \wseven{} \citep{larson11,komatsu11}. 
The agreement is especially remarkable given that the smaller angular scales probed by the SPT data are sensitive to new physical effects that are unimportant on the larger
angular scales probed by \wmap.  
The combination of SPT and \wseven{} data can be used to exploit these additional effects in order to improve constraints on the \LCDM{} model and possible
extensions.

Measurements of the damping tail of the CMB power spectrum are sensitive to 
the low-redshift universe through the gravitational lensing of the
CMB as well as the angular-diameter distance to the last scattering surface. 
The lensing sensitivity introduces a second constraint on the late-time expansion rate that breaks the geometric degeneracy \citep{bond97,zaldarriaga97c} that exists in large-scale CMB data. 
Measurements of CMB lensing, like that recently reported by S12, can be used to
constrain the mean curvature of the observable universe.

More importantly, the SPT data expand the angular range over which the 
CMB temperature power spectrum is probed, leading to improved measurements of the damping scale due to photon diffusion \citep{silk68} 
and the positions of acoustic peaks.  Past CMB experiments have found evidence for a trend of decreasing power at high multipoles, 
relative to the predictions of the \LCDM{} model conditioned on CMB
measurements at $\ell \la 1000$ \citep{hamann10, dunkley11,
  keisler11}, a trend that persists with the inclusion of the full-survey SPT bandpowers.  This trend may be thought of as a scale-dependent tilt in the CMB power 
spectrum, and we explore five one-parameter extensions to \LCDM{} that effectively allow for such a scale-dependent tilt.
These extensions are: allowing massive neutrinos (\sumnu), introducing  ``running'' of the spectral index of the primordial 
power spectrum (\nrun), varying the effective number of neutrino species (\neff), allowing the helium abundance (\yp) to depart 
from the predictions of big bang nucleosynthesis (BBN), and allowing non-zero early dark energy (\omegae).  A unifying theme in this 
paper will be a preference for model extensions that can accommodate a scale-dependent tilt that becomes increasingly red at higher multipoles.

The CMB damping tail measurements do not add sensitivity to the dark energy equation of state $w$, since dark energy becomes important at late times and this parameter has little effect on the lensing amplitude. 
Therefore, we do not present constraints on \LCDM+$w$ in this paper, although we sometimes include $w$ when exploring the effect of parameter degeneracies on the constraints for other parameters.

We explore two 2-parameter extensions to the \lcdm{} model which are
physically and theoretically motivated: \lcdm+\neff+\yp{} and
\lcdm+\neff+\sumnu{}.  The first case is an interesting consistency
test of BBN.  Both \yp{} and \neff{} are primarily constrained by
measurements of the damping scale, however both can be
constrained simultaneously since varying \neff{} has the secondary effect of inducing a phase shift in the acoustic oscillations. 
The second extension, \lcdm+\neff+\sumnu{}, is a test of the possible
existence of sterile neutrinos, which have been postulated with masses in the eV range in order to explain some neutrino oscillation anomalies (e.g., \citealt{aguilar01, aguilar-arevalo10, mention11}).

While all of these extensions can be constrained with
CMB data alone, low-redshift measurements add constraining power and
provide consistency checks to the CMB data.  We incorporate
measurements of the Hubble constant (\ho), the baryon acoustic oscillation (BAO) feature of the
matter power spectrum,  luminosity distances from supernovae (SNe), the
matter power spectrum inferred from a redshift catalog of luminous red
galaxies (LRGs), and the abundance of Sunyaev-Zel'dovich (SZ)-selected clusters.  
We pay particular attention to the BAO and \ho{}
measurements, specifically the \citet{anderson12} BAO measurement at
$z=0.57$ and the \citet{riess11} determination of \ho, because they
are highly precise constraints that interact in an interesting way
with those from the CMB.  We quantify the level of consistency between
the CMB, BAO, and \ho{} datasets in the context of the \LCDM{} model,
and we study which model extensions are preferred by the CMB-only and
combined CMB+BAO+\ho{} datasets.

Throughout this work,  we make a point of explaining the physical mechanisms behind the constraints we find.  That is, we address questions such as, ``{\em How}
are the neutrino masses constrained?''  and ``{\em What} features in the data explain the preference for nonzero masses?" 
We also clarify what other extensions might explain the same features in the data.  

The structure of the paper is as follows.  
We briefly describe the analysis methodology and data from the SPT and other
experiments used in this analysis in \S~\ref{sec:prelim}. 
The physical origins of the CMB constraints on the standard \LCDM{} cosmology are presented in \S~\ref{sec:lcdm}.   
We discuss the consistency between datasets, and the
evidence for extensions to the \LCDM{} model  in \S~\ref{sec:consistency}. 
We study the physical origins of the constraint on non-zero curvature in \S~\ref{sec:curvature}.
We consider  massive neutrinos in \S~\ref{sec:mnu} and running of the scalar index in \S~\ref{sec:nrun}.
Other single-parameter extensions -- \neff, \yp, and \omegae{} -- 
 are discussed in \S~\ref{sec:early}.  
We consider  2-parameter extensions in \S~\ref{sec:two}.
Finally, we conclude in \S~\ref{sec:conclusions}.

\section{Preliminaries}
\label{sec:prelim}

Here we introduce the datasets that will be used in the
cosmological parameter fitting. 
We also describe the Markov Chain Monte Carlo (MCMC) methods that will be used to fit cosmological parameters. 

\subsection{CMB Power Spectrum Measurements}
\label{subsec:cmb}

\begin{figure*}
\begin{center}
   \includegraphics[width=0.90\textwidth, trim=1cm 12.5cm 1cm
2.5cm]{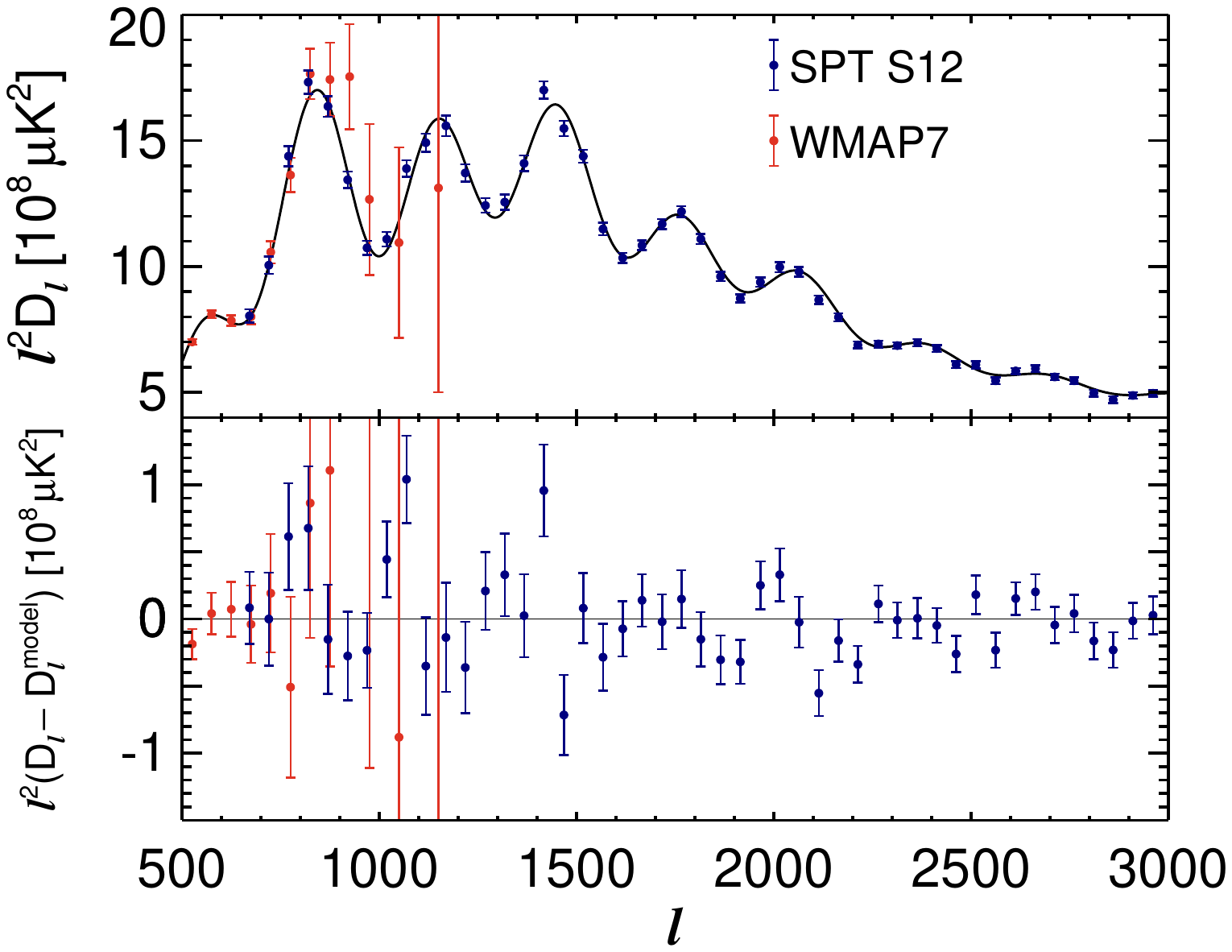}
\caption{\textit{Upper:} the SPT bandpowers from S12 (dark blue) and \wseven{} TT bandpowers (red) with $\ell \in [500, 3000]$.  
The solid curve shows the best-fit \LCDM{} model including the best-fit extragalactic foreground components with priors for the combination of SPT and \wseven{} data.  An additional normalization factor of $\ell^2$ has been applied to enhance the feature of the CMB damping tail.  \textit{Lower:} the residual between the data and the best-fit \LCDM{} model with extragalactic foregrounds.  As discussed in S12, the data are consistent with the standard cosmology.}
\label{fig:w7s12power}
\end{center}
\end{figure*}

In this work, we will use CMB bandpowers from \wmap{} and SPT, and will refer to the combined dataset as the ``CMB'' dataset. 
The \wmap{} data we use are the measurements of the CMB power spectrum at $2 \le \ell \le 1200$
from the \WMAP{} 7-year release \citep{larson11}.\footnote{A few days prior to submission, the WMAP collaboration released the 9-year data \citep{hinshaw12, bennett12}.  Due to time constraints and the consistency of the 9-year data with the 7-year data, we chose not to incorporate the new data into this work.} 
We will focus on the new SPT power spectrum results presented by S12, hereafter referred to as the SPT dataset.  
The SPT bandpowers  represent a
two-fold improvement over the best previous measurement \citep[K11]{keisler11} of the CMB power
spectrum at $\ell > 800$.    We briefly summarize
the analysis of S12 and review the modeling of the contributions from
extragalactic foregrounds.

The full SPT-SZ  survey was completed in
November 2011 and covers $\sim$2500\,\degs{} in three frequency bands at 95, 150, and $220\,$GHz.  
The SPT bandpowers used in this work are based on the
150\,GHz data from the complete survey.  As in earlier SPT power spectrum
analyses, the bandpowers are calculated using a pseudo-\cl{} method. 
The bandpowers cover the range $650 < \ell < 3000$ and are sample-variance limited at $\ell
< 2900$.  The covariance matrix contains terms for the sample and noise variances
as well as beam and calibration uncertainties.  Beams are estimated through a
combination of observations of planets and bright synchrotron sources in the survey fields.  The
bandpowers are calibrated with an accuracy of 2.6\% in power by comparing the
SPT and \WMAP{} 7-year power over the multipole range $\ell \in [650,1000]$.

We model contributions from extragalactic foregrounds to the SPT bandpowers exactly as in S12.  Using the
notation $D_\ell \equiv \ell (\ell+1) C_\ell /(2\pi) $ these are given by
\begin{equation}
D_\ell^{\rm{fg}} = D^{\rm PS}_{3000} \left(\frac{\ell}{3000}\right)^2 + D^{\rm
CL}_{3000} \left(\frac{\ell}{3000}\right)^{0.8} + D^{\rm SZ}_{3000}
f^{\rm{SZ}}_{\ell}
\end{equation} where
$D^{\rm PS}_{3000}$,
$D^{\rm CL}_{3000}$, and $D^{\rm SZ}_{3000}$ are the amplitudes at $\ell=3000$
for the Poisson point source power, clustered point source power, and total
SZ power spectrum respectively, and $f^{\rm{SZ}}_\ell$ gives the $\ell$-dependence
of the SZ power spectrum as modeled in \citet{shaw10}.  We apply Gaussian
priors on these amplitudes of $D^{\rm PS}_{3000} =
19.3 \pm 3.5\, \microKsq$, $D^{\rm CL}_{3000} = 5.0 \pm 2.5 \,\microKsq$, and
$D^{\rm SZ}_{3000} = 5.5 \pm 3.0\, \microKsq$ \citep{shirokoff11}. 
We include the SZ term in the \wseven{} likelihood calculation, but we do not include the two point source terms since the  \wseven{} likelihood code already marginalizes over appropriate point source terms. 
We have tested that all parameter constraints are insensitive to loosening the
assumed extragalactic foreground priors.

As was argued by S12, we expect galactic foregrounds to have negligible impact on the SPT bandpowers 
and do not include a contribution from them in the bandpower modeling.  The SPT and \wseven{} bandpowers are shown in the upper panel of Figure~\ref{fig:w7s12power} plotted with the best-fit \LCDM{} model, including the best-fit extragalactic foreground model consistent with the prior presented above.  The residual between the data and this best-fit model is shown in the lower panel.

\subsection{External Datasets}
\label{subset:external}

Low-redshift observations can substantially inform the constraints on some cosmological parameters, 
particularly those parameters with impact at late
times such as the mean curvature of the observable universe.  We
include these low-redshift datasets in many of the cosmological
constraints presented in this work.  These datasets are described
below.

\subsubsection{Baryon Acoustic Oscillations} 

BAO measurements use the spatial correlation of galaxies to constrain the comoving size of the sound horizon at the epoch when baryons and photons decouple, $r_s(z_{\rm drag})$.
This redshift, $z_{\rm drag}$, is defined by $\tau_b(\eta(z_{\rm drag}))=1$, where 
\begin{equation}
\tau_b(\eta) \equiv \int^{\eta_0}_{\eta} a \sigma_T n_e d\eta'/(1+\frac{3\rho_b}{4\rho_\gamma}) \,.
\end{equation}
Here $\eta$ is conformal time with $\eta_0$ the conformal time today, $a$ is the scale factor,
$\rho_b$ and $\rho_\gamma$ are the baryon and photon densities, $n_e$ is the number density of free electrons,
and $\sigma_T$ is the Thomson cross section.  
The BAO feature in the galaxy correlation function can be observed as a function of both angular and redshift separations.
Thus galaxy surveys can, in principle, constrain both
$D_A(z)/r_s(z_{\rm drag})$ and $H(z) r_s(z_{\rm drag})$.  
However, the current generation of BAO observations typically reports constraints on a hybrid quantity,
$r_s(z_{\rm drag})/D_V(z_{\rm eff})$ (henceforth $r_s/D_V$), where $D_V^3(z) = D_A^2(z)
cz/H(z)$, $c$ is the speed of light and $z_{\rm eff}$ is the effective redshift
for the population of galaxies in the survey.  

We use measurements of $r_s/D_V$ from the
following experiments: SDSS at $z_{\rm eff} = 0.35$
\citep{padmanabhan12}, BOSS at $z_{\rm eff}=0.57$ \citep{anderson12},
and WiggleZ at $z_{\rm eff} = 0.44, 0.6$, and $0.73$ \citep{blake11}.
All of these experiments report $r_s/D_V$ (or its inverse) using the
fitting formula for $z_{\rm drag}$ presented by \citet{eisenstein98}.  We
numerically calculate $r_s(z_{\rm drag})$ with CAMB \citep{lewis00}.
To account for the difference between $z_{\rm drag}$ as calculated by
the fitting formula with $z_{\rm drag}$ as calculated by CAMB, \citet{hamann10b}
find it necessary to rescale $r_s$ by a factor 154.66/150.82.
\citet{mehta12} show this constant rescaling adequately accounts for
the differences in $z_{\rm drag}$ over the range of baryon density and
matter density allowed by \wseven. 
We therefore apply
this rescaling to the reported $r_s/D_V$ data for all parameter fitting.

\subsubsection{The Hubble Constant}

In contrast to the BAO approach to measuring cosmological distances, which
starts from a known distance scale at $z
\sim 1100$, measurements of the Hubble constant are based on the classic
distance ladder which is built outwards from the solar system. 
\citet{riess11} use 253 Type 1a SNe to reach out into the Hubble flow and calibrate their
luminosity distances with over 600 Cepheid variable stars in the host galaxies
of eight nearby Type 1a SNe.  The Cepheids are calibrated in turn with three
different methods, including 13 with trigonometric parallax.  
Accounting for both statistical and systematic sources
of uncertainty, \citet{riess11} conclude that
$H_0 = 73.8 \pm 2.4$ km/s/Mpc.

\citet{freedman12} recently released a new analysis of the HST Key Project, with
a {\it Spitzer}-based calibration of the Cepheid distance scale, in which they find
$H_0 = 74.3 \pm 2.1$ km/s/Mpc. 
As we had already completed many of the Markov chains used in this work, we
chose not to rerun with this slightly improved constraint.
We note that the tension in the \lcdm{} model between BAO and \ho{} discussed later would mildly
increase with this new data since the measured \ho{} value increases by 
$0.2\,\sigma$, and the uncertainty decreases by  $\sim$\,10\%. 

\subsubsection{Luminosity Distances from Supernovae}

Type Ia SNe, employed as standardizable candles, remain the most precise means
to extend the distance ladder to redshifts that are sensitive to cosmic
acceleration.  
We use a sample of 427 SNeIa assembled by \citet{conley11} to
constrain the luminosity distance. 
We follow the treatment in \citet{sullivan10} to handle the dependence of supernovae absolute magnitude on host galaxy stellar mass. 
We follow the approach of \citet{conley11} to marginalize over systematic uncertainties related to the uncertain intrinsic properties of supernovae.

\subsubsection{Matter Power Spectrum from SDSS DR7 LRG}

\citet{reid10} use a catalog of luminous red galaxies from the seventh data
release of the Sloan Digital Sky Survey (SDSS DR7) to reconstruct the dark
matter halo density field  at $z \la 0.4$ over nearly 8,000\,\degs{}.
They use the reconstructed density field to determine the matter power spectrum over the range $0.02 < k < 0.2\, h$Mpc$^{-1}$.  
We use their published likelihood code to incorporate this information in cosmological constraints. 
The information is primarily from the shape of the spectrum since the unknown halo masses, and therefore unknown biases, prevent the
amplitude of the measured matter power spectrum from being used.

\subsubsection{SPT SZ-selected Clusters}

\citet[R12]{reichardt13} present a catalog  of 224 galaxy cluster candidates selected by their SZ
significance from the first 720 \degs\ of sky surveyed by the SPT. 
As in R12, we use only the 100 cluster candidates at $z > 0.3$ with a significance
greater than five. The optical properties of these systems are detailed by \citet{song12b}. 
X-ray $Y_X$ measurements of 14 of the clusters are also used in the
cosmological constraints; these measurements have been described by
\citet{andersson11} and \citet{benson13}.
For brevity, we only include the cluster data in models showing a clear correlation between the model extension and the quantity $\sigma_8(\Omega_m/0.25)^{0.47}$. \citet{vikhlinin09} have shown that this quantity is well constrained by cluster abundance data.  Given a CMB+BAO+$H_0$ prior, we find such a correlation, and thus include the cluster data, for two extensions, $\sumnu$ and $N_{\rm eff}$.

\subsection{Parameter Estimation Methodology}
\label{subsec:method}

Posterior probability distributions for the 
parameters are calculated using Markov Chain Monte Carlo (MCMC) methods
\citep{christensen01} with the publicly available {\textsc
CosmoMC}\footnote{http://cosmologist.info/cosmomc - January 2012 version}
package \citep{lewis02b}.  We have modified the code to include the foreground
terms from \S~\ref{subsec:cmb} and the extensions to \LCDM{} detailed
in later sections; these modules and instructions for compiling them are
available at the SPT
website\footnote{http://pole.uchicago.edu/public/data/story12/ }.
As can be inferred from Figure~\ref{fig:w7s12power}, the \wseven{} bandpowers are noise dominated on the multipoles measured by the SPT. The correlation between the two datasets is thus negligible, and we treat the two datasets as independent in this work.

We use the
high\_accuracy\_default option in {\textsc CosmoMC}, and by default call the
January 2012 version of CAMB \citep{lewis00} to calculate the CMB and matter
power spectra.  Unless otherwise noted (for instance in \S~\ref{sec:mnu_constraints}), we will report parameter constraints
based on the median value of the posterior distribution with $1\,\sigma$ error
bars defined based on the interval that includes the central 15.85\% to 84.15\% of the MCMC samples.
For cases when we compare a parameter to some fiducial value (e.g. an extension parameter to its \lcdm-consistent value), we report--unless otherwise noted--a significance in $\sigma$ based on the Gaussian z-score of the probability above (or below) the fiducial value.  

We will present parameter constraints on a spatially flat,
gravitationally lensed \LCDM{} model along with selected one- and 
two-parameter extensions to this model.  The properties of the 
standard \LCDM{} model are specified by six parameters: the optical depth to reionization ($\tau$), the
baryon density ($\Omega_b h^2$), the cold dark matter density ($\Omega_c
h^2$), the angular scale of the sound horizon at last scattering
($\theta_s$), the scalar spectral index ($n_s$), and the amplitude of
primordial scalar perturbations at wavenumber $k = 0.05$\,Mpc$^{-1}$
($\Delta_R^2$).  Throughout this paper, we will use the notation
$\omega_x \equiv \Omega_x h^2$.  Since $\Omega_x \equiv \rho_x/\rho_c$,
where $\rho_c = 3H_0^2/(8\pi {\rm G})$ is the critical density,
$\omega_x$ is independent of the Hubble constant.  We define the total
matter density as the sum of baryons and cold dark matter, so that
$\omega_m = \omega_b + \omega_c$.  All densities are evaluated at the
present epoch.

To speed up parameter estimation, we make use of the Boltzmann code
emulator PICO \citep{fendt07a,fendt07b}.  PICO works by empirically modeling the
CMB and matter power spectra as sums of polynomials, with the (cosmological
parameter-dependent) polynomial coefficients determined by fitting them to
output from CAMB.  We have trained PICO for a 10-parameter model that includes
the six \LCDM{} parameters as well as $Y_{\rm P}$, \neff{},
$\sumnu$, and $dn_s/d\ln k$.  We use PICO instead of CAMB when working with
any subset of this model space.  Evaluating the power spectrum with PICO is
several orders of magnitude faster than doing so with CAMB, and is even more
accurate due to its training on CAMB runs with higher numerical accuracy
than the high\_accuracy\_default setting.  The PICO code and results of this
training are available at the PICO
website.\footnote{\url{https://sites.google.com/a/ucdavis.edu/pico}}

\section{\LCDM{} results}
\label{sec:lcdm}

\begin{figure}
\begin{center}
   \includegraphics[width=0.48\textwidth, trim=4.2cm 12.8cm 5.8cm
4.8cm]{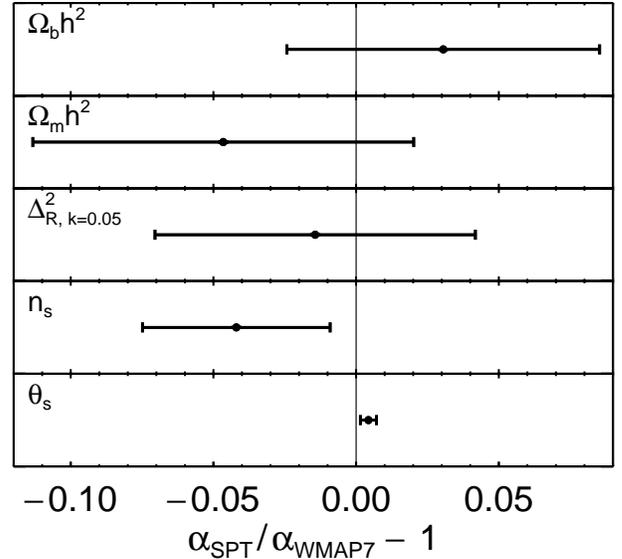}
\caption{This figure illustrates the agreement between parameter values derived
from the SPT and \wseven{} bandpowers. 
For each \LCDM{} parameter, we plot the fractional change in the parameter,
defined by ($\alpha_{\rm SPT}/\alpha_{WMAP7} - 1$), and the error bar for this
quantity.  Following S12, the amplitude of scalar perturbations,
$\Delta_R^2$, is evaluated at $k=0.05$ Mpc$^{-1}$.
The optical depth to reionization is not plotted, as a \wseven-based
prior is used in the fits to the SPT bandpowers.  The most significant shift is to the
sound horizon scale which increases by $1.5\,\sigma$.
}
\label{fig:pardiff}
\end{center}
\end{figure}

In this section, we consider the SPT-only and \wseven-only constraints on the \LCDM{} model parameters. 
As shown by S12 and Figure~\ref{fig:pardiff}, the parameters estimated from the two
datasets are consistent.  
Here we focus on the physical origins of the baryon and matter density
constraints derived from different angular scales probed by \wseven{}
and SPT.  
For each case, we begin by briefly reviewing the origins of the \wseven{}
constraints (see, e.g.,~\citealt{hu02b}) and then discuss the origins of the SPT constraints. 
As we will see in later sections, understanding these mechanisms
provides a foundation from which to explore other datasets and model
extensions. 

The \wseven{} constraint on the matter density $\omega_m$ derives from
differences in the acoustic oscillations during the radiation-dominated 
and matter-dominated epochs.  Modes that ``enter the
horizon''\footnote{The horizon expands with time; a Fourier mode
  enters the horizon when its wavelength equals the horizon size.}
during radiation domination receive a boost to their amplitude from
the decay of the gravitational potential that occurs during their
first compression into potential wells.  In an expanding universe, the
plasma would need to collapse at the free-fall (pressureless) rate to
preserve the potential.  The pressure support of the plasma slows the
collapse, leading to a decaying gravitational potential.  The decay
boosts the oscillation amplitude since modes gain more energy during
the compression into the initially larger potential well than they
later lose during the expansion out of the much reduced potential well.
Modes entering during matter domination do not see such a boost, since
the potential is predominantly sourced by pressureless dark matter.
The difference in the oscillation amplitudes allows us to determine
the angular scale separating modes that entered the horizon during
matter domination from those that entered during radiation domination,
$\theta_{\rm eq}$, which can be related to the redshift of
matter-radiation equality, $z_{\rm eq}$.  Because the radiation
density, $\omega_{\rm rad}$, is completely specified in the \LCDM{}
model by the well-known temperature of the microwave background today
and assumptions about neutrino production, this measurement of $z_{\rm
  eq}$ corresponds to a determination of $\omega_m$ through the
relation $1 + z_{\rm eq} = \omega_m/\omega_{\rm rad}$.

The angular scales probed by the SPT bandpowers correspond to modes that entered the horizon in the radiation-dominated era, since the transition occurs around $\ell_{\rm eq} = \pi/\theta_{\rm eq} \simeq 434$. 
Thus the SPT bandpowers are largely insensitive to $\omega_m$ through the mechanism described above. 
To confirm this, we have tested removing the largest scales (third peak, $\ell < 1000$) from the SPT bandpowers and find only a minimal (5\%) degradation of the SPT-only $\omega_m$ constraint. 
The SPT data are instead sensitive to the amplitude of gravitational lensing which in turn depends on $\omega_m$. 
Gravitational lensing of the CMB smooths out the peak-trough structure of the
CMB power spectrum, allowing lensing to be detected in the SPT power spectrum.
The magnitude of the lensing potential power spectrum increases with  $\omega_m$. 
To support the claim that the SPT bandpowers determine $\omega_m$ through its
effect on the amplitude of the lensing power spectrum, we have tested removing
the lensing amplitude information.  We accomplish this by rescaling the lensing
power spectrum, $C_\ell^{\phi\phi}$, by a free parameter,  $A_L$,  defined by:
\begin{equation}
C_\ell^{\phi\phi} \rightarrow A_{L} C_\ell^{\phi\phi}.
\label{eq:A_L}
\end{equation}
Marginalizing over $A_L$ doubles the uncertainty on $\omega_m$, degrading the constraint from $\omega_m = 0.1286\pm0.0071$ to
$\omega_m = 0.129\pm0.013$.  
When SPT and \wseven{} data are combined, $\omega_m$ is 
primarily constrained through its impact on $\theta_{\rm eq}$.

The \wseven{} constraint on $\omega_b$ derives from the relative heights of the
even and odd acoustic peaks.  The baryon density affects the plasma's pressure,
which impacts the acoustic oscillations.  Changing the
pressure has opposite effects on compression into gravitational potential
wells (odd peaks) versus compression onto gravitational potential hills (even
peaks).  The net result is that increasing $\omega_b$ raises the odd peaks
relative to the even peaks.  

The SPT constraint on baryon density is primarily derived from the same even/odd peak modulation. 
Due to the decay of the gravitational potentials, the even/odd modulation of peak heights is suppressed on smaller scales. 
The dominant constraining power in the SPT data is from the height of the third acoustic peak. 
Increasing $\omega_b$ at fixed $z_{\rm eq}$ and $\theta_s$ increases the height of the third acoustic peak relative to higher peaks.
If we remove the SPT bandpowers over the third acoustic peak (dropping all bandpowers at $\ell < 1000$), the SPT-only constraint on $\omega_b$ degrades by nearly a factor of 2 from $\omega_b = 0.0230\pm0.0011$ to $\omega_b = 0.0236\pm 0.0020$.

The damping scale provides a second (weaker) lever arm to constrain the baryon density from the SPT data. 
Photons are imperfectly
coupled to baryons in the primordial plasma, and the resulting photon diffusion damps the acoustic oscillations. 
This photon diffusion is described by the diffusion length
$r_d$, the root mean squared comoving distance a photon has traveled over
the time in which the scale factor grows from 0 to $a_*$ (the scale factor at decoupling).  The diffusion
length is given approximately by
\be
\label{eqn:rd}
r_d^2 = \pi^2\int_0^{a_*} \frac{da}{a^3\sigma_T n_e
  H}\left[\frac{R^2 + \frac{16}{15}\left(1+R\right)}{6(1+R^2)}\right].
\ee
Here $n_e$ is the number density of free electrons, $\sigma_T$ is
the Thompson cross-section, and $R = 3\rho_b/(4\rho_\gamma)$.  The factor
in square brackets is due to the directional and polarization
dependence of Thompson scattering \citep{kaiser83,zaldarriaga95}.  The
diffusion length is affected by $\omega_b$ through the number
density of free electrons, $n_e(z) \propto X_e(z) \omega_b \left(1-\yp
\right)$, where $X_e(z)$ is the fraction of Hydrogen atoms that are
ionized and \yp{} is the primordial fraction of baryonic mass in helium.

We quantify the importance of the SPT damping scale measurement in constraining the baryon density by introducing \yp{} as a free parameter.
At fixed $\omega_b$, varying \yp{} only impacts the free electron density and hence the diffusion scale.  
When we allow $\yp$ to vary freely, the uncertainty on $\omega_b$ slightly increases from $0.0011$ to $0.0013$. 
Thus, the primary sensitivity of the SPT data to the baryon density is from the measurement of the third acoustic peak rather than damping scale.

The SPT constraints on the other \lcdm{} model parameters are based on 
 similar physical mechanisms to the \wseven{} constraints.  The
angular size of the sound horizon, $\theta_s$ is constrained via the
peak locations, while the primordial parameters \ns{} and $\Delta_{\rm
  R}^2$ are constrained through their impact on the amplitude and tilt
of the CMB power spectrum.  As mentioned in S12, the SPT bandpowers do
not constrain $\tau$; we instead apply a prior based on the \wseven{}
polarization measurements.

We note that there is a trend in the CMB data for decreasing power at higher multipoles, relative to fits to the \LCDM{} model using data from $\ell \la 1000$.  In the \lcdm{} model space, this
deficit manifests as the downward shift in \ns{} between \wseven{} and SPT
seen in Figure~\ref{fig:pardiff}.  As we shall see in later sections, model extensions
that can accommodate this trend are somewhat preferred by
the CMB data.

\section{Consistency of CMB, BAO, and \ho{} Measurements and Evidence for Extensions to the Standard Cosmological Model}
\label{sec:consistency}

Combining the BAO and $H_0$ measurements with the CMB will have important implications for the model extensions we consider. 
Here we investigate the consistency of these datasets with one another.  
Tension between two sets of data in the context of a \LCDM{} model can be caused by (1) statistical fluctuations in the measured values, 
(2) an inaccurate or incomplete accounting of the  uncertainties (or biases) in one or more of the datasets, or 
(3) the inadequacy of the cosmological model.  
We discuss the consistency between datasets in \S~\ref{subsec:dataconsistency} and discuss any evidence for extensions to the \LCDM{} model in \S~\ref{sec:extensions}.

\subsection{Consistency between datasets}
\label{subsec:dataconsistency}

\begin{figure*}
\begin{center}
    \includegraphics[width=0.33\textwidth, trim=2.0cm 13.0cm 7.1cm
2.5cm]{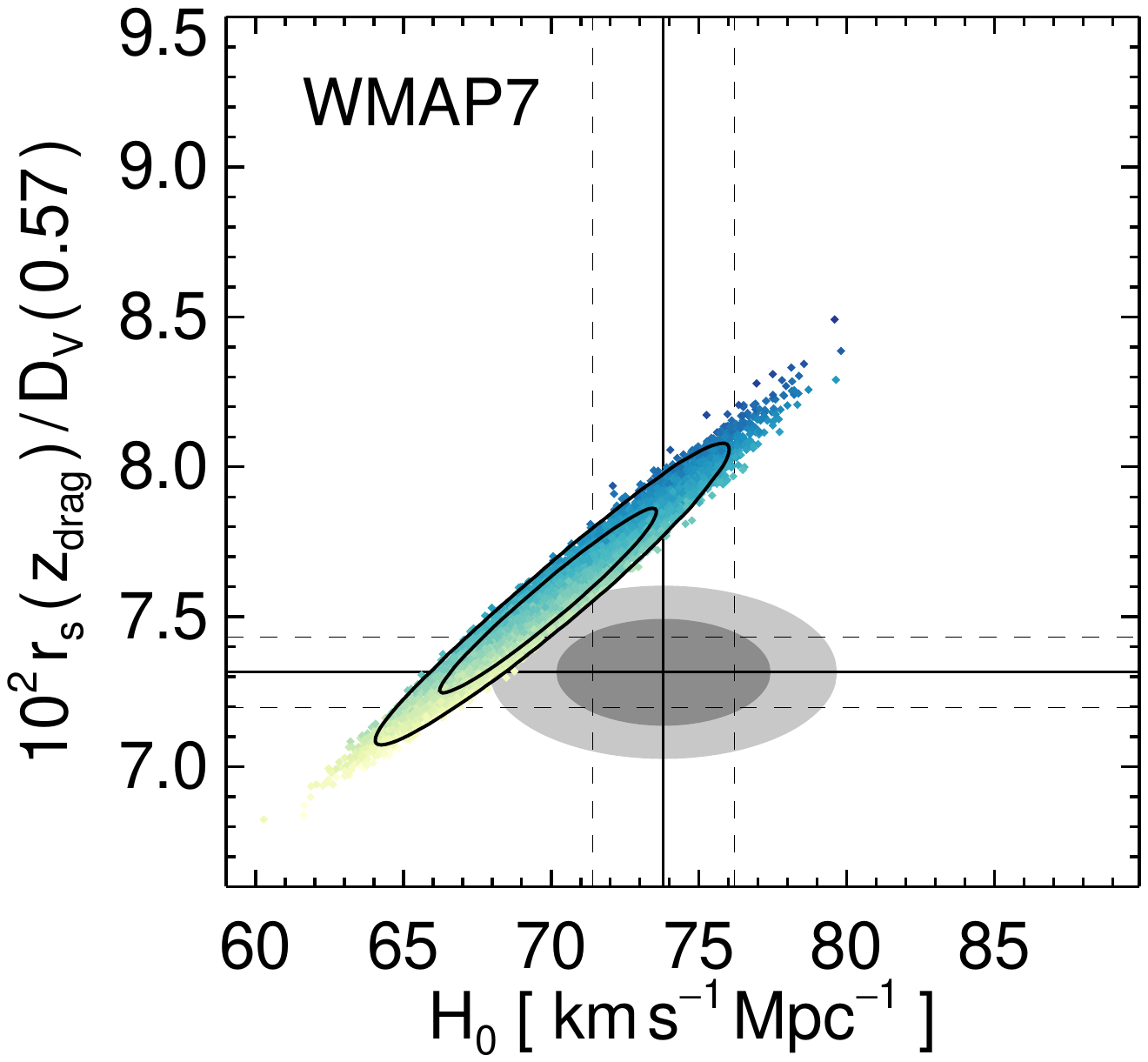}
\includegraphics[width=0.33\textwidth, trim=4.02cm 13.0cm 5.08cm
2.5cm]{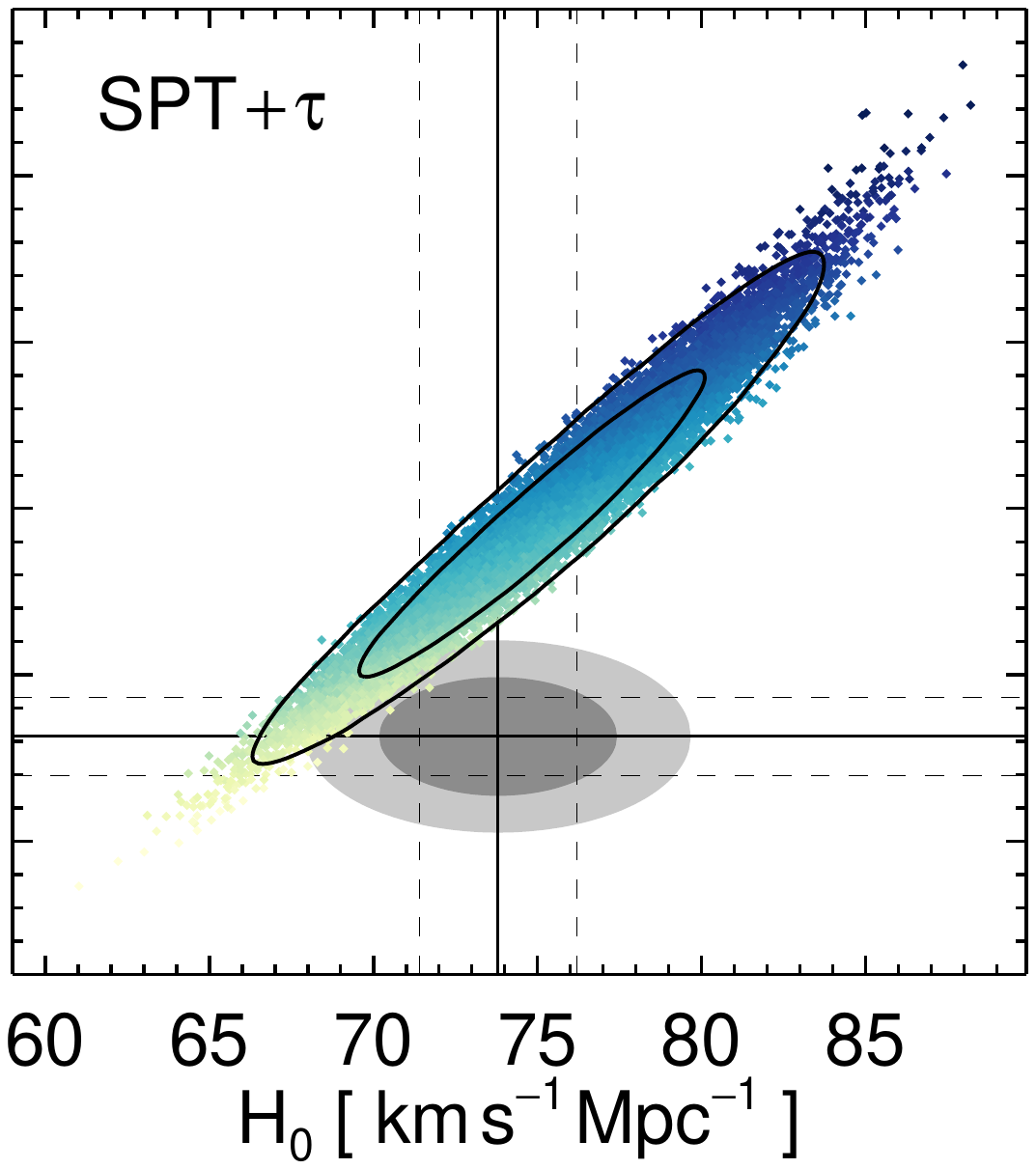}
\includegraphics[width=0.33\textwidth, trim=6.05cm 13.0cm 3.05cm
2.5cm]{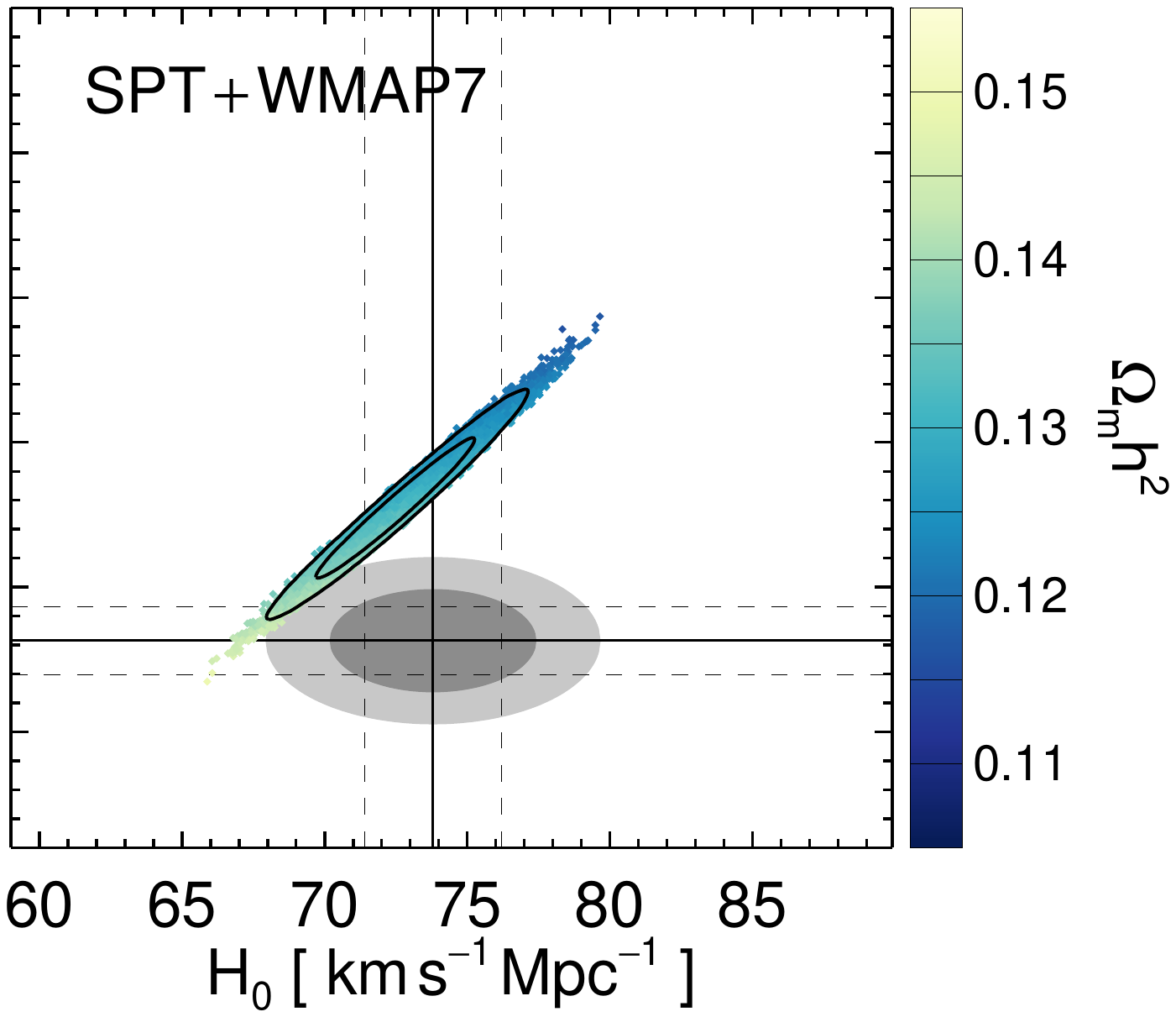}
\end{center}
\caption{
Here we illustrate the degree of consistency between the CMB data and the two external datasets which are in the most tension: BOSS and \ho. 
The solid lines mark the 1 and $2\,\sigma$ contours from the CMB data in the $H_0 - r_s/D_V(0.57)$ plane, while the color encodes the value of $\Omega_m h^2$, as shown in the color scale on the right. 
From left to right, the CMB data used is \wseven, SPT+$\tau$, and SPT+\wseven. 
The horizontal solid and dashed lines mark the central value and $1\,\sigma$ region for the BOSS BAO measurement, while the vertical lines do the same for \ho{}. 
The joint 1 and $2\,\sigma$ likelihood regions for  BOSS+$H_0$ measurements are denoted by the dark and light grey shaded contours. 
}
\label{fig:lcdm_baoh0_ommh2}
\end{figure*}

\begin{table}
\begin{center}
\begin{threeparttable}
\caption[Relative Consistency Between Datasets in the \LCDM{} Model]{Relative Consistency Between Datasets in the \LCDM{} Model}
\begin{tabular}{ | l | c | c | c |}
\hline
 Dataset Comparison & Add. & $\Delta\chi^2$ & corrected \\
 & dof & & PTE \\
\hline
(CMB) $vs$ (+\ho{}) & 1 & 0.23 & -- \\
(CMB) $vs$ (+BAO$_{\rm SDSS}$) & 1 & 1.8 & -- \\
(CMB) $vs$ (+BAO$_{\rm WiggleZ}$) & 3 & 1.7 & -- \\
(CMB) $vs$ (+\boss) & 1 & 5.1 & 0.090 \\
\hline
(CMB+\ho) $vs$ (+BAO) & 3 & 7.4 & -- \\
(CMB+BAO) $vs$ (+\ho) & 1 & 3.5 & 0.12 \\
\hline
\end{tabular}
\label{tab:consistency_lcdm}
\begin{tablenotes}
\item {\bf Top 4 rows:} 
the relative consistency between the CMB and the individual low-redshift measurements.
For each model, we calculate the $\Delta\chi^2$ between the $\chi^2_{\rm min}$ for the CMB dataset and the $\chi^2_{\rm min}$ for the CMB+(each low-redshift experiment).
The least consistent, \boss{}, is shown in row four, along with the probability to get a larger $\Delta\chi^2$ given four dataset combinations, specifically: (corrected~PTE)~$=~1-(1-PTE)^4$. \\
{\bf Bottom 2 rows:}
the $\Delta\chi^2$ between the CMB+\ho{} and CMB+BAO+\ho{} datasets combinations, and the CMB+BAO and CMB+BAO+\ho{} combinations. For the least consistent comparison, (CMB+BAO) $vs$ (+\ho), we show the probability to get a larger $\Delta\chi^2$ given two chances.
\end{tablenotes}
\end{threeparttable}
\end{center}
\end{table}

In the context of the \lcdm{} model, we quantify the consistency between two datasets, $x$  and
  $y$, with the use of a symmetric $\Delta \chi^2$ statistic.  We calculate
$\Delta \chi^2 = \chi^2_{x+y} - \chi^2_x - \chi^2_y$ where $\chi^2_{x+y}$ is
the minimum value of the $\chi^2$ obtained for the joint ($x+y$) dataset in the model $M$ with a similar definition for $\chi^2_x$ and $\chi^2_y$. 
We use a modified simulated annealing minimizer \citep{goffe94} to find the minimum $\chi^2$  point. 
The $\Delta \chi^2$ is proportional
to a likelihood ratio statistic, and is also connected to a Bayesian
evidence ratio in the case of normal probability distributions, as
shown by \citet{marshall06}. 
The expected distribution of $\Delta\chi^2$ is a $\chi^2$ distribution with degrees of freedom (dof) $N_{\Delta\chi^2} = N_{x+y} - N_{x} - N_{y}$.  
Here $N_{x}$ is the dof for dataset x. 
We then compare the observed value of $\Delta\chi^2$
with its expected distribution and find the probability to exceed (PTE)
this observed value.  

For example, to compare the consistency of the CMB dataset with the
$H_0$ ``dataset,'' we calculate $\chi^2_{\rm min, [CMB+\ho{}]} -
\chi^2_{\rm min, [CMB]} - \chi^2_{\rm min, [H_0]}$=0.23.  
Because $H_0$ has only one datapoint while the model has six free parameters, we have $N_{\ho} = 0$, $N_{CMB}  = X$, and $N_{CMB+\ho} = X+1$. 
Thus $\Delta \chi^2$ in this comparison should be drawn from a
$\chi^2$ distribution with 1 dof, and we find a PTE of 0.63.  

As we are considering the consistency between the CMB and multiple datasets, we have a set of size $N$ of these PTEs, where $N$ corresponds to the number of dataset comparisons in the test. 
The minimum PTE should yield the strongest evidence of possible tension. 
We therefore compare the observed minimum PTE to the expected distribution of the minimum of a set of $N$ uniformly distributed random variables to derive a ``corrected PTE". 
Specifically, the corrected PTE $\alpha$ is derived from the uncorrected PTE $\alpha_0$ by $\alpha = 1-(1-\alpha_0)^N$. 
The corrected PTE for $\Delta \chi^2$  represents a measure of tension between datasets under the assumption that the given model is correct, and the difference between datasets is due to random scatter in the measurements.
A high PTE indicates the data are remarkably consistent, and low PTE indicates the presence of tension between the measurements.
We note however that if the model is poor, the corrected PTE may not be a good representation of tension with the model.

In the top four rows of Table~\ref{tab:consistency_lcdm}, we report the $\Delta\chi^2$
between the CMB and each individual low-redshift dataset in the \lcdm{} model.  
Among the BAO datasets, the BAO$_{\rm WiggleZ}$ dataset has a 
low $\Delta\chi^2$ of 1.7 (3 dof), while the \boss{} dataset has an
unusually high $\Delta\chi^2$ of 5.1 (1 dof), driven in part by the high precision of this data point.
The probability of getting a $\Delta\chi^2$ this large given 1 dof (un-corrected PTE) is 2.3\%; in the case of 4 random draws (corresponding to the 3 BAO plus 1 \ho{} datasets) from this $\Delta\chi^2$ distribution the probability of at least one being this high (corrected PTE) is 9.0\%.  

When the three BAO measurements are combined into a single dataset,
the $\Delta\chi^2$ contributions from BAO$_{\rm WiggleZ}$ and \boss{}
balance out to give a more reasonable total $\Delta\chi^2$. 
Thus the CMB+BAO versus \ho{} case has a lower PTE than the CMB+\ho{} versus BAO case. 
The $\Delta\chi^2$ between CMB+BAO and \ho{} is 3.5 for 1 dof, 
and the corresponding probability of getting a lower PTE given two random draws is 12\%.

From Table~\ref{tab:consistency_lcdm}, we see some degree of tension
between the CMB and the BOSS BAO data point, and between CMB + BAO and
$H_0$.  Both of these tensions exist (to a lesser degree) with the
\wseven{} data alone and have been noted by others \citep{anderson12, mehta12}.  
\citet{anderson12} highlighted the importance of $\omega_m$ for
consistency between \wseven{} and BAO measurements, noting that a $1\,\sigma$
increase in $\omega_m$  from its \wseven-inferred value
would bring them into near-perfect agreement.  Instead, the
SPT+\wseven{} data prefer a slight {\em downward} shift in $\omega_m$
which, together with the decreased uncertainty, leads to the increased
level of tension between BAO and the CMB in the \lcdm{} model.  At the
same time, the shift in $\omega_m$ improves agreement of the CMB data 
with the \ho{} measurement.

The relationship between $\omega_m$ and the CMB, BAO, and \ho{} measurements can be seen
in Figure~\ref{fig:lcdm_baoh0_ommh2}.  This figure shows the parameter space \ho{} 
vs. $r_s/D_V(z=0.57)$ (the characteristic BAO parameter at the redshift reported by BOSS), 
which is the plane in parameter space that is constrained by the \ho{} and \boss{} datasets.  
When combined with any of the
three combinations of SPT and \wseven{} data shown in this figure, the
\ho{} data prefers lower values of $\omega_m$, while the BAO data
prefers higher $\omega_m$.  

We find the apparent tension significant enough in some model spaces,
including \LCDM{}, to suggest caution in interpretation of the
results.  However, in no model spaces is the significance sufficient to
rule out statistical fluctuations, and we have no evidence for either
systematic biases or underestimated uncertainties in the data.

\subsection{Evidence for Extensions}
\label{sec:extensions}

\begin{figure*}
\begin{center}
   \includegraphics[width=0.48\textwidth, trim=2.7cm 14.3cm 3.0cm
3cm]{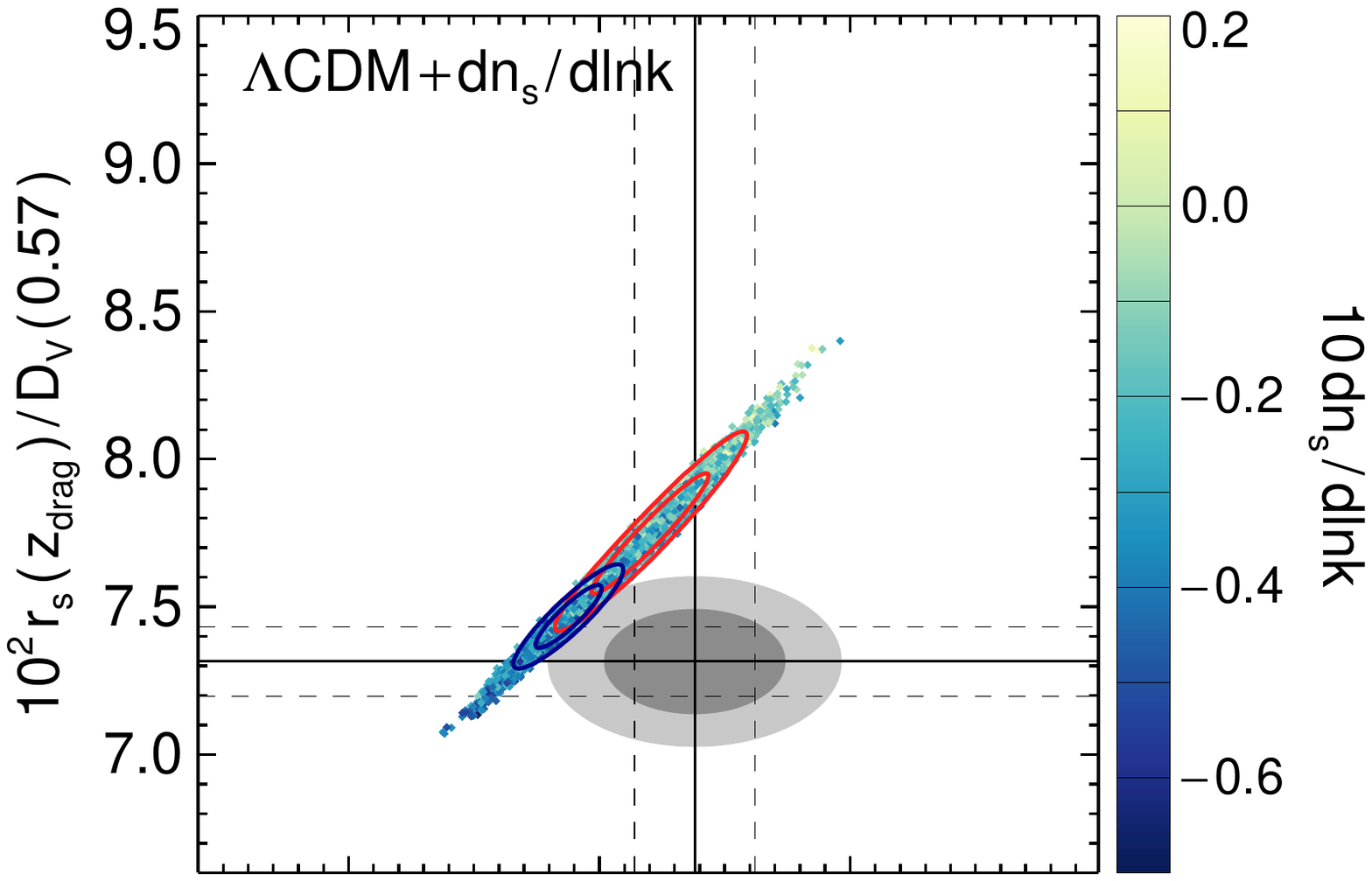}
   \includegraphics[width=0.48\textwidth, trim=2.3cm 14.3cm 3.4cm
3cm]{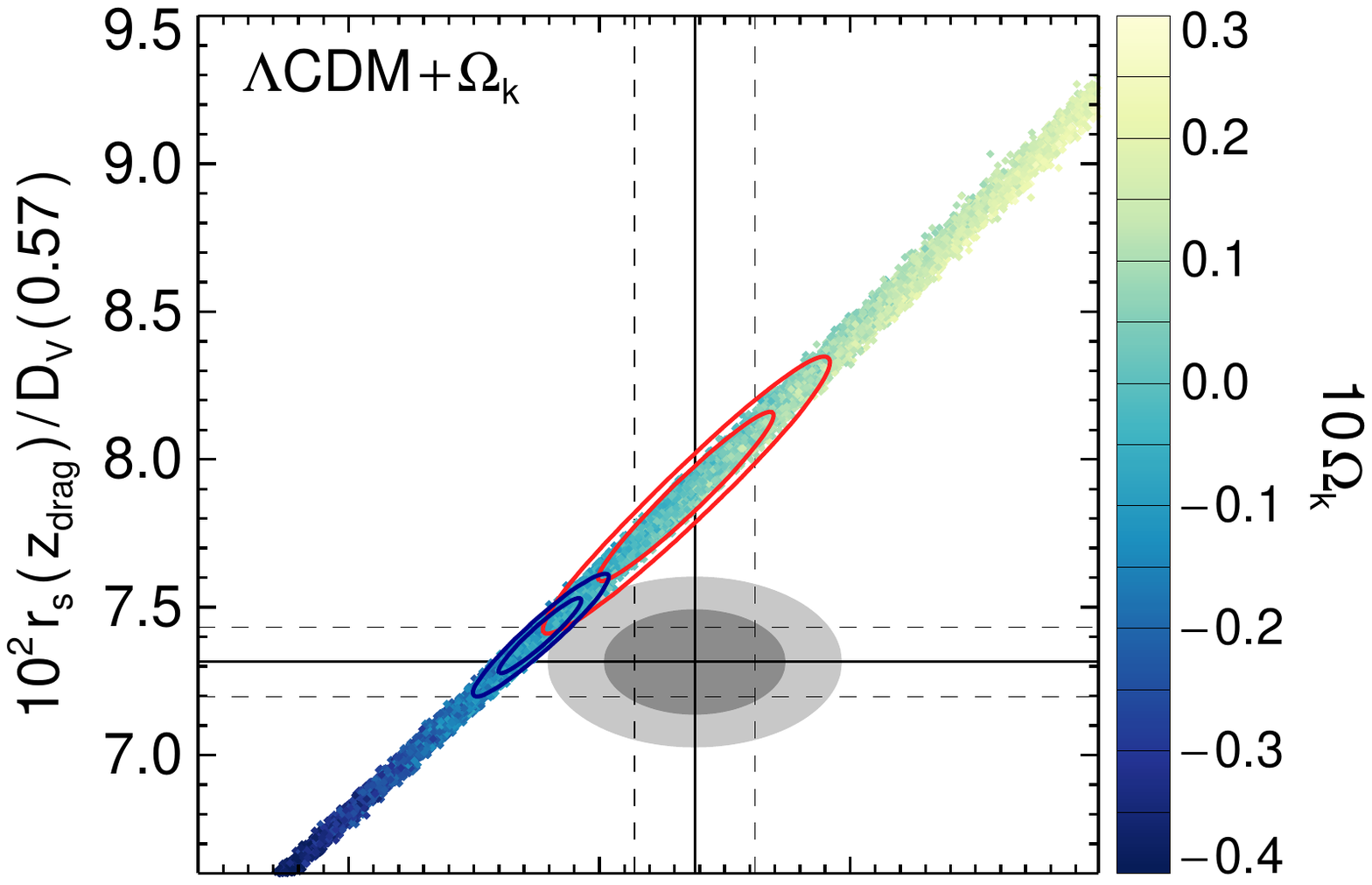}
   \includegraphics[width=0.48\textwidth, trim=2.7cm 14.3cm 3.0cm
3cm]{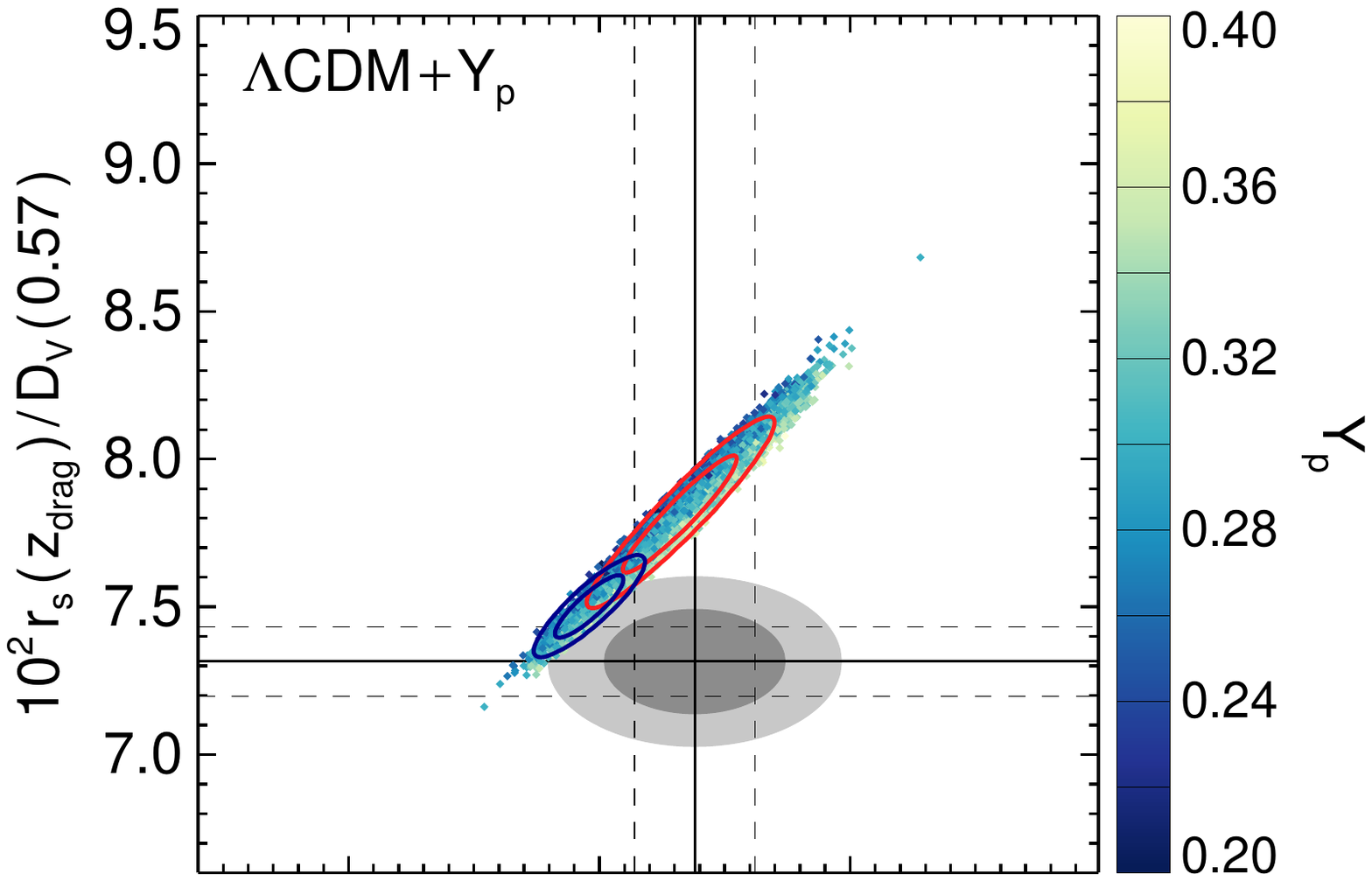}
   \includegraphics[width=0.48\textwidth, trim=2.3cm 14.3cm 3.4cm
3cm]{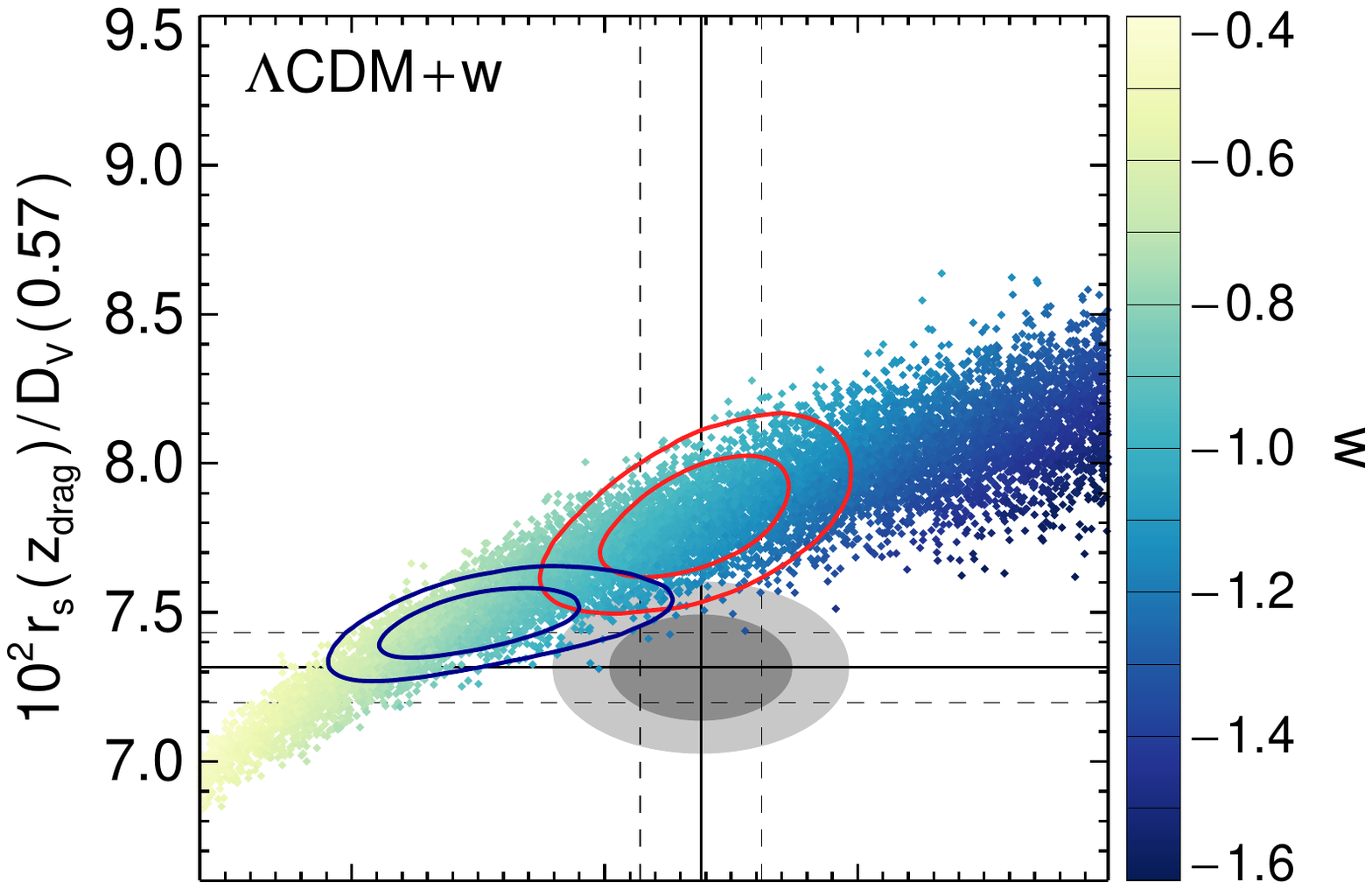}
   \includegraphics[width=0.48\textwidth, trim=2.7cm 13cm 3.0cm
3cm]{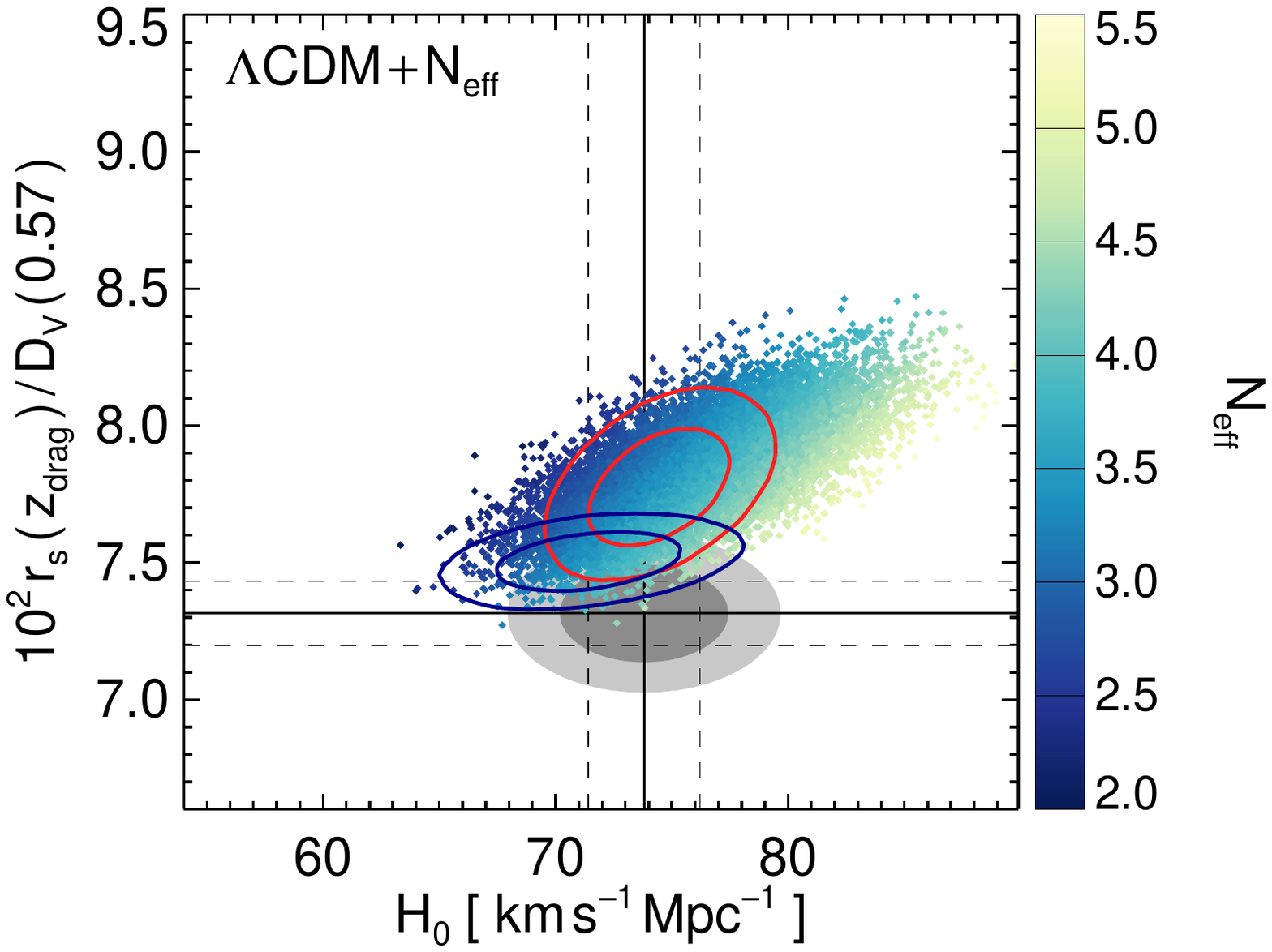}
   \includegraphics[width=0.48\textwidth, trim=2.3cm 13cm 3.4cm
3cm]{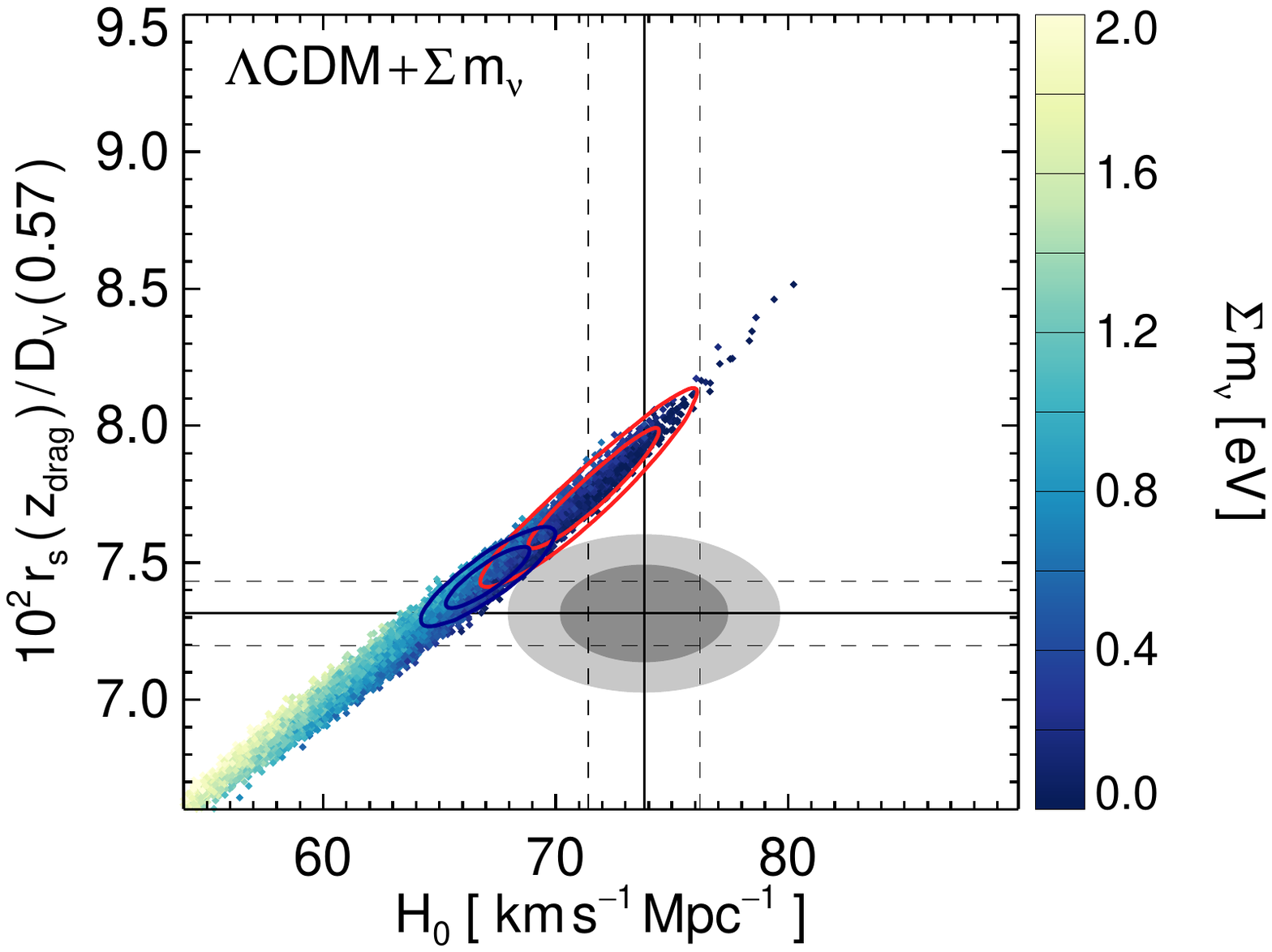}
\end{center}
\caption{
This figure expands Figure~\ref{fig:lcdm_baoh0_ommh2} to investigate the consistency between the CMB, \boss{}, and \ho{} datasets in one-parameter extensions to \LCDM. 
Each panel shows the $\ho - r_s/D_V(0.57)$ plane for a different extension. 
The colored dots are samples drawn from the SPT+\wseven{} MCMC chain, with the color coding reflecting the value of the extension parameter, as shown in the color scales on the right. 
The horizontal solid and dashed lines mark the central value and $1\,\sigma$ region for the \boss{} measurement, while the vertical lines do the same for \ho{}. 
The joint 1 and $2\,\sigma$ likelihood region for  \boss+$H_0$ measurements is denoted by the dark and light grey shaded contours. 
 The blue (red) contours overlaid show the 68\% and 95\% confidence regions for CMB+BAO
(CMB+$H_0$). 
Varying the effective number of neutrino species leads to the best agreement between the CMB, \boss{}, and \ho{}. 
}
\label{fig:bao_vs_h}
\end{figure*}

Two natural questions are: (1) which extensions to the \LCDM{} model
are preferred by the data, and (2) do these extensions ease or increase
the tension between the datasets?
Here we answer these two questions, then give physical explanations for the results.

We define a slightly different $\Delta\chi^2$ statistic to quantify how strongly the data prefer a given extension $\phi$ over the \LCDM{} model:   $\Delta \chi^2 = \chi^2_{\LCDM} - \chi^2_{\LCDM+\phi}$.
A large reduction in $\chi^2_{\rm min}$ (equivalently a large $\Delta\chi^2$) relative to the baseline \LCDM{} model indicates that the extended model is a much better fit to the data.
For instance, if $\Delta\chi^2 = 4$ (6.2) for a given extension with one (two) degrees of freedom, this means that the data favor this extension at  2\,$\sigma$.
We consider models which are favored by more than 2\,$\sigma$  to be ``preferred'' by the data.
However, it is worth noting that we do not consider all possible physically motivated extensions, and also that a single feature in the data may lead to an apparent preference for multiple extensions with degenerate effects. 
The results of this statistic for the  extensions we consider in this work are presented in Table~\ref{tab:dchisq_ext}.

The CMB dataset shows a $>$\,2\,$\sigma$ preference  for the extensions \LCDM+\nrun and \LCDM+\yp{}. 
The $\Delta\chi^2$ for \LCDM+\nrun and \LCDM+\yp{} is 4.9 and 4.4 respectively. 
Allowing running or varying the helium abundance allows for an increasingly red, scale-dependent tilt, and provides a better match to the observed CMB bandpowers than does the \LCDM{} model.
All other single-parameter extensions are not significantly favored by the CMB data.  
The combined CMB+BAO+\ho{} data show even stronger preference for these same model extensions, with $\Delta\chi^2$ for \LCDM+\nrun and \LCDM+\yp{} of 7.4 and 5.3, respectively.

The CMB alone prefers neither two-parameter extension at $>$\,2\,$\sigma$. 
Although the $\Delta\chi^2=5.5$ for the \LCDM+\neff+\yp{}  model is higher than that of any single parameter extension given the CMB data, this is for two, instead of one, additional dof. 
The \LCDM+\neff+\sumnu{} model is preferred by the  CMB+BAO+\ho{} combination, with $\Delta\chi^2 = 7.9$. 
As we will discuss later in this paper in \S~\ref{sec:neff_mnu} and Figure~\ref{fig:neff_mnu_scatter}, this model space also improves the consistency between the CMB, BAO, and \ho{} datasets. 

The \LCDM+$w$ model is not significantly preferred by either the CMB or the CMB+BAO+\ho{} combination.
Additionally, we find that adding the SPT bandpowers do not
significantly improve $w$ constraints.
Therefore, we do not present constraints on \LCDM+$w$ in this paper, although we sometimes include $w$ when exploring the effect of parameter degeneracies on the constraints for other parameters. 

Next we explore the second question: how do these extensions affect the aforementioned tension between the CMB and low-redshift (BAO and \ho) measurements?
This question is most clearly explored in the context of Figure~\ref{fig:bao_vs_h}.
All the panels of this figure have the same x- and y-axes as Figure~\ref{fig:lcdm_baoh0_ommh2}.  
The \boss{} dataset has been singled out for two reasons.  
First, it is the most precise low-redshift BAO measurement; second, it has the most tension with the CMB and \ho{} data in the \LCDM{} model. 
Each panel of the figure shows color-coded samples from a CMB-only MCMC for a different cosmological model.
The color coding reflects values of the extension parameter, thus showing how the chosen parameter moves  within the CMB constraints in this parameter space.

Among the 1-parameter extensions, \LCDM+\neff{} is most effective at reconciling the CMB, BAO, and \ho{} datasets.
A thorough discussion of the physical and observational effects of \neff{} is reserved for \S~\ref{subsec:neff}, however, we briefly preview the important aspects here.
The expansion of the CMB likelihood volume towards the \boss+\ho{} constraint arises
because increasing \neff{} increases the expansion rate at early times which reduces the sound horizon length $r_s$. 
To match the measured CMB acoustic peak locations, one must also decrease $D_A$ by increasing \ho{}.  
The net result, as discussed in \citet{hou13a}, is an increase in \ho{} with minimal change to $r_s/D_V$.  
Thus, the contours for the \LCDM+\neff{} case resemble those for \LCDM{} except expanded horizontally in \ho{}. 

The extensions which become important primarily at late-times (\omk, $w$, and \sumnu{}) have a smaller effect on $r_s$ and instead influence the late-time geometry and thus the inverse distance measures $1/D_V$ and \ho.  
As a result, these extensions stretch the CMB contours along lines with positive slope in the \ho{}-$r_s/D_V$ plane, as seen in the right column of Figure~\ref{fig:bao_vs_h}. 
Low-redshift datasets are especially important for constraining these extensions.


In summary, the CMB data prefer the \nrun{} and \yp{} extensions to \LCDM{} because these extensions allow for an increasingly red tilt at higher multipoles.
The preference for these two extensions remains in the CMB+BAO+\ho{} combination. 
Although the quality of fit improves less for the \LCDM+\neff{} model, this model maximizes the consistency between the three datasets. 
The CMB+BAO+\ho{} data also prefer the \lcdm+\neff+\sumnu{} model, although this preference is less significant (given the 1 extra dof) than that for running. 
The three datasets are completely consistent within this two-parameter extension.

\begin{table*}[htp]
\begin{center}
\begin{threeparttable}
\caption{Improvements to the quality of fit for extensions to the \LCDM{} model}
\begin{tabular}{  |l  | c c c c c c | c c |}
 \hline
& \LCDM & \LCDM & \LCDM & \LCDM & \LCDM & \LCDM & \LCDM & \LCDM \\
Dataset  & +\nrun & +\neff & +\yp & +\sumnu & +\omk & +$w$ & +\neff+\sumnu & +\neff+\yp \\
 \hline
&&&&&&&&\\
CMB  & 4.9 & 1.1 & 4.4 & 2.4 & 0.3 & 0.0 & 2.6 & 5.5 \\
CMB+BAO+\ho{}  & 7.4 & 3.4 & 5.3 & 2.8 & 2.0 & 0.2 & 7.9 & 5.4 \\
\hline
\end{tabular}
\label{tab:dchisq_ext}
\begin{tablenotes}
\item This table shows how extensions to the \LCDM{} model change the quality of the fit to the data for the CMB and the datasets combination of CMB+BAO+\ho{}.
For each dataset we report $\Delta \chi^2$ defined as the reduction in $\chi^2$ from the best-fit \LCDM{} model to the best-fit \LCDM+(extension) model.
A large $\Delta\chi^2$ relative to the baseline \LCDM{} model indicates that the extended model is a much better fit to the data.
For instance, if $\Delta\chi^2 = 4$ (6.2) for a given extension with one (two) degrees of freedom, this means that the data favor this extension at  2\,$\sigma$.
Of the one-parameter model extensions considered here, both datasets have the strongest preference for non-zero running.
Of the two-parameter model extensions considered here, the combined data prefer the \LCDM+\neff+\sumnu{} model most strongly.
\end{tablenotes}
\end{threeparttable}
\end{center}
\end{table*}

\section{Curvature}
\label{sec:curvature}

\begin{figure}
\begin{center}
    \includegraphics[width=0.48\textwidth, trim=1.8cm 13cm 2.3cm
2.5cm]{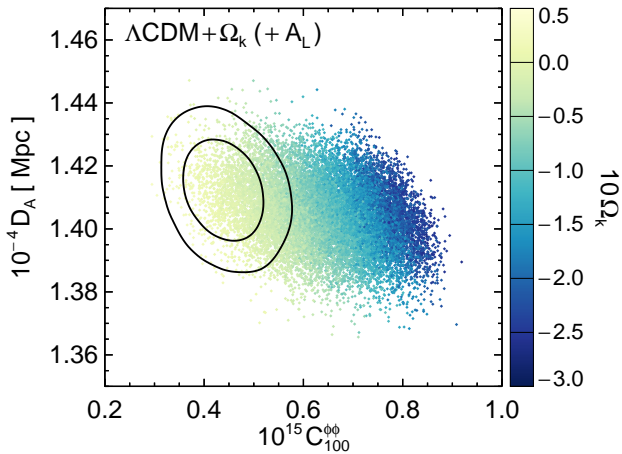}
\end{center}
\caption{
The impact of the angular diameter distance to last scattering, $D_A$, and the lensing information on the curvature constraint from the CMB. 
The contours show the 68\% and 95\% C.L.\ contours when lensing information is included.
The colored points are samples from the posterior distribution after marginalzing over $A_L$, which effectively removes all lensing information.  
The allowed range of $C_{100}^{\phi\phi}$ values remains finite in the absence of lensing information, because  the CMB still places (weaker) constraints on the \LCDM{}+$\Omega_k$ model parameters. 
The color reflects the curvature value as indicated by the color bar on the right side of the plot. 
The lensing sensitivity of the data is clearly crucial to the curvature constraint from the CMB data.
}
\label{fig:cont_clpp}
\vskip 10 pt
\end{figure}

The SPT+\wseven{} constraint on the mean curvature of the universe has been presented by \citet{story13}.
In brief, the addition of the SPT bandpowers tightens the \wseven{} constraint by a factor of five 
 to $\Omega_k =
-0.003^{+0.014}_{-0.018}$.
Here we briefly present the physical underpinnings of how the SPT bandpowers 
tighten the constraint on $\Omega_k$.

The constraint on curvature tightens due to the sensitivity
of the SPT bandpowers to low-redshift information through
gravitational lensing.  
S12 used the combination of SPT and \wseven{} bandpowers to report an
$8.1\,\sigma$ detection of gravitational lensing of the CMB.
Without the lensing information, the CMB constraints exhibit a strong
degeneracy between curvature and the dark energy density.  In essence,
the primary CMB anisotropy is exquisitely sensitive to the
angular-diameter distance to last scattering, but this is only one
number.  Therefore, if there are $n$ parameters that only affect
late-time geometry and structure, the CMB data will only tell us about
one direction in this $n$-dimensional space.  This is insufficient to
distinguish between the two late-time parameters $\Omega_k$ and $\Omega_\Lambda$.  Lensing adds a
second late-time measurement, breaking this degeneracy.

We demonstrate the effect of including lensing on curvature
constraints from the CMB in Figure~\ref{fig:cont_clpp}.  The axes in
the figure are the two late-time quantities constrained by the CMB
data:  the angular-diameter distance to last scattering, $D_A$, and the angular power spectrum of the
lensing potential $\phi$ evaluated at $\ell=100$, $C_{100}^{\phi \phi}$.  
The color-coded points are drawn from an SPT + \wseven{} MCMC chain
with lensing information removed by marginalization over $A_L.$\footnote{The
  current CMB data are insensitive to the small shape variations in
  $C_\ell^{\phi\phi}$ and hence marginalizing over a scaling parameter
  removes all significant lensing information.}  We can see the
impact of the geometric constraint, confining the samples to a narrow
range of $D_A$ values.  Also, despite removing all lensing
information, we see that the \lcdm{}+$\Omega_k$ model predicts a finite range of values
of the lensing power.  
\comment{
When we include the lensing information (by
fixing $A_L = 1$) we get the probability contours shown by the black lines.
The viable range for the lensing power and curvature shrinks significantly.  
}

As Figure~\ref{fig:cont_clpp} shows, varying $\Omega_k$ leads to
significant variation in the  $C_{100}^{\phi \phi}$ direction in the
$D_A - C_{100}^{\phi \phi}$ plane.  The direction of the response in
this plane is an important aspect of why lensing plays a significant
role in constraining $\Omega_k$ whereas it does not for $\sumnu$. 
The effect of lensing information (i.e.,~requiring $A_L = 1$) is shown by the black
contours.  The observed lensing amplitude rules out the negative
curvature tail, and as mentioned above, tightens the curvature
constraint from the CMB by a factor of five.

Better constraints on curvature are possible by including low-redshift probes. 
For instance, CMB+BAO+$H_0$ leads to a constraint on curvature of $\Omega_k =
-0.0061\pm 0.0040$.  
Even with low-redshift datasets included, SPT data remains important.
Without the SPT bandpowers (and lensing information therein), the uncertainty would be roughly 15\% larger: $\Omega_k = -0.0019\pm 0.0047$ for \wseven{}+BAO+$H_0$.

\section{Massive Neutrinos}
\label{sec:mnu}

We now consider allowing non-zero sum of neutrino masses. 
As we saw in \S~\ref{sec:consistency}, the CMB data provide only  a weak preference for
this extension.
 However, once we include datasets sensitive to late times, the
support for non-zero neutrino mass is fairly robust to the particular choice of
the datasets and additional extensions of the \LCDM{} model.

The range of possible neutrino masses is currently constrained from both above and below.
Neutrino oscillation experiments place a lower bound on the sum of the neutrino
masses of $\gtrsim 0.058\,{\rm eV}$ \citep{robertson08}. 
Neutrino oscillations are insensitive to the sum of the neutrino masses above this minimum. 
The best laboratory upper bound on any single neutrino mass is $0.3-0.6$\,eV (90\% CL) depending on the nuclear matrix elements adopted for the electron neutrino from the Kamland-Zen double beta decay experiment \citep{gando12}.
Therefore, the sum of the neutrino masses must lie between the limits $0.06\ {\rm eV} <\sumnu < 1.8\ {\rm eV}$.

Cosmological observations can provide significantly stronger upper limits on the sum of the neutrino masses.
The strongest constraints on \sumnu{} come from combining CMB measurements with low-redshift information.
In this section, we include low-redshift measurements of the BAO feature, \ho{}, the halo power spectrum derived from SDSS luminous red galaxies (LRGs) from \citet{reid10}, and the cluster abundances from the SPT SZ-selected cluster catalog (\sptcl{}) described by \citet{reichardt13}.  
The LRG, \ho, and \wseven{} data have been used previously to set an upper limit on the neutrino masses of $\sumnu < 0.44$\,eV (95\% CL) \citep{komatsu11}. 
A somewhat tighter limit of $\sumnu < 0.32$\,eV (95\% CL) has been placed using SZ-selected galaxy clusters, CMB, BAO, and \ho{} by \citet{benson13}.
Similar limits of about 0.3\,eV come from combining CMB+BAO+\ho{} with galaxy clustering data \citep{thomas10, riemer-sorensen12a, deputter12} or X-ray cluster abundance and cluster gas mass fraction measurements \citep{mantz10c}.  
As we note in \S~\ref{sec:mnu_constraints}, the inclusion of the new
BAO data (particularly the BOSS point)  increases the preference for nonzero mass.

In this section, we present updated constraints on the sum of the neutrino masses incorporating the new SPT bandpowers. 
We first discuss how measurements of the CMB, structure growth, and geometry constrain massive neutrinos. 
We then present constraints for combinations of the SPT bandpowers and other datasets. 
We finally consider potential degeneracies between the sum of the neutrino masses and other parameter extensions such as running
of the spectral index.

\subsection{Cosmological effects of massive neutrinos}

\begin{figure}
\begin{center}
    \includegraphics[width=0.48\textwidth, trim=1.5cm 13.0cm 2.cm
2.7cm]{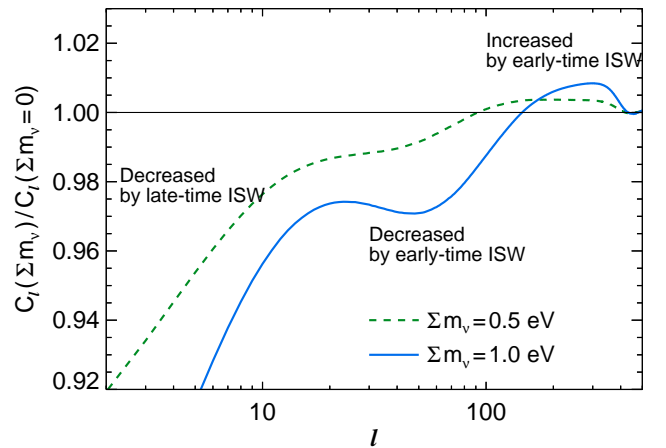}
\end{center}
\caption{This figure shows the effects of massive neutrinos on the CMB power
spectrum.  The curves show the ratio of models with 
$\sumnu = 0.5$ (\textbf{dashed green}) and 1.0\,eV (\textbf{solid blue})
respectively to the best-fit 
\LCDM{}   (\sumnu=0\,eV)
model spectrum for SPT+\wseven.  When increasing $\sumnu$ we adjust
$\Omega_\Lambda$ downward to keep $\theta_s$ fixed.  On large
scales, we see a reduction in  the power added by the
late-time ISW effect. 
 On intermediate scales below the neutrino free-streaming length, we see a reduction in 
 the power contributed by the early ISW effect. 
 On scales smaller than the neutrino free-streaming length,  the
more rapid decay of  gravitational potentials boosts the
early ISW power.  
The amplitude of the early ISW effect is damped at $\ell \ga 500$ by averaging over multiple positive and negative contributions. 
}
\label{fig:mnu_power}
\end{figure}

In this subsection, we give a physical description of the effects of the 
massive neutrinos  on cosmology.
We start with the effects massive neutrinos have on the  CMB. 
We note that with the current experimental precision, information from CMB lensing is not important to the CMB constraints on $\sumnu$.
We then move on to consider other effects in the low-redshift universe.

To understand neutrino mass constraints from CMB data, we
must understand how the predicted CMB power spectrum changes with neutrino mass. 
In the standard thermal history of the universe, massless
neutrinos have a temperature corresponding to $\sim 0.17\,{\rm eV}$ at the epoch of
last scattering.  
The scale at which masses start to have an appreciable effect is set by this temperature to be $\sumnu \approx 3 \times 0.17\,{\rm eV}$.\footnote{For  simplicity we assume three families of neutrinos with degenerate
  masses. }
Neutrino masses well below this value have no impact on primary CMB anisotropy.
 \citet{hu02b} and \citet{ichikawa05} study  in detail the impact of higher masses on the CMB and find the dominant impact is due to the ISW effect.

In a matter-dominated universe with zero mean curvature, gravitational
potentials remain constant to first order in linear perturbation
theory.  Adding  components that do not cluster, while keeping the
curvature fixed to zero, increases the expansion rate which causes the gravitational potentials to decay. 
 As photons traverse these
decaying potentials on their way toward the observer, new anisotropies
are created by what is called the Integrated Sachs-Wolfe (ISW)
effect.  The ISW anisotropy is generated in the \lcdm{} model both at
early times, as photons free stream immediately after decoupling through a
not-completely-matter-dominated universe (the early ISW effect) and
at late times after the cosmological constant becomes important (the late ISW effect).

We illustrate how the ISW effect changes with neutrino mass in Figure~\ref{fig:mnu_power}. 
In this figure, we plot the ratio of $C_{\ell}$ at either \sumnu = 0.5 or 1.0\,eV relative to a fiducial \lcdm+\sumnu{} model $C^{\rm fid}_{\ell}$ with $\sumnu = 0.0$. 
The baryon density $\omega_b$, cold dark matter density $\omega_c$, and the sound horizon scale $\theta_s$ are fixed between the three models -- $\sumnu$ and $\Omega_\Lambda$ vary. 
Three regimes are labeled in the figure: a reduction of power due to the late-time ISW effect at $\ell\lesssim20$, a reduction of power due to the early ISW effect at $20 \lesssim \ell \lesssim 100$, and an increase in power due to the early ISW effect at $100 \lesssim \ell \lesssim 500$. 
We briefly explain these three regimes in the next paragraphs.

As $\sumnu$ increases with $\omega_b+\omega_c$ fixed, the expansion rate
increases at early times.  Therefore, $\Omega_\Lambda$ must decrease (increasing $D_A$) to
keep $\theta_s$ fixed.  Without this adjustment to $\Omega_\Lambda$, 
$\theta_s$ would change, primarily due to the change in $D_A$.  With
this adjustment, we find that in the mass range of
interest, $H(z)$ increases relative to the $\sumnu = 0$ model at $z \ga 1$ and decreases at $z \la 1$.
The decreased expansion rate at $z \la 1$ results in 
less decay of the gravitational potential on very large scales, and
therefore a reduction in the contribution to the power from the
late-time ISW effect.  The net effect is less power at $\ell\lesssim20$.  
However, the large cosmic variance at these low multipoles makes the CMB data largely insensitive to the reduced power.

On scales shorter than the neutrino free-streaming length,
the increased expansion rate just after photon decoupling enhances the decay of gravitational
potentials and thus enhances the early ISW effect.  
On scales longer than the free-streaming length, the early ISW effect is suppressed;
the clustering of neutrinos prevents the potential from decaying more rapidly and,  due to the increased expansion rate,  there is less time for the early ISW effect to accumulate. 
The dividing line in multipole space between these two regimes increases with $\sumnu$. 
The magnitude of the ISW effect decreases with increasing $\ell$ as
cancellations between an increasing number of positive and negative contributions washes out the
signal, becoming negligible by $\ell \sim 500$. 

\begin{figure}
\begin{center}
    \includegraphics[width=0.48\textwidth, trim=2cm 13cm 1.5cm
1.8cm]{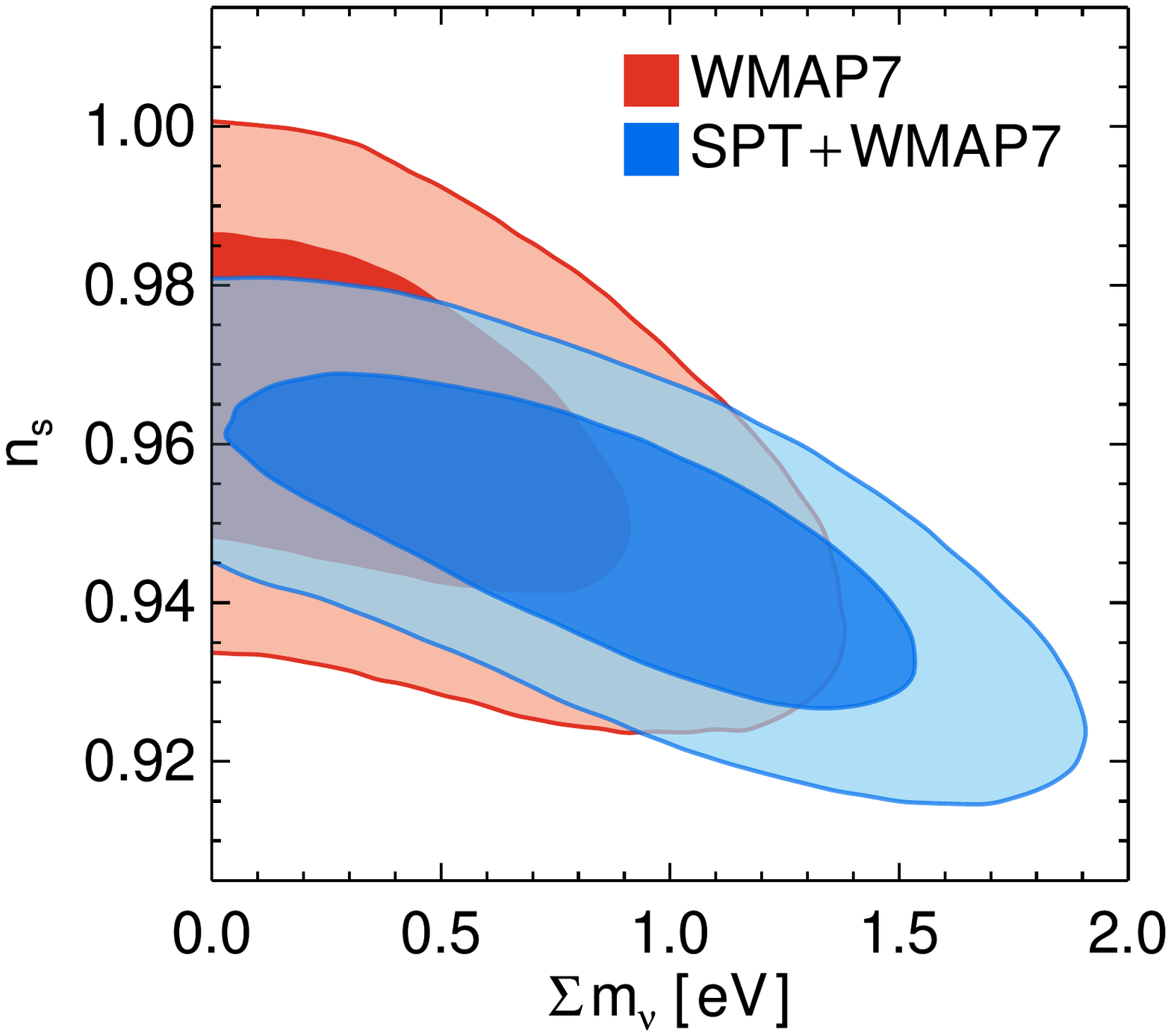}
\end{center}
\caption{This figure illustrates the degeneracy between $n_s$ and \sumnu, and its role in the SPT+\wseven{} preference for nonzero neutrino masses. 
The contours are the 68\% and 95\% confidence intervals in the $\Lambda$CDM+$\sumnu$ parameter space for \wseven{} (red) and
SPT+\wseven{} (blue).
The SPT data prefer a lower value of $n_s$ than \wseven, which leads the CMB data to prefer higher \sumnu. }
\label{fig:mnu_ns}
\end{figure}

The reduction of power at $\ell \lesssim 100$ and increase of power at $\ell \gtrsim 100$ shown in Figure~\ref{fig:mnu_power} can be mimicked by an increase in $\omega_b$ and \ns.  
As a result, \sumnu{} is anti-correlated with  $\omega_b$ and \ns{}.  
Figure~\ref{fig:mnu_ns} shows this anti-correlation between \sumnu{} and \ns{}.

\begin{figure}
\begin{center}
    \includegraphics[width=0.48\textwidth, trim=2.cm 13.5cm 1.8cm
2.7cm]{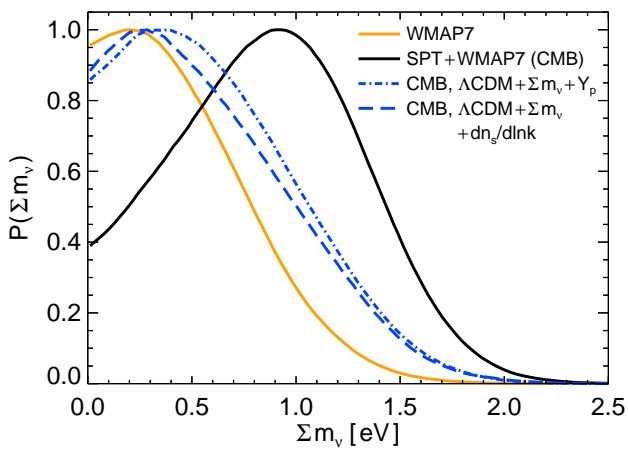}
\end{center}
\caption{
The constraints on \sumnu{} from the CMB data degrade significantly when we decouple the tilt of the primordial power spectrum on small and large scales.
The marginalized one-dimensional posteriors of \sumnu{} from \wseven{} (SPT+\wseven) data are shown with the orange solid (black solid) line. 
The SPT data prefer a lower value of \ns than \wseven, which leads the CMB data to prefer higher \sumnu.
If the tilt of the primordial power law on small and large scales is decoupled, whether directly by introducing running (blue dashed line) or indirectly by freeing the damping scale with the introduction of  \yp{} (blue dot-dashed line), the shift towards higher neutrino masses in the CMB likelihood is substantially reduced. 
}
\label{fig:mnu_like_margi}
\end{figure}

Although the ISW effect is limited to angular scales covered by \wmap, SPT bandpowers contribute significantly to the neutrino mass constraints by alleviating these
degeneracies between \sumnu{}, \ns{}, and $\omega_b$. 
In particular, SPT alone prefers a lower value of \ns relative to \wseven{} (see
Figure~\ref{fig:pardiff}), which causes the preferred value of
\sumnu{} to increase when SPT data are combined with \wseven. 
This shift in the SPT-preferred value of \ns{} disappears with a
freely varying \yp{} or \nrun, thus the shift in \sumnu{} for the
combination of SPT+\wseven{} disappears as well when these parameters
are added. 
These
 shifts in the preferred value of \sumnu{} are seen clearly in
Figure~\ref{fig:mnu_like_margi}.


Next, we examine the other observable consequences that massive
neutrinos have on the geometry and the growth of structure at late times.
The decreased expansion rate at $z \la 1$ directly affects the late-time geometry, and
geometric observables. 
BAO measurements constrain the late-time
expansion rate through $r_s(z_{\rm drag})/D_V(z)$. 
Although increasing \sumnu{} from zero to  1.5\,eV has some effect on $r_s(z_{\rm drag})$, the dominant effect is the increase in $D_V(z)$ as $\Omega_\Lambda$ decreases. 
Therefore $r_s(z_{\rm drag})/D_V(z)$ and \ho{}
respond similarly, decreasing with increasing neutrino mass as shown in
Figure~\ref{fig:mnubaoh0}.

For $z \ga 1$, the increased expansion rate suppresses growth on scales
below the neutrino free-streaming length ($\sim$\,$140$\,Mpc for a
100\,meV neutrino).   On larger scales the neutrinos can cluster,
which counteracts the suppression.  For $z \la 1$, the reduced
expansion rate boosts the rate of structure growth on all scales.  We
hasten to point out, though, that the net effect on integrated growth to
$z = 0$ on small scales is that of suppression. 

In summary, the summed neutrino mass is constrained by the CMB
primarily through the early ISW effect.  The SPT data improve \sumnu{}
constraints indirectly via an improved determination of \ns and 
the anti-correlation of \sumnu{} with \ns.
Adding probes of low-redshift
geometry and structure growth further improve constraints on
\sumnu{}.  We discuss these constraints in the next section.

\subsection{Cosmological constraints on massive neutrinos}
\label{sec:mnu_constraints}

\begin{table*}
\begin{center}
\begin{threeparttable}
\caption{$\Lambda$CDM+$\sum m_{\nu}$ results from various combinations of datasets}
\footnotesize
\begin{tabular}{c | c c c c c c c c c c}
\hline
Datasets & \multicolumn{3}{c}{$\sum m_{\nu}\,[\rm eV]$} & $10^2\Omega_{b}h^2$ & $n_s$ &
$H_0$ & $\sigma_8$ \\
Combinations  & 68\% CL & 95\% CL & Peak  &  & &
$[\mathrm{km\;s^{-1}\;Mpc^{-1}}]$ &  \\
\hline
\\
CMB  & $[0.41, 1.34]$ & $[0, 1.60]$ & $  0.93$ & $2.184\pm 0.041$ & $0.948\pm 0.014$ &
$62.8^{+5.3}_{-4.0}$ & $0.658^{+0.075}_{-0.061}$ \\
CMB+BAO  & $[0.29, 0.68]$ & $[0.11, 0.88]$ & $  0.48$ & $2.207\pm 0.034$ & $0.957\pm 0.008$ & $67.1\pm
1.2$ & $0.712\pm 0.048$ \\
CMB+$H_0$ & $[0, 0.24]$ & $[0, 0.48]$ & $  0.01$ & $2.233\pm 0.035$ & $0.965\pm 0.009$ &
$71.6\pm 1.8$ & $0.758^{+0.031}_{-0.039}$ \\
CMB+LRG & $[0.17, 0.61]$ & $[0, 0.78]$ & $  0.39$ & $2.204\pm 0.036$ & $0.956\pm 0.009$ &
$66.4\pm 2.5$ & $0.731\pm 0.048$ \\
CMB+\sptcl & $[0.28, 0.80]$ &$[0, 1.03]$ & $  0.55$ & $2.196\pm 0.040$ & $0.954\pm
0.011$ & $65.1^{+3.8}_{-3.4}$ & $0.702^{+0.045}_{-0.038}$ \\
CMB+BAO+$H_0$ & $[0.16, 0.51]$ & $[0, 0.66]$ & $  0.33$ & $2.219\pm 0.034$ & $0.959\pm 0.008$ & $68.4\pm 1.0$ & $0.741\pm 0.046$ \\
CMB+BAO+$H_0$+ \sptcl & $[0.21, 0.43]$ & $[0.10, 0.54]$ & $  0.32$ & $2.221\pm 0.034$ & $0.960\pm
0.008$ & $68.3\pm 1.0$ & $0.739\pm 0.027$ \\
\hline
\end{tabular}
\label{tab:lcdmmnu}
\begin{tablenotes}
\item This table shows the results for the \lcdm + \sumnu{} model.
  The confidence level intervals shown indicate the region of highest probability
  density that contains either 68\% or 95\% of the probability. The
  `Peak' value is the peak of the posterior distribution.  We include
  some other parameters of particular interest for this extension.
  Note in particular that high peak values of $\sumnu$ correspond to low
  values of $n_s$, low values of $H_0$ and low values of $\sigma_8$. 
\end{tablenotes}
\end{threeparttable}
\end{center}
\end{table*}

\begin{figure}
\begin{center}
    \includegraphics[width=0.48\textwidth, trim=0.5cm 15.4cm 2cm
1.7cm]{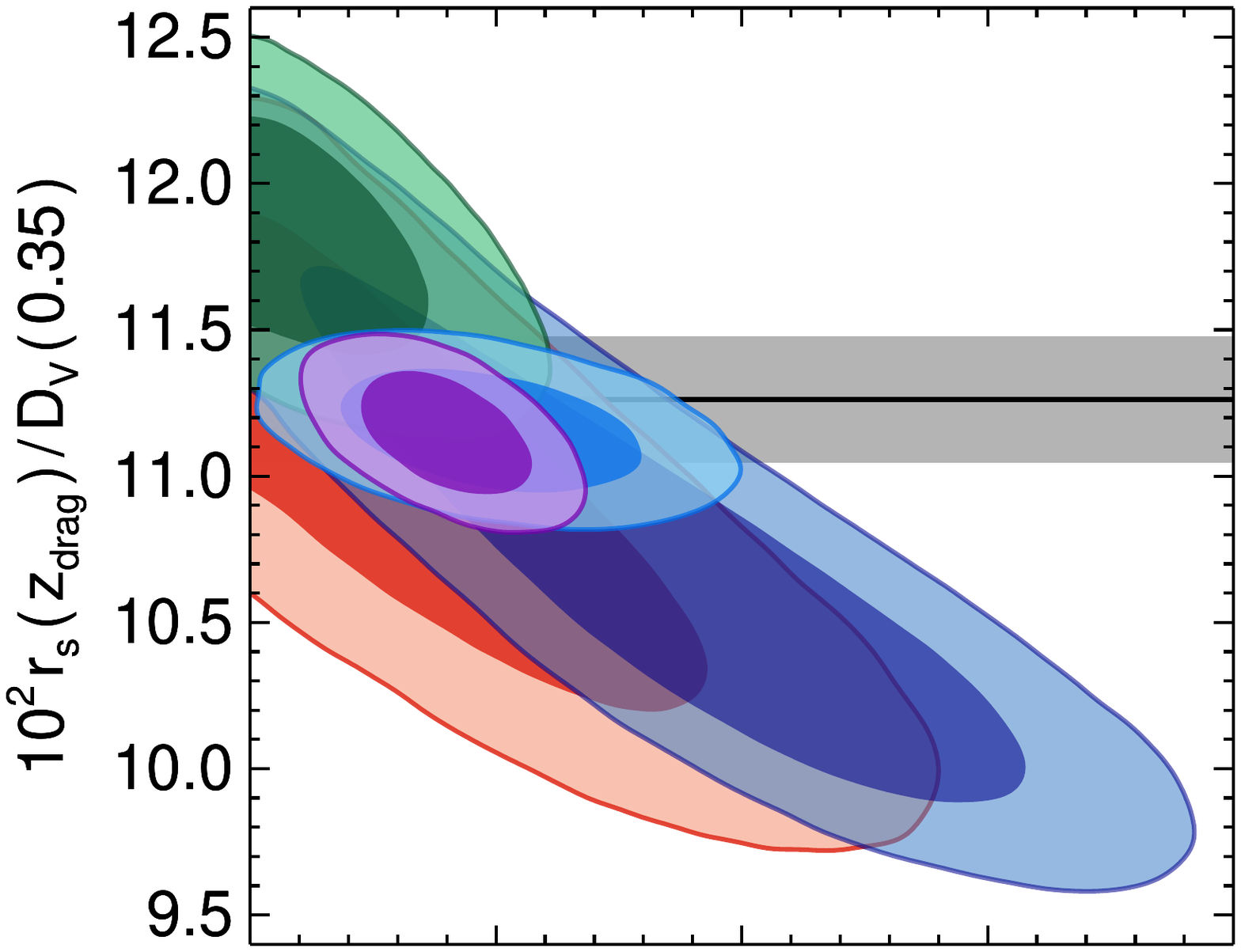}
    \includegraphics[width=0.48\textwidth, trim=0.5cm 16.7cm 2cm
1.8cm]{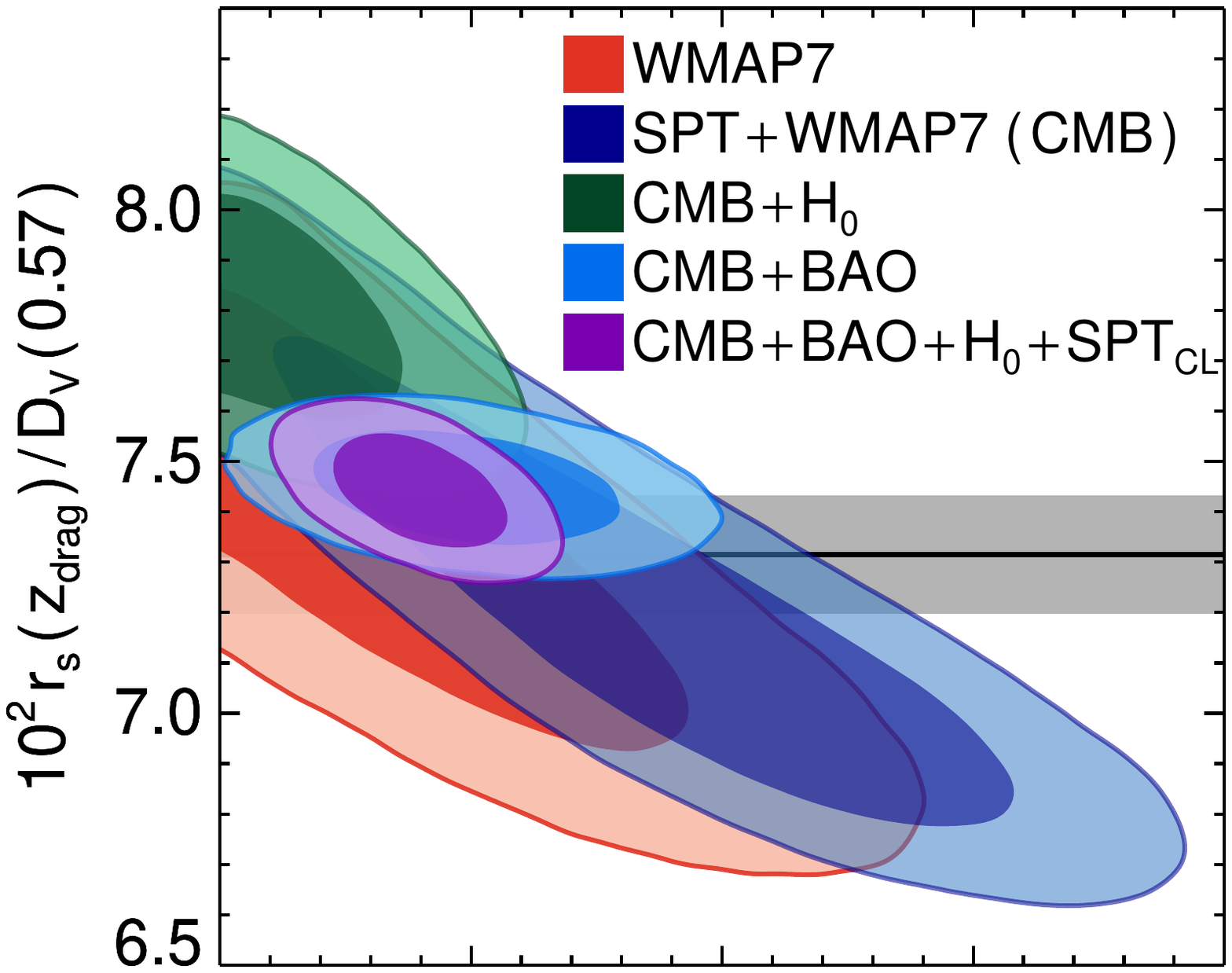}
    \includegraphics[width=0.48\textwidth, trim=0.5cm 13.5cm 2cm
0.5cm]{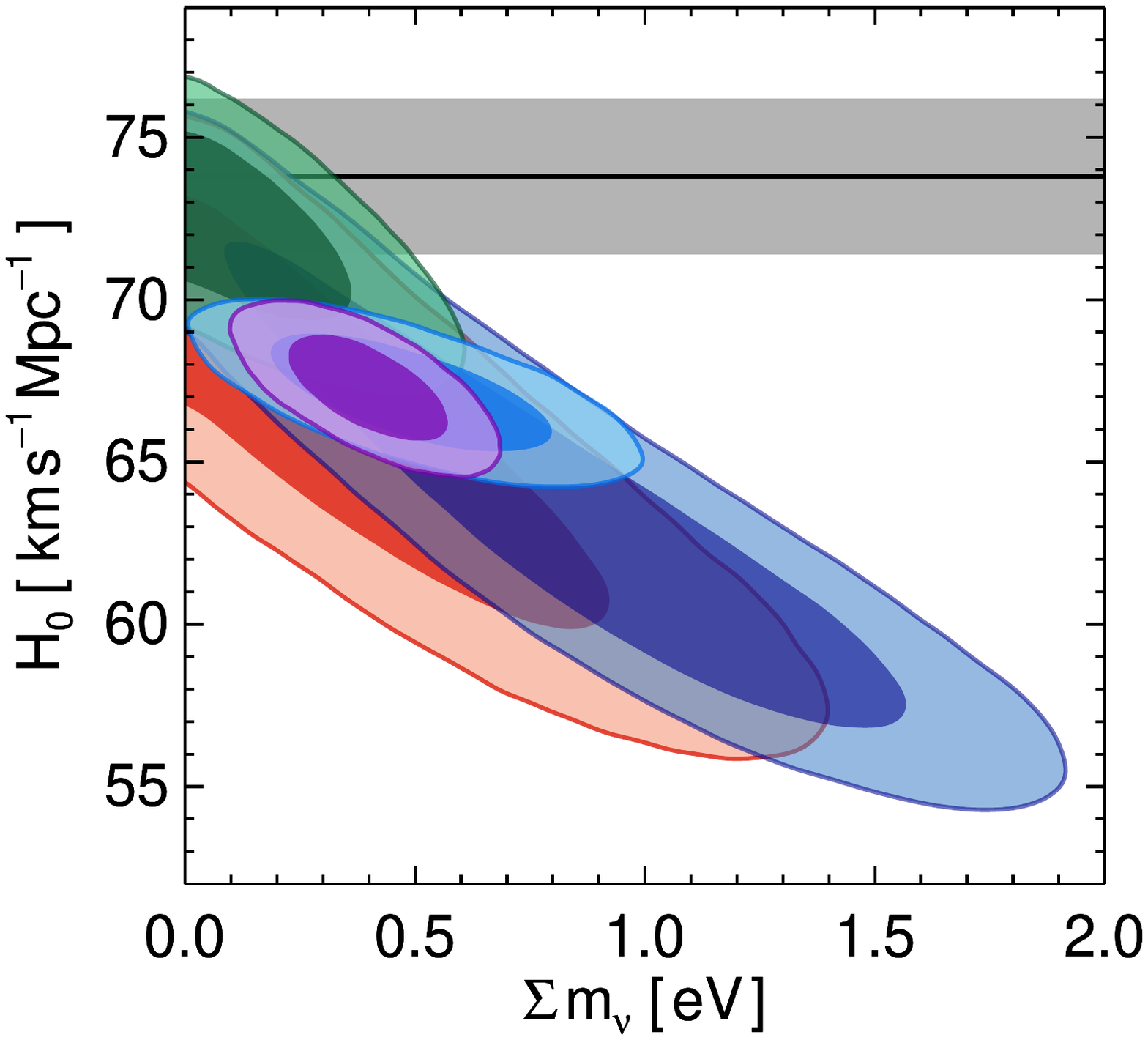}
\end{center}
\caption{This figure shows how low-redshift datasets affect the neutrino mass constraints.
In the planes of \sumnu{} and three quantities inferred at low redshifts, 
we show the marginalized two-dimensional posteriors from different combinations of datasets shown by the legend in the central panel. 
\textit{Top}: constraints are plotted in the $\sumnu - r_s(z_{\rm drag})/D_V(0.35)$ 
plane with the SDSS BAO data point shown by the grey band
as the $1\,\sigma$ region.
\textit{Middle}: similar to the left panel but showing $r_s(z_{\rm drag})/D_V(0.57)$ 
with the BOSS BAO data point for the gray band as the
$1\,\sigma$ region.
\textit{Bottom}: similar to the top and middle panels but showing \ho{} with the
grey band showing the $1\,\sigma$ region.  The tension between BAO and \ho{} can
be seen in the three panels.
The CMB+BAO preference for massive neutrinos is clearly visible.}
\label{fig:mnubaoh0}
\end{figure}

We now look at the cosmological constraints on neutrino mass. 
In order to discuss changes in the posterior probability density for \sumnu{} near the prior at zero,  
we will sometimes quote a  different confidence interval than the rest of the paper. 
In Table~\ref{tab:lcdmmnu} and this section, we report the estimated confidence interval 
$\sumnu \in [x_1, x_2]$~(68\% CL) such that
\be
\int_{x_1}^{x_2} P(\sumnu) d(\sumnu) = 0.68,
\ee
where $P(x_1) = P(x_2)$ and $P(\sumnu)$ is the normalized neutrino
mass posterior probability density. 
When we find $P(0) > P(x_2)$, we report the confidence interval as $[0,x_2]$; this is an upper limit. 
We will also report the peak of $P(\sumnu)$. 
For cases with a non-zero $95\%$ confidence lower limit as defined above, we will continue to report the median and $1\,\sigma$ error which is used  in the rest of the paper and defined  in \S~\ref{subsec:method}.

The SPT data prefer lower values of \ns than those preferred by \wseven, which translates into a preference for higher neutrino masses. 
For the SPT+\wseven{} data, we find the position of the peak moves to 0.93\,eV with a 68\% confidence interval of $\sumnu \in [0.41, 1.34]$\,eV. 
The 95\% CL upper limit from the CMB is $\sumnu < 1.60$\,eV.

\begin{figure*}
\begin{center}
    \includegraphics[width=0.48\textwidth, trim=2.2cm 13.5cm 2cm
3.0cm]{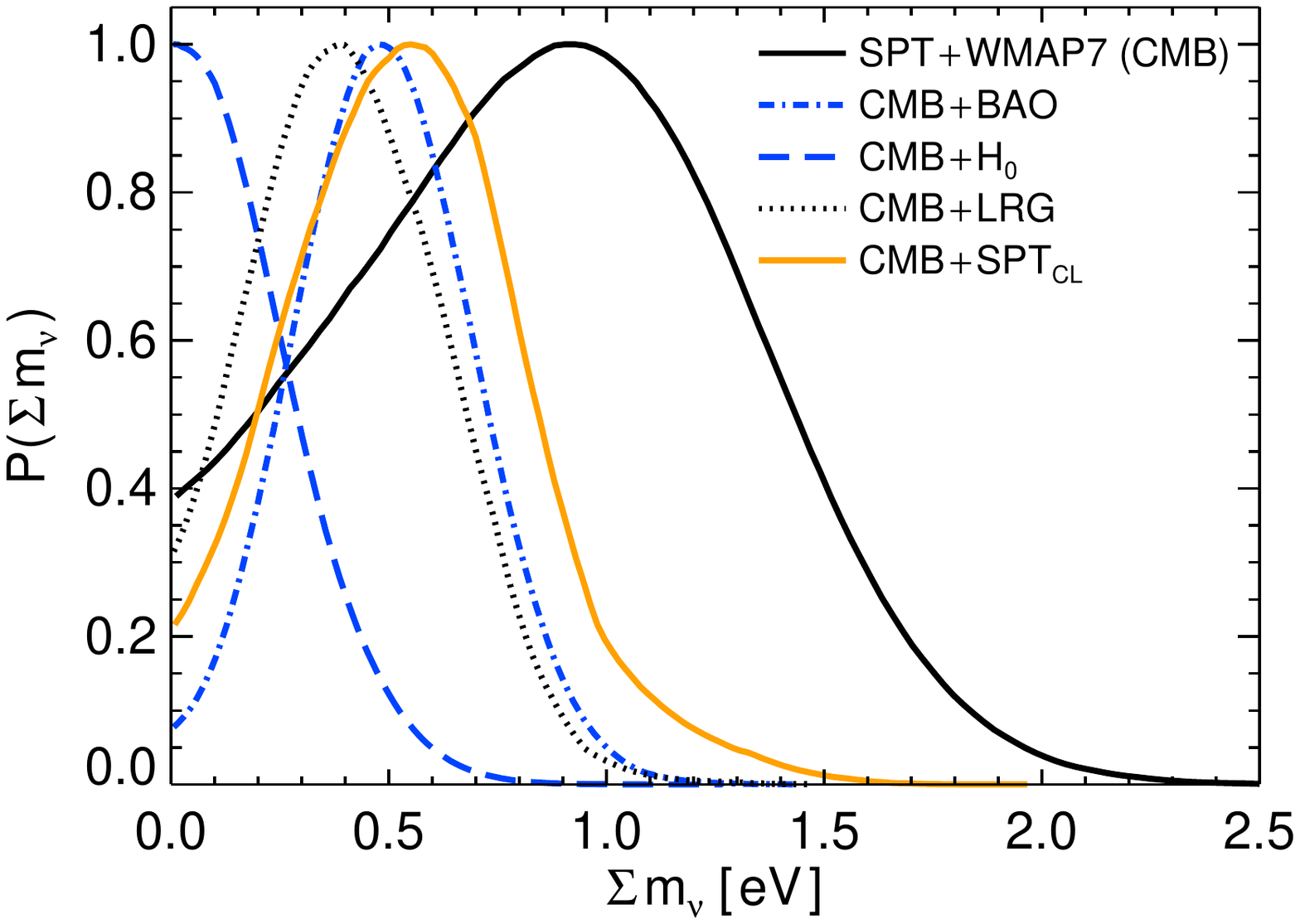}
    \includegraphics[width=0.48\textwidth, trim=2.2cm 13.5cm 2cm
3.0cm]{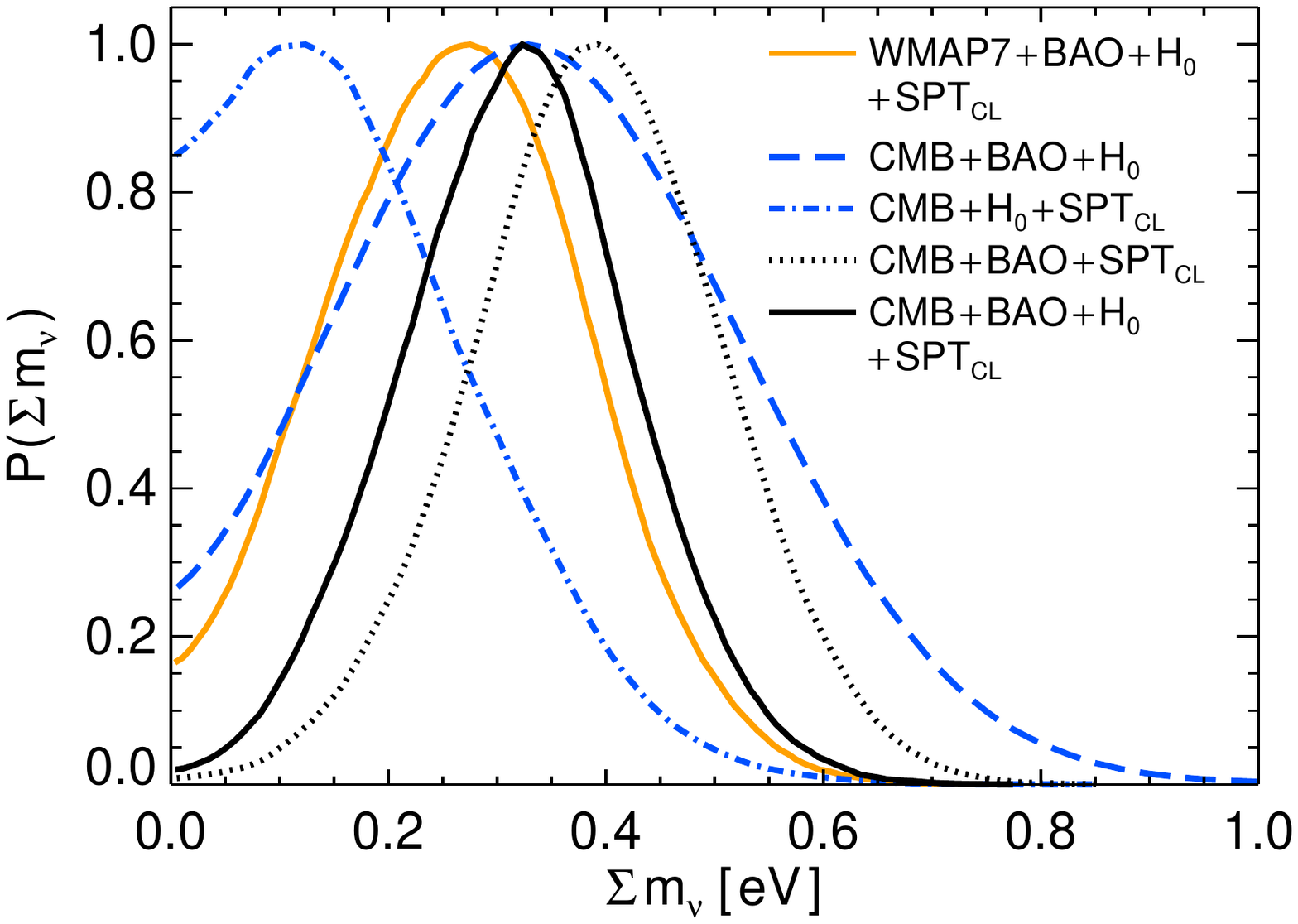}
\end{center}
\caption{
\textbf{\textit{Left panel:}} This panel demonstrates how the marginalized one-dimensional posterior distribution of \sumnu{} changes when a single external dataset is \textit{added} to the CMB dataset.
The CMB constraint is shown by the black solid line. 
The datasets added are the BAO data (blue dot-dashed line), \ho{} measurement (blue dashed line),  LRG sample (black dotted line), and SPT cluster data (orange solid line).
\textbf{\textit{Right panel:}} This panel demonstrates how the marginalized one-dimensional posterior distribution of \sumnu{} changes when a single external dataset is \textit{removed} from the 
combination of CMB+BAO+\ho+\sptcl{}.
The datasets dropped are the SPT bandpowers (orange solid line), \sptcl{} (blue dashed line), BAO data (blue dot-dashed line), and \ho{} measurement (black dotted line). 
The marginalized posterior for the combined CMB+BAO+\ho+\sptcl{} dataset is shown by the black solid line. 
}
\label{fig:mnu_like}
\end{figure*}

We now consider low-redshift probes of geometry: BAO and \ho.
As mentioned above, these low-redshift geometric probes can strengthen inferences of neutrino mass through the impact on the expansion rate.
With massive neutrinos, the CMB data is compatible with either the BAO or \ho{} measurements (see Figures~\ref{fig:bao_vs_h} and \ref{fig:mnubaoh0}), 
though the resulting constraints on \sumnu{} are quite different. Adding BAO data to the CMB data tightens the neutrino mass constraint
significantly; for the combination of CMB+BAO, we obtain a $2.4\,\sigma$ preference for
nonzero neutrino masses with $\sumnu=(0.49 \pm 0.20)\,{\rm eV}$.  
The 95\% confidence interval is $\sumnu \in [0.11, 0.88]$\,eV. 
On the other hand, the \ho{} dataset prefers higher values of \ho, corresponding to lower values of \sumnu{}.
The likelihood for CMB+\ho{} peaks near zero neutrino mass, with an
upper limit of $\sumnu<0.48$\,eV (95\% CL).  
Combining all three datasets produces a neutrino mass constraint that lies between the CMB+BAO and CMB+\ho{} constraints.
For the combination of CMB+BAO+\ho{}, we find the posterior peaks at 0.33\,eV with a 68\% confidence interval of $\sumnu \in [0.16, 0.51]$\,eV. 
The 95\% CL upper limit is $\sumnu<0.66$\,eV. 
The constraints from these combinations are shown in Figure~\ref{fig:mnu_like}.  
As an aside,
we note that as in the CMB-only case, the small-scale information from SPT data is a significant contributor to the suggestion of nonzero neutrino masses.  
Without the SPT bandpowers, the neutrino mass probability density for \wseven+BAO+\ho{} peaks just above zero mass with a 95\% CL upper limit of $\sumnu<0.48$\,eV.

We next turn to low-redshift probes of structure growth: LRGs and galaxy clusters.
For the CMB+LRG dataset, we find the posterior probability peaks at 0.39\,eV with a 68\% confidence interval of $\sumnu \in [0.17, 0.61]$\,eV.  
The 95\% confidence interval is an upper limit at  $\sumnu<0.78\,\rm{eV}$ (95\%\,CL).
Instead, adding the SPT galaxy cluster sample\footnote{Note that the interpretation of the measured galaxy number counts depends on the Tinker mass function \citep{tinker08}.  
Although the Tinker mass function was not originally calculated for massive neutrinos, 
later papers \citep{marulli11, ichiki11} have shown that if rescaled to the new value of $\sigma_8$, 
the Tinker mass function remains accurate for massive clusters ($M>10^{14}h^{-1}M_{\sun}$) -- which includes all SPT galaxy clusters.}
 (\sptcl{}) to the CMB leads the posterior probability to peak at slightly higher masses 0.55\,eV with a 68\% confidence interval of $\sumnu \in [0.28, 0.80]$\,eV.  
 The 95\% CL upper limit is $\sumnu<1.03\,\rm{eV}$. 
Both the CMB+LRG and CMB+\sptcl{} combinations show a preference for non-zero neutrino mass at
just under $2\,\sigma$.

\begin{figure}
\begin{center}
    \includegraphics[width=0.48\textwidth, trim=2.cm 13.5cm 1.8cm
2.7cm]{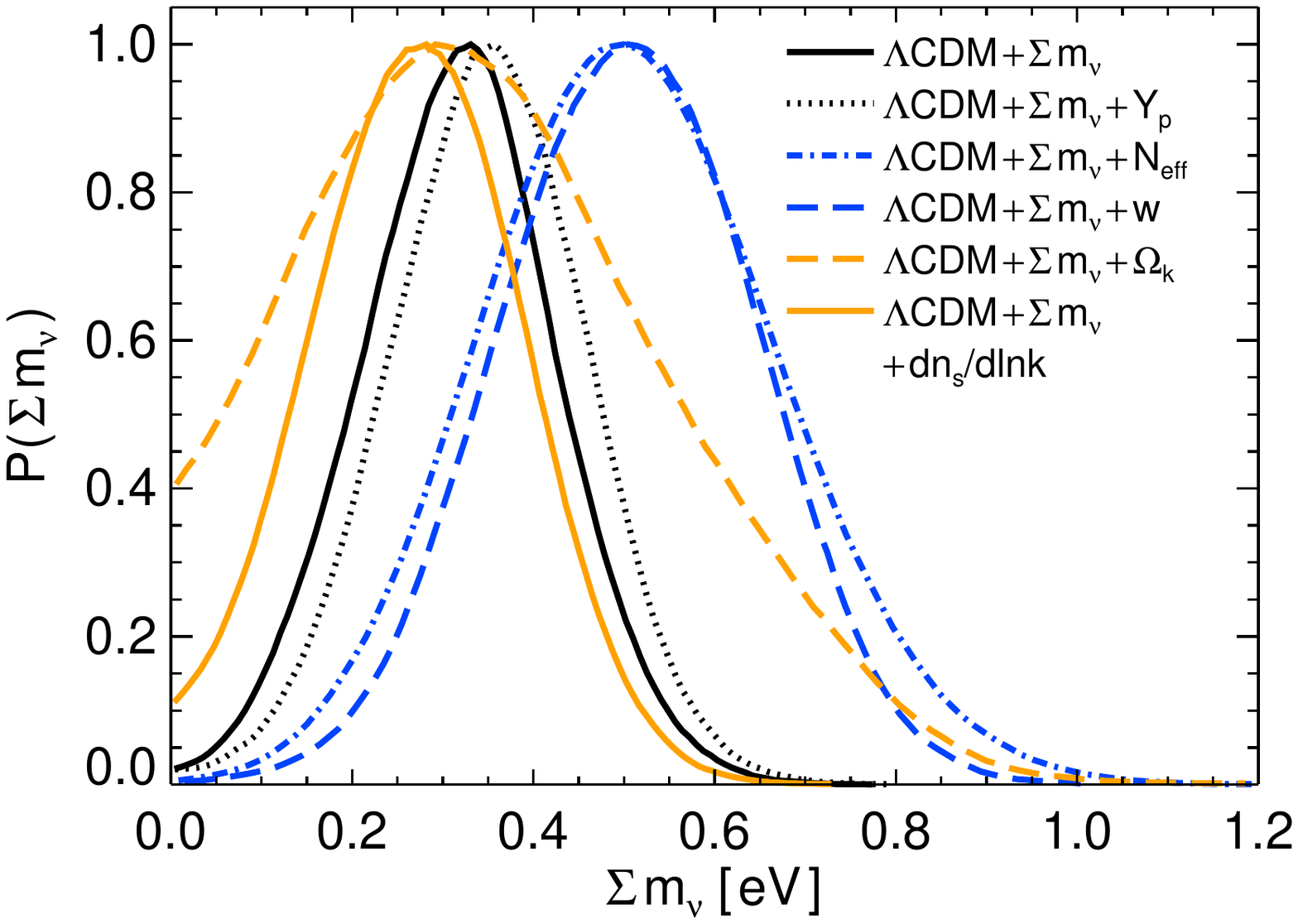}
\end{center}
\caption{This figure illustrates the robustness of the preference for positive neutrino masses 
to other parameter extensions.  
The marginalized one-dimensional posteriors for $\sumnu$ are shown for
two-parameter extensions to \LCDM{} for the combined
CMB+BAO+$H_0$+\sptcl{} data sets (for $w$, SNe are used instead of
$H_0$).  Allowing significant curvature or running can significantly
reduce the preference for nonzero neutrino masses (to $1.7$ and
$2.4\,\sigma$ respectively).  Other extensions increase the preference
for positive neutrino masses.  }
\label{fig:mnu_like_nuisance}
\end{figure}
 
Finally, we combine the CMB, geometrical, and large-scale structure observations.  
Because the SDSS LRG and SDSS BAO
galaxy samples overlap, we do not include the LRG dataset in this combination.
In this case, we find (for CMB+BAO+$H_0$+$\rm{SPT}_{\rm CL}$):
\begin{equation}
\label{eq:mnu_cmbbaoh0szc}
	\sumnu = (0.32 \pm 0.11)\,{\rm eV}, 
\end{equation}
\begin{equation}
\label{eq:mnu_cmbbaoh0szc2}
    \sumnu \in [0.01, 0.63]\,{\rm eV}\,\left(99.7\%\,{\rm CL}\right). 
\end{equation}
This data combination gives  a 3\,$\sigma$
preference for positive neutrino masses.
Without BAO, the preferred mass drops significantly while the uncertainties
increase; the constraint is consistent with no mass at $68\%$ confidence, and we end up with an upper
limit of $\sumnu<0.39\,{\rm eV}\,(95\%\,\rm{CL})$.   
Without \ho{}, the likelihood shifts to higher masses without degrading the
uncertainties, $\sumnu = (0.40\pm 0.11)\,{\rm eV}$.    

The are several drivers behind the \sumnu{} constraint (Equation~\ref{eq:mnu_cmbbaoh0szc}) derived from the dataset combination of CMB+BAO+$H_0$+$\rm{SPT}_{\rm CL}$:
\begin{itemize}
\item CMB: As discussed above, the CMB data contribute to this
  constraint by enforcing the height of the first and second acoustic
  peaks (from \wseven{} data) and the slope of the damping tail (from
  SPT data).  The latter leads to a lower \ns{}, with  $\sumnu$ thus
  increasing to compensate for the changes to the power in the \wseven{} region.
The resulting posterior distribution from the CMB peaks around $\sumnu\sim1$\,eV.
These CMB constraints contribute $\sim 1\,\sigma$ to the combined \sumnu{}
constraint. 
If we keep only the optical depth prior from \wseven, we find a confidence interval of $\sumnu \in [0.07, 0.53]$\,eV (95\% CL). 
Without the SPT bandpowers (and \ns information), the confidence interval is $\sumnu \in [0.03, 0.48]$\,eV (95\% CL). 
In both cases, there remains a greater than $2\,\sigma$ preference for nonzero masses. 

\item BAO: The BAO measurement is an important driver of the combined neutrino mass constraint.
As seen in Figure~\ref{fig:mnu_like}, the BAO measurements push up the preferred value of \sumnu{} and tighten the constraints considerably relative to the other datasets.
The preference for massive neutrinos is common between the three BAO datasets.
To confirm the robustness of the BAO preference, we have run three MCMC chains, dropping one of the BAO datasets in each chain. 
When dropping the SDSS, WiggleZ, or BOSS dataset, we find, at $95\%$ CL, $\sumnu \in [0.10, 0.55],~[0.10, 0.53],~{\rm or}~[0.03, 0.47]$\,eV respectively. 
In short, the BAO data prefer massive neutrinos, and the results are robust against dropping any single BAO observation.  
The largest shift of $0.6\,\sigma$ is introduced by dropping the BOSS measurement.  
However as previously noted, if all 
BAO measurements are removed and \ho{} information is included, the preference for massive neutrinos disappears.

\item \sptcl{}: Adding the \sptcl{} galaxy cluster sample reduces the neutrino mass uncertainties by a factor of 1.6 without significantly changing
the median value, as can be seen by comparing the value without clusters -- $\sumnu = (0.34 \pm 0.18)$\,eV --
 to the value with clusters -- $\sumnu=(0.32\pm 0.11)\,{\rm eV}$. 
In principle, this result depends on the accuracy of the cluster mass calibration, which was determined through an 
X-ray based scaling relation.  
To test this dependency, we have doubled the uncertainty on the normalization of the X-ray scaling relation and find little change in the distribution, $\sumnu = (0.34\pm 0.12)\,{\rm eV}$.  
We conclude that the reduction in uncertainties is insensitive to the specific prior on the mass scaling relation. 
\end{itemize}

As shown in Table~\ref{tab:lcdmmnu}, allowing for non-zero neutrino masses decreases the value of $\sigma_8$ inferred by the SPT and \wseven{} ($\sigma_8=0.658^{+0.075}_{-0.061}$) from the value obtained within the \LCDM{} model ($\sigma_8 = 0.795 \pm 0.022$) by roughly $2\sigma$.  Inferring $\sigma_8$ from the CMB is quite indirect, and allowing for neutrino masses introduces a large degeneracy which, when compared to \LCDM{}, significantly enlarges the uncertainty on $\sigma_8$.  The preference for positive neutrino mass discussed above shifts the median of $\sigma_8$ down.   Within the $\Lambda{\rm CDM}+\sumnu$ model, the constraints from adding SPT data are consistent with the constraints inferred from \wseven{} alone, $\sigma_8=0.685^{+0.079}_{-0.078}$ \citep{komatsu11}.  This is also consistent with the low redshift probe on $\sigma_8$.  For example, the constraint on the quantity $\sigma_8 (\Omega_m/0.25)^{0.47}$ from X-ray clusters gives $\sigma_8 (\Omega_m/0.25)^{0.47}=0.813\pm 0.013\,(\rm stat) \pm 0.024\,(\rm sys)$ \citep{vikhlinin09}, and  $\sigma_8 (\Omega_m/0.25)^{0.47} = 0.785 \pm 0.050$ from CMB within the same model - consistent at $0.5\,\sigma$.

We conclude that the preference for nonzero neutrino masses is coming from
three independent sources and remains at $>$\,$2\,\sigma$ significance
as long as we keep cluster abundances and at least one of the two most precise BAO
measurements.

\subsection{Degeneracies with other extensions}
\label{subsec:mnu_margi}

A final question is to what extent the neutrino mass constraint is weakened by introducing additional free parameters to the 7-parameter \LCDM+$\sumnu$ model.
To address this, we show the marginalized posteriors for the sum of the neutrino masses in 8-parameter models in Figure~\ref{fig:mnu_like_nuisance}.
We consider several possible additions: $\yp$, $\neff$, $dn_s/d\ln{k}$, $\Omega_k$, or $w$. 
All curves are for the combination of CMB+BAO+$H_0$+SPT${_{\rm CL}}$.  

Of these five extensions, two (\nrun and $\Omega_k$) decrease the preference for positive \sumnu, two ($w$ and \neff) increase the preference, 
and one (\yp) has minimal effect. 
Allowing both curvature and massive neutrinos significantly increases the
uncertainties on both parameters. 
The peak of the neutrino mass
likelihood remains nearly unchanged at $0.34\,{\rm eV}$ with the 95\% confidence interval expanded to $\sumnu \in [0, 0.70]$\,eV. 
The curvature is  consistent with zero.
Allowing for running of the spectral index reduces the preference for nonzero neutrino masses, in this case by shifting the median down; 
 the neutrino mass constraint with running is  $\sumnu = (0.27\pm 0.11)\,{\rm eV}$. 
On the other hand, allowing the dark energy equation of state or number of
neutrino species to vary increases the preference for positive neutrino masses.  
The summed neutrino mass is correlated with \neff{} (see \S~\ref{sec:neff_mnu} and \citealt{ichikawa05}) and anti-correlated with $w$ \citep{zhao12}. 
Freeing either of these parameters increases the median masses, although the uncertainties also increase.
For a wCDM+$\sumnu$ model, the mass constraint is $\sumnu = (0.51\pm
0.14)\,{\rm eV}$, $3.7\,\sigma$ above zero.
For a \LCDM+$\neff$+$\sumnu$ cosmology, the mass constraint is $\sumnu = (0.51\pm
0.15)\,{\rm eV}$.
This last case is discussed in detail in \S~\ref{sec:neff_mnu}.  
We find the CMB+BAO+$H_0$+SPT${_{\rm CL}}$ preference for neutrinos remains at greater than $95\%$ 
confidence when we add additional model parameters, with the exception of the mean curvature of the universe.  
Fitting the posterior with a Gaussian, we find curvature reduces the preference for massive neutrinos to $1.7\,\sigma$.  

\section{Running of the spectral index}
\label{sec:nrun}

The difference in preferred \ns{} from SPT and \wseven{} data suggests a  scale dependence in the power spectrum of the primordial fluctuations.
To test this, we allow the primordial power spectrum to deviate from a pure power law by introducing a logarithmic dependence on scale $k$, a so-called ``running spectral index'' \citep{kosowsky95}:
\be 
n_s(k) = n_s(k_0) + \nrun \ln\left(\frac{k}{k_0}\right).
\ee
Here, $k_0$ is a defined pivot point and \nrun is the running parameter.  
Throughout this section, we will define this pivot point to be $k_0 =
0.025$\,Mpc$^{-1}$, which projects to $\ell \simeq 350$. 
This pivot point is chosen to decorrelate the uncertainties on \ns{} and \nrun{} in the SPT+\wseven{} data. 

The running parameter \nrun{} is predicted to be undetectable by most inflationary theories, and a detection of non-zero \nrun{} could provide information about the inflationary potential \citep{kosowsky95}, or point to models other than inflation. 
There have been a number of recent CMB constraints on running. 
\citet{komatsu11} obtain $-0.084<\nrun<0.020\,(95\%\,{\rm CL})$
from \wseven.  
\citet{dunkley11} find $\nrun=-0.034\pm0.018$ from \wseven+ACT.  K11 use the combination of \wseven{} data and the first 790 
$\deg^2$ SPT survey data to obtain $\nrun = -0.024\pm0.013$, a preference for negative spectral running at 1.8\,$\sigma$.

\subsection{Constraints on \nrun}

\begin{figure*}
\begin{center}
    \includegraphics[width=0.40\textwidth, trim=5cm 13cm 1cm
1.8cm]{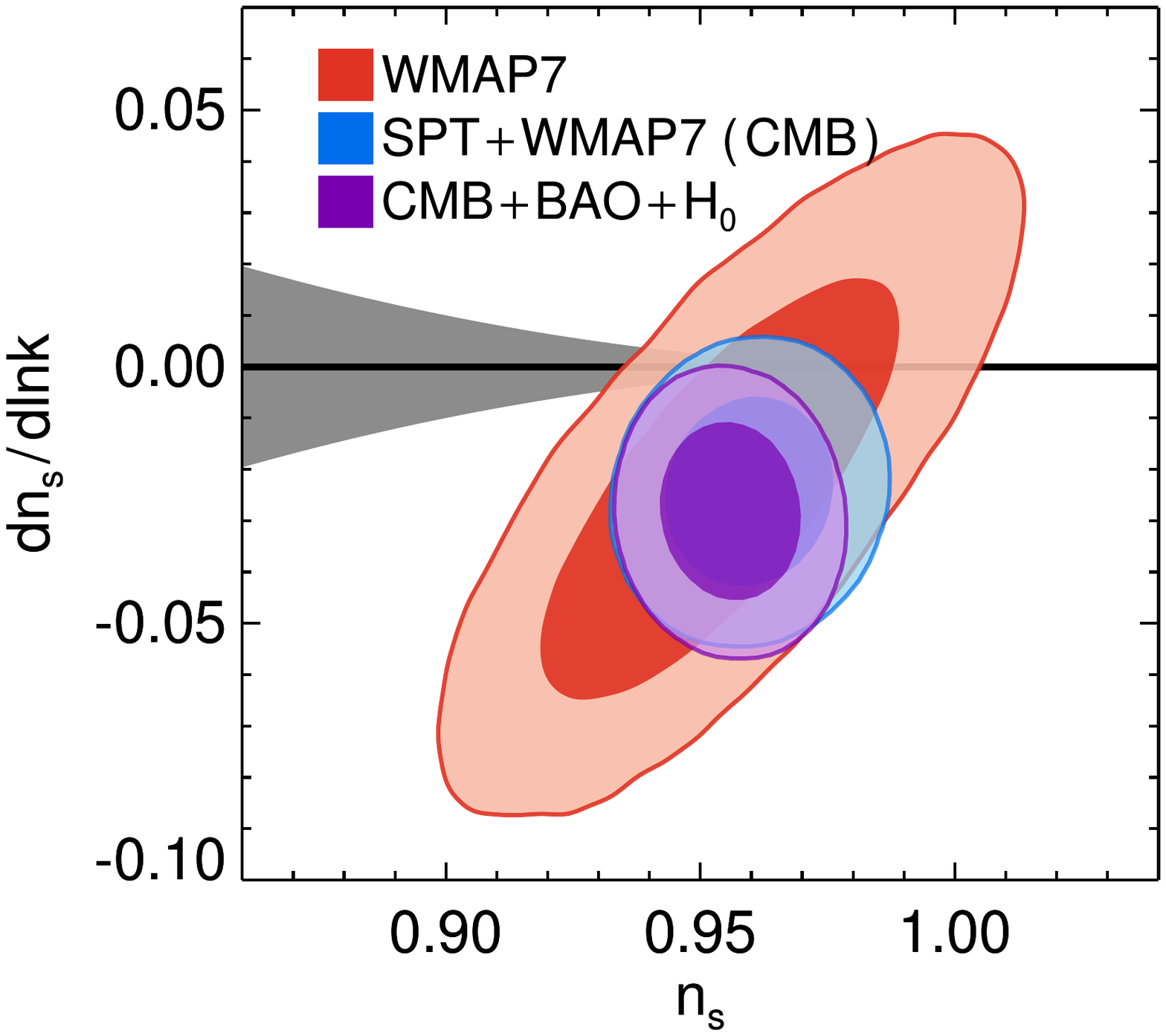}
    \includegraphics[width=0.40\textwidth, trim=2cm 13cm 4cm
1.8cm]{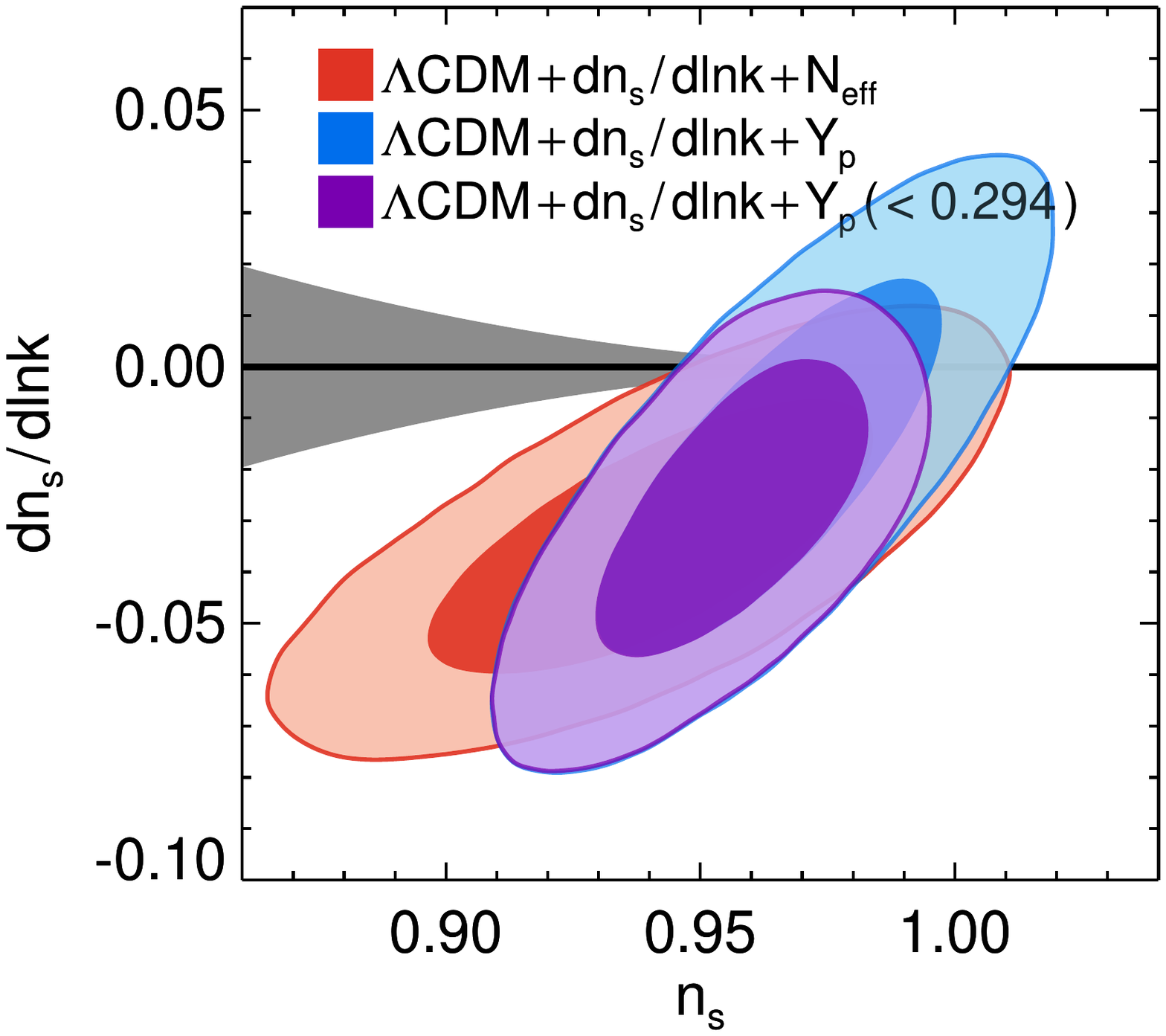}
\end{center}
\caption{
The data prefer negative running; this preference remains when additional parameters are freed.
We show the marginalized two-dimensional posteriors for \ns (at $k = 0.025$ Mpc$^{-1}$) and \nrun.  
\textbf{\textit{Left panel:}} the \ns - \nrun
joint distribution in the $\Lambda$CDM+\nrun model for different combinations of the
datasets. \textbf{\textit{Right panel:}} the \ns - \nrun joint distribution obtained by
marginalizing $\neff$ (red) and $\yp$ (blue) for \wseven+SPT data.  The purple
filled contours show the case with $\yp$ marginalized with the prior
$\yp < 0.294$, the $2\,\sigma$ upper limit of the solar initial helium
abundance by \citet{serenelli10}.  We show the region where $|\nrun| <
(1-\ns)^2$ in gray.  Inflation models with slow-roll expansion that can be
terminated at second order make predictions in this region.
}
\label{fig:ns_nrun}
\end{figure*}

We now look at the  constraints on \nrun from the SPT bandpowers. As shown in the left panel of Figure~\ref{fig:ns_nrun}, adding the SPT bandpowers to \wseven{} dramatically reduces the allowed likelihood volume, and leads to a preference for negative running. 
For SPT+\wseven{}, we find:
\be
\nrun = -0.024 \pm 0.011. 
\ee
The probability of negative running with the CMB data is $P(\nrun \le 0) = 98.6$\%, equivalent to a $2.2\,\sigma$ Gaussian preference.
In the appendix, we examine the dependence of the preference for negative running on the multipole range, and beam or foreground priors. We conclude that the preference does not depend strongly on the SPT experimental beam uncertainty or foreground modeling.

Adding BAO and \ho{} data marginally improves the constraints and shifts the median to more negative values. 
The combination CMB+BAO+\ho{} constrains \nrun to be
\be
\nrun = -0.028 \pm 0.010. 
\ee
The small shift from the CMB constraint is driven almost entirely by the BAO data, which prefers smaller values of \ho{}.
The probability of negative running with the CMB+BAO+\ho{} data is $P(\nrun \le 0) = 99.7\%$, equivalent to a $2.7\,\sigma$ Gaussian preference.

\begin{figure}
    \includegraphics[width=0.46\textwidth, trim=2.2cm 13.0cm 5.8cm
2.5cm]{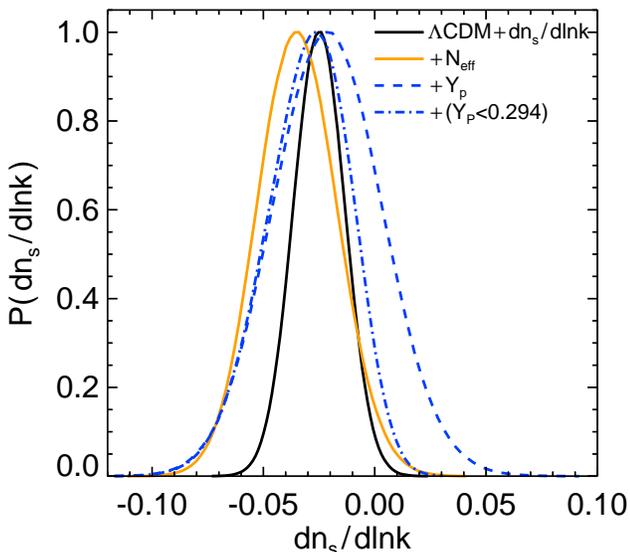}
\caption{
The preference of the SPT+\wseven{} data for negative running remains as additional parameters are freed. 
The marginalized one-dimensional posteriors for \nrun from the CMB in the \LCDM+\nrun model is shown by the black solid line. 
The preference for negative running is reduced when ${\rm Y_p}$ is allowed to
vary (blue dashed line), although this depends on extremely high helium abundances.
If an upper limit based on solar abundances is set on \yp{}, the preference for
negative running is mostly restored (blue dot-dashed line).
Allowing \neff{} to vary (orange solid line) increases the preference for negative running.
}
\label{fig:1Dnrun}
\end{figure}

We next consider how robust the preference for negative running is to other model extensions. 
Introducing tensor perturbations (by making the tensor-to-scalar ratio $r$ a free parameter) is a particularly natural extension in the context of inflationary models. 
However, the preference for running cannot be reduced by non-zero tensor perturbations.  
Tensor modes would increase the power at large scales without affecting small and intermediate scales.  
Adding tensor modes, therefore, increases the preference for negative running. 
 In a \LCDM+$r$+\nrun cosmology, we
obtain \nrun$ = -0.045\pm0.016$ for the CMB data and \nrun$ = -0.046\pm0.015$
for the CMB+BAO+$H_0$.  

More generally, the preference for negative running might be reduced by any extension that also effectively allows for an increasingly red, scale-dependent tilt.
We thus consider how the running constraint changes for the CMB data alone when freeing \sumnu{}, \neff{}, or \yp. 
For \nrun+\sumnu, we find $\nrun = -0.020 \pm 0.012$, similar to the $\sumnu = 0$\,eV constraint  of $\nrun = -0.024 \pm 0.011$. 
For \nrun+\neff, we find a shift to more negative running and larger uncertainties, $\nrun = -0.034 \pm 0.017$. 
We find the largest increase in the  uncertainties for \nrun+\yp, $\nrun = -0.020 \pm 0.024$. 
Therefore, the significance of the preference for running is most reduced by freeing \yp{} and, to a lesser extent, \neff{}, and we plot constraints for these extensions in the right panel of Fig.~\ref{fig:ns_nrun} as well as Fig.~\ref{fig:1Dnrun}. 

As will be discussed in \S~\ref{sec:early}, the primary effect of \yp{} and \neff{} is on the damping scale -- which obviously mimics  running. 
This degeneracy expands the uncertainty on \nrun. 
The shift in the ISW effect due to massive neutrinos does not mimic the scale dependence of running very well and has minimal effect. 
Freeing \neff{} and \yp{} yield different results because, as will be discussed in \S~\ref{sec:two}, \neff{} also changes the locations of the acoustic peaks -- which running and \yp{} do not. 
This reduces the degeneracy between the parameters and also explains the shift in the preferred value for running. 
With \nrun+\neff{} free, the model is free to move to lower values of \neff{} to better match the observed peak locations while compensating for the decreased damping with negative running. 

Among the models considered, the preference for negative running is only removed by freeing the helium abundance.
However, the inferred helium abundance is in tension with the $2\,\sigma$ upper limit of the protosolar measurement that will be discussed in \S~\ref{subsec:yp}. 
Forcing \yp{} to be less than this $2\,\sigma$ upper limit of $\yp < 0.294$ degrades the effectiveness of \yp{} in removing the preference for negative running. 
This point is illustrated by the purple contours in the right panel of Figure~\ref{fig:ns_nrun} and the blue dot-dashed line in Figure~\ref{fig:1Dnrun}. 
Excluding values of \yp{} above 0.294, the preference for running is robust against other model extensions. 

\subsection{Implications for Inflationary Models}
We now explore the implications of the constraint on running for models of inflation.
Although we expect \ns{} to have some scale dependence, \nrun is predicted to be undetectably small with current data in single-field, slow-roll inflation models. 
These models predict \nrun to be second order in the slow roll parameters, and thus of order $(1-\ns)^2$ unless the potential experiences significant jerk in the observable range \citep{chung03,finelli10}.
A significant detection of running is a potential problem for both small and large-field inflation models. 
Small-field inflation models predict $r \simeq 0$, thus one can interpret the \ns{}--\nrun constraints shown in Figure~\ref{fig:ns_nrun} as a direct test of small-field models. 
The observed value of \nrun disfavors single-field small-field inflation with negligible jerk at the $\sim 2\, \sigma$ level ($\sim 2.7\,\sigma$ for CMB+BAO+\ho). 
Large-field inflation models, which predict non-negligible $r$ (and thus more negative running), are even more disfavored.

Models with significant jerk at observable scales can allow for larger absolute values of running \citep{easther06},  
however for  $n_s \lesssim 1$, these models also predict a number of
e-foldings $N \lesssim 30$.\footnote{Allowing for fourth or higher derivatives
can increase the number of e-foldings, as is the case with large-field chaotic
models with sinusoidal modulations \citep{kobayashi11} or extra-dimensional
natural inflation models \citep{feng03}.} 
Here $e^N$ gives the increase in the scale factor between the time when the observable scale leaves the horizon and the end of inflation.
Such small values of $N$ are incompatible with a standard cosmological history \citep{liddle03}, as they imply an energy scale of inflation which is below  the electroweak phase transition \citep{easther06}.
 We therefore treat the prediction that \nrun is of order $(1-\ns)^2$
 as a fairly robust one for single-field slow-roll models.

Finally, we mention that the choice of a constant \nrun{} as a departure from power law behavior in the primordial spectrum is not the only possible choice. 
 It is also possible that the running is significant over only a small range of wavenumber $k$; \citet{cline06} argue that such a parametrization fits more easily into the slow-roll framework and allows inflation to continue for a longer number of e-foldings. 
 
As we conclude this section on running, it is worth reiterating that the statistical preference for running is less than $3\sigma$.
However, the other model extensions we consider do not significantly weaken this preference for running.
If the constraints from future experiments confirm this preferred value of running, it would represents an important clue
about the earliest epochs of the universe.

\section{Constraints on other one-parameter extensions}
\label{sec:early}


In this section, we consider other physically motivated models that can affect the scale-dependence of the tilt of the CMB power spectrum.  Specifically, we test the effect of  
(1) varying the effective number of neutrino species (\neff{}), a probe of the standard model of particle physics, and (2) an increase in 
the primordial fraction of baryonic mass in helium ($\yp$), a possible signature of non-standard BBN.  
These extensions can increase the predicted damping due to photon diffusion relative to \lcdm, which preferentially decreases small-scale power.  
The data therefore favor these extensions.
Conversely, the data will disfavor extensions that exacerbate the discrepancy between the observed and predicted small-scale power.
Examples of this type of extension include the tensor-scalar ratio $r$, which as discussed in S12 leads to larger values of $n_s$, and the effect of early dark energy, which we consider below.

\subsection{\neff}
\label{subsec:neff}

In the standard thermal history, 
the radiation density in the early Universe is given by
\be
\rho_r = \rho_\nu + \rho_\gamma = (1 + 0.227 \,N_{\rm eff}) \rho_\gamma,
\ee
where $\rho_\gamma$ is the photon density, a quantity measured
extremely well by {\it COBE}/FIRAS \citep{fixsen96b},  
and \neff{} is an effective number of neutrino species. 
The factor of 0.227 is calculated assuming all the entropy of $e^+$, $e^-$ annihilation goes into the
photons. 
This assumption is not perfect; the small correction due to the neutrinos gaining some entropy has traditionally been incorporated into \neff{}. 
For  the standard model of 
three neutrino species,  $N_{\rm eff} = 3.046$ \citep{dicus82, lopez99, mangano05}. 
Although generally discussed in the context of extra neutrino species, \neff{} parametrizes the total non-photon contribution to the radiation density and thus
includes contributions from any weakly/non-interacting,
relativistic particles.

\begin{figure*}
\begin{center}
    \includegraphics[width=0.48\textwidth, trim=2.2cm 13.5cm 5.8cm
2.5cm]{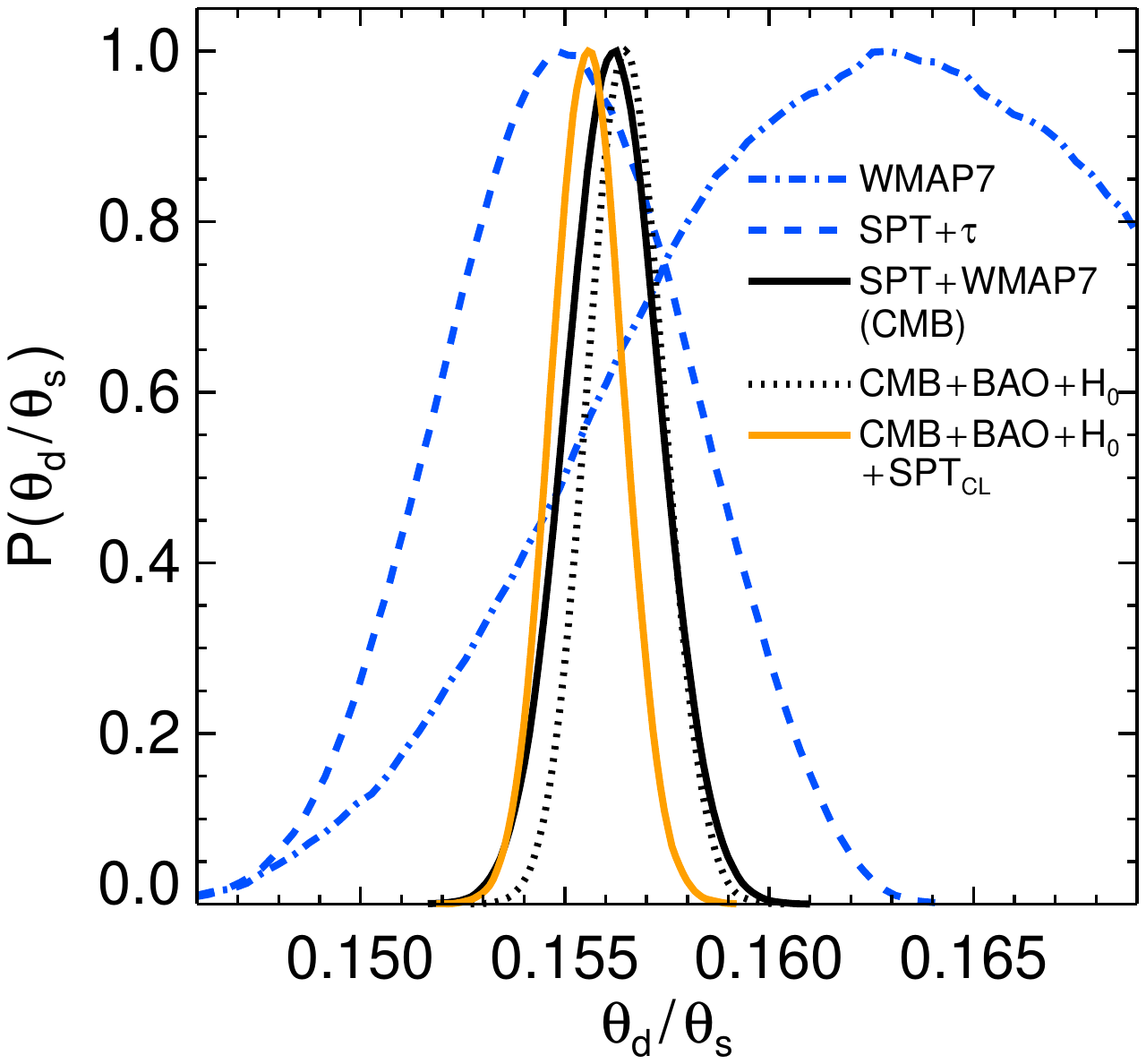}
    \includegraphics[width=0.48\textwidth, trim=2.2cm 13.5cm 5.8cm
2.5cm]{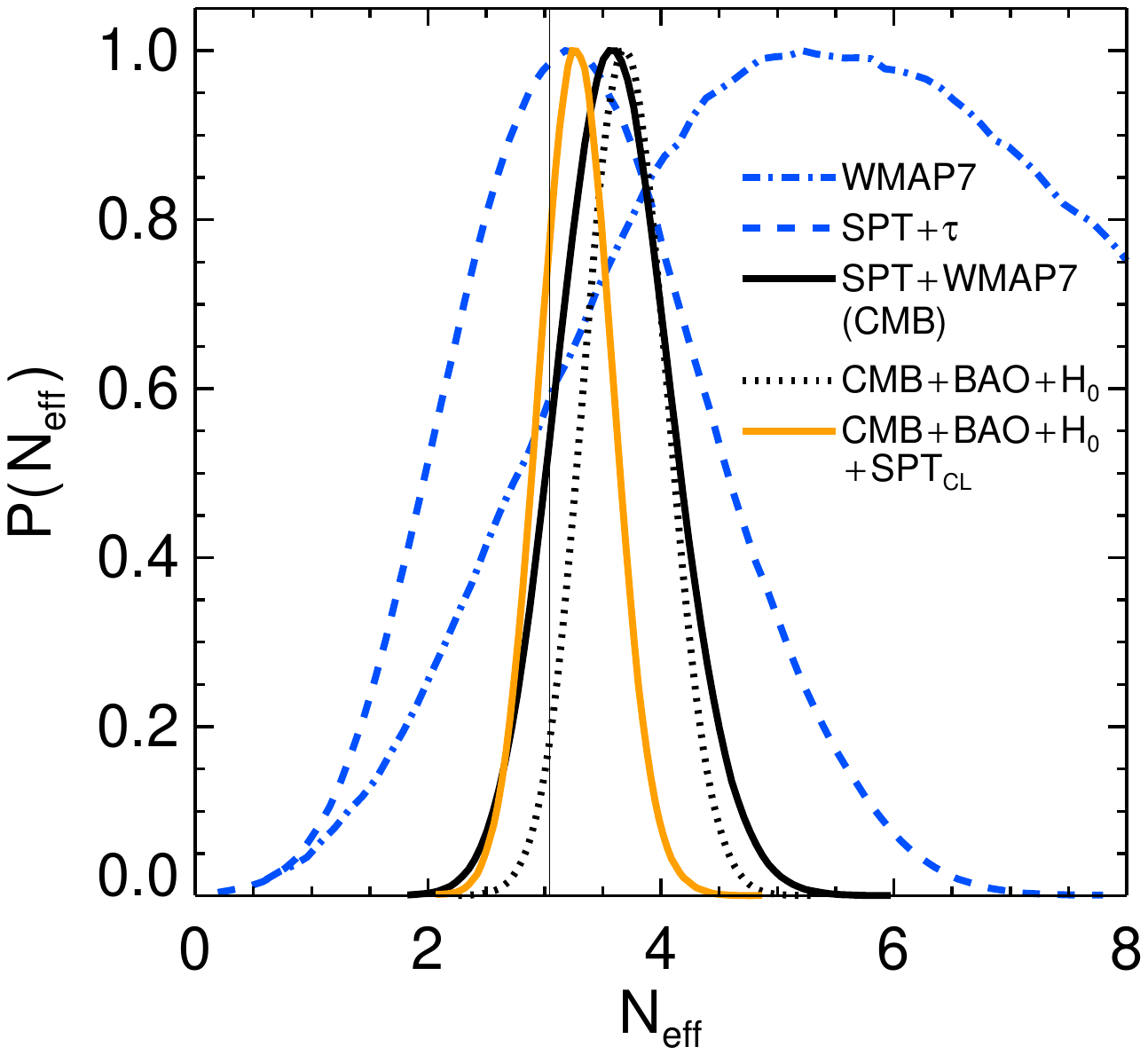}
\end{center}
\caption{The marginalized one-dimensional posteriors for $\theta_d/\theta_s$ (\textit{left}) and
\neff{} (\textit{right}) from various combinations of datasets.
The constraint on $\neff$ in the $\Lambda$CDM+$\neff$
model for different combinations of datasets can be interpreted from the corresponding $\theta_d/\theta_s$ posterior.
The shift in $\theta_d/\theta_s$ between broad \wseven{} (dot-dashed blue curve)
and SPT (dashed blue curve) implies a preference for lower $\neff$ with SPT than \wseven{}.  The $\neff$ posterior of 
SPT+\wseven{} is shown by the solid black curve.  The tighter constraints on $\neff$ can be obtained by adding BAO and $H_0$ (black dotted curve)
and combining CMB, BAO, $H_0$ and SPT SZ-selected clusters (solid orange curve).}
\label{fig:neff}
\end{figure*}

As described by \citet{bashinsky04} and \citet{hou13a}, the SPT+\WMAP{} data
constrain \neff{} through its effects on the expansion rate at early times.
Both the sound horizon $r_s$ and the square of the damping length $r_d$ depend on an integral
with an integrand proportional to $1/H(z)$.  
However, the angular power spectrum is sensitive to the corresponding angular scales
$\theta_d = r_d/D_A$ and $\theta_s = r_s/D_A$, where $D_A$ is the
angular-diameter distance to recombination.  
Since the angular-diameter distance is not known, measurements of $\theta_d$  or $\theta_s$ alone do not constrain the expansion rate well.
The dependence on the unknown angular-diameter distance can be removed by looking at the ratio $r_d/r_s = \theta_d/\theta_s$.  
As the effect \neff{} has on the expansion rate is roughly redshift independent, its effect can be pulled out of the integral, so in this extension $\theta_d/\theta_s \propto H^{1/2}$.  
A second effect of \neff{} is on the phase of the acoustic oscillations.
Constraints from this phase shift are subdominant in the \lcdm+\neff{} model, but become more important when \yp{} is also allowed to vary; we discuss this effect in that context in \S~\ref{sec:two}.
The \neff{} constraint in the \lcdm+\neff{} model can thus be interpreted through its effect on the $\theta_d/\theta_s$\ ratio.

Here we hold \yp{} to the value predicted by BBN, which is a function of both \neff{} and $\omega_b$.
These \yp{} predictions are implemented in CosmoMC through interpolating over a table produced using PArthENoPE v1.00 \citep{pisanti08}.  
We look at the effect of varying \yp{} independent of BBN constraints later in this section, and we consider jointly varying \neff{} and \yp{} independently in \S~\ref{sec:two}.

The left panel of Figure~\ref{fig:neff} shows the CMB constraints on 
$\theta_d/\theta_s$ in the \LCDM{}+\neff{} model space.  The
parameter $\theta_s$ is well-measured by both SPT and \wseven, and the
fractional shift in preferred $\theta_s$ is small between the two datasets.
As a result, the uncertainty in \neff{} is primarily due to the
uncertainty in $\theta_d$.  In \lcdm, $\theta_d$ is primarily
constrained from the \wseven{} determinations of $\omega_b$ and
$\omega_m$ -- not by the damping tail measurement.  Freeing \neff{}
greatly broadens the \wseven{} constraint on $\theta_d$ due to a
degeneracy between \neff{} and $\omega_m$.  With \neff{} free, the SPT determination of
$\theta_d$ from the damping tail becomes tighter than that from \wseven{}.

Constraints on \neff{} are shown in the right panel of Figure~\ref{fig:neff}.  
The dot-dashed
blue curve marks the broad \wseven{} posterior; \citet{komatsu11}
find $\neff>2.7$ (95\% CL) using \wseven{} alone.  As expected from the
observed shift in $\theta_d / \theta_s$ between \wseven{} and SPT, the
SPT data
prefer lower values of \neff{} than \wseven.  Adding the SPT data to
\wseven{} markedly improves the measurement as shown by the black curve.
The joint SPT+\wseven{} constraint is:
\begin{equation*}
 \neff = 3.62\pm 0.48,
\end{equation*}
representing a $20\%$ reduction in uncertainty from the constraint determined from the \wseven +K11 bandpowers.  
For the CMB data, the probability that $N_{\rm eff} > 3.046$  is 89\%.  In Table~\ref{tab:lcdmneff}, we report the 
constraint on \neff{} from the datasets discussed in this section, including constraints on other parameters of particular
interest.

We now turn to the addition of the late-time BAO and \ho{} data.  When
\zeq{} is held fixed at its well-measured value from the CMB,
increasing the effective number of neutrino species results in an
increased expansion rate, which decreases $r_s$.  Since $\theta_s$ is
well constrained, this results in a decrease in the angular-diameter
distance $D_A$, and thus $D_V$.  The end result is that, for the BAO
observable $r_s/D_V$, the changes largely cancel.
The CMB+BAO
constraint is $\neff=3.50\pm 0.47$.  
The direct \ho{} measurement is more sensitive to the value of \neff{}. 
However, the \ho{} and CMB datasets prefer similar values of \ho and the preferred value of \neff{} 
hardly moves when \ho{} data is added. 
The resulting CMB+\ho{} constraint is $\neff=3.46\pm 0.35$.  The significance of the
preference for $\neff > 3.046$ is largely unchanged between these
three cases: the CMB, CMB+BAO, and CMB+\ho{} prefer $\neff >
3.046$ at $1.2\,\sigma$, $1.0\,\sigma$, and $1.2\,\sigma$,
respectively.

Though adding BAO or \ho{} data individually to the CMB data slightly reduces the preferred value of \neff{}, 
 combining all three datasets has the opposite effect of shifting the distribution towards slightly  higher \neff{}. 
The joint CMB+BAO+$H_0$ constraint is $\neff = 3.71\pm 0.35$, 
a $1.9\,\sigma$ preference for $\neff > 3.046$. 
The reasons for this shift can be seen in Figure~\ref{fig:bao_vs_h}.
Increasing \neff{} moves the CMB-predicted \ho{} and BAO quantities
towards the measured values.  This ability of high \neff{} to
reconcile the CMB with BAO and \ho{} has been noted previously by,
e.g., \citet{anderson12}.  The inclusion of the BAO and \ho{}
information leads to a small upward shift in the value of \neff{};
a higher value of \neff{} is required to reduce the tension
between these datasets than that inferred from the CMB
determination of $\theta_d/\theta_s$.

Prior to the decay of the gravitational potentials, the additional radiation density from extra neutrino species deepens potential wells, thus boosting structure growth at very early times.  
This increases the predicted abundance of massive galaxy clusters today. 
We therefore combine the SPT$_{\rm CL}$ galaxy cluster counts with CMB, BAO and $H_0$ data to obtain the tightest constraint on the number of effective neutrino species,
\begin{equation}
N_{\rm eff} = 3.29 \pm 0.31.
\end{equation}
This is consistent at less than 1 sigma with the standard model of 3 neutrino species.

The effective number of neutrino species is partially degenerate with other extensions to \lcdm{}, in particular, \nrun{}, \yp{}, and \sumnu{}. 
Including \nrun{} shifts the preferred value of \neff{} down by $1\,\sigma$ to $\neff=2.98\pm0.38$; the effect of including \yp{} is nearly identical. 
Including \sumnu{} has the opposite effect and moves the preferred value of \neff{} up by $1\,\sigma$ to $\neff=3.86\pm0.37$.  
The slight preference for $\neff > 3$ disappears when either \nrun{} or \yp{} are freed, but is strengthened when freeing \sumnu.

In summary, CMB data primarily constrain \neff{} through measurements of the ratio $\theta_d/\theta_s$.
Adding \ho{} and BAO measurements tightens this constraint and leads to a modest (1.9\,$\sigma$) preference for more than 3 neutrino species.  
This preference is driven by tension between the BAO and $H_0$ datasets.  
The tightest constraints come from the combination of the CMB, BAO, $H_0$ and cluster abundances and are consistent with 3 neutrino species at 0.8\,$\sigma$. 
The \neff{} constraints are sensitive to the addition of further model extensions.

\begin{table*}
\begin{center}
\begin{threeparttable}
\caption{\LCDM + $\neff$ Results from SPT, CMB, CMB+BAO, CMB+$H_0$ and
CMB+BAO+$H_0$}
\footnotesize
\begin{tabular}{c | c  c c c c c c c}
\hline
Datasets & $\neff$ & $\yp\footnote{As calculated from BBN theory}$ & $10^2\Omega_{b}h^2$ &
$H_0\;[\mathrm{km\;s^{-1}\;Mpc^{-1}}]$ & $z_{\rm eq}$ & $10\;\theta_d/\theta_s$ \\
\hline
\\
SPT+$\tau$ & $3.36^{+1.16}_{-1.03}$ & $0.252\pm 0.014$ & $2.307\pm 0.122$ & $77.1^{+9.0}_{-8.1}$ & $3053\pm 192$ & $1.552\pm 0.030$ \\
CMB  & $3.62\pm 0.48$ & $0.255\pm 0.006$ & $2.268\pm 0.049$ & $75.9\pm 3.4$ & $3136\pm 96$ & $1.562\pm 0.011$ \\
CMB+BAO  & $3.50\pm 0.47$ & $0.254\pm 0.006$ & $2.232\pm 0.045$ & $71.4\pm 2.5$ & $3326\pm 50$ & $1.561\pm 0.011$ \\
CMB+$H_0$ & $3.46\pm 0.35$ & $0.253\pm 0.005$ & $2.253\pm 0.038$ & $74.5\pm 1.9$ & $3161\pm 84$ & $1.559\pm 0.009$ \\
CMB+BAO+$H_0$ & $3.71\pm 0.35$ & $0.256\pm 0.004$ & $2.247\pm 0.038$ & $72.6\pm 1.8$ & $3322\pm 49$ & $1.565\pm 0.009$ \\
CMB+BAO+$H_0$+\sptcl & $3.29\pm 0.31$ & $0.251\pm 0.004$ & $2.223\pm 0.037$ & $71.2\pm 1.7$ & $3268\pm 43$ & $1.556\pm 0.009$ \\
\hline
\end{tabular}
\label{tab:lcdmneff}
\begin{tablenotes}
\item This table shows constraints on the \lcdm + $\neff$ model.
   In addition to $\neff$, we include constraints on other parameters
   of particular interest for this extension.  In this model space,
   $\yp$ is related to $\Omega_b h^2$ and $\neff$ by a BBN consistency relation.
\end{tablenotes}
\end{threeparttable}
\end{center}
\end{table*}

\subsection{$\yp$}
\label{subsec:yp}

\begin{figure*}
\begin{center}
    \includegraphics[width=0.45\textwidth, trim=2cm 13cm 2.5cm
1.8cm]{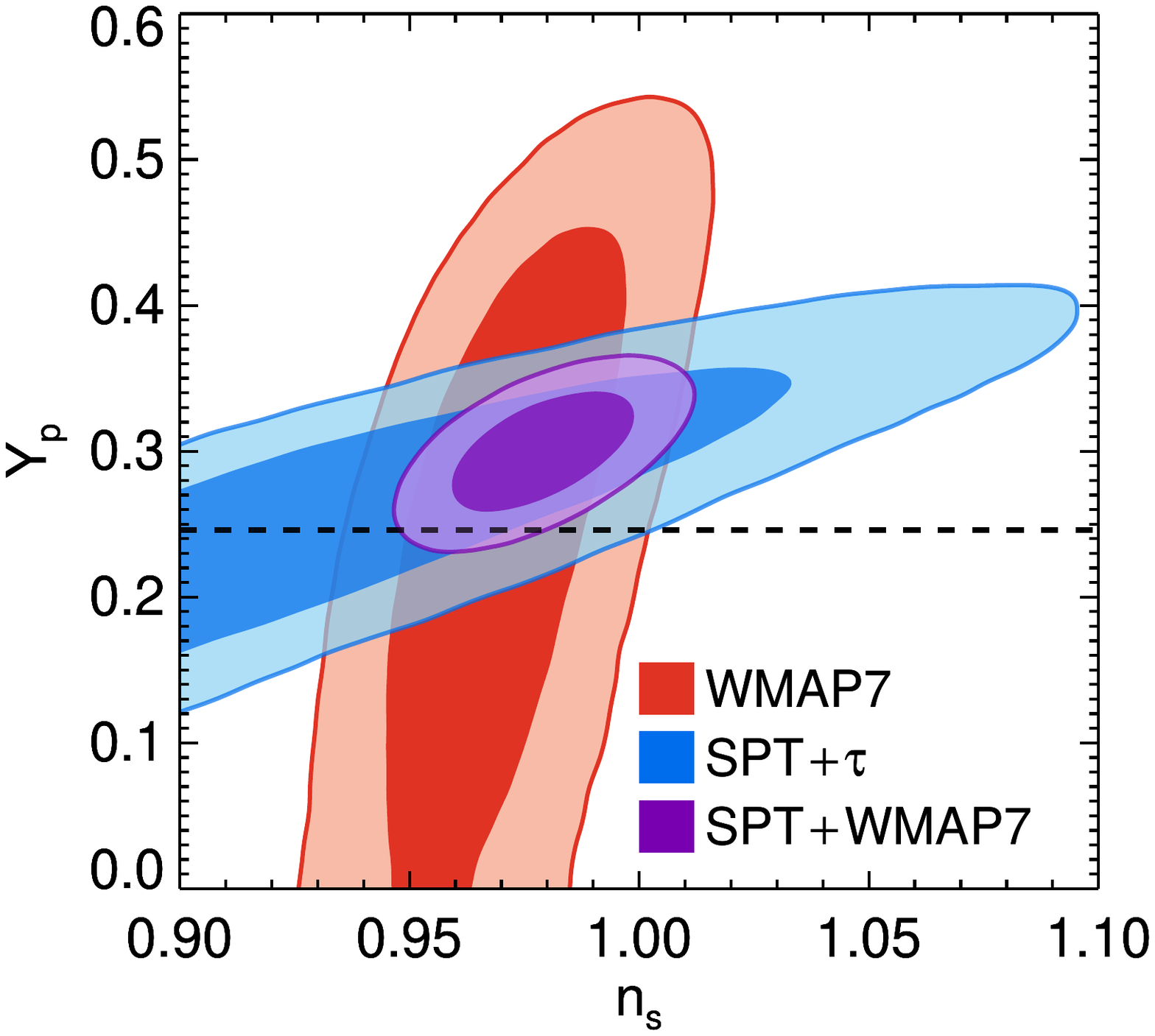}
    \includegraphics[width=0.45\textwidth, trim=2cm 12.95cm 6.3cm
2.0cm]{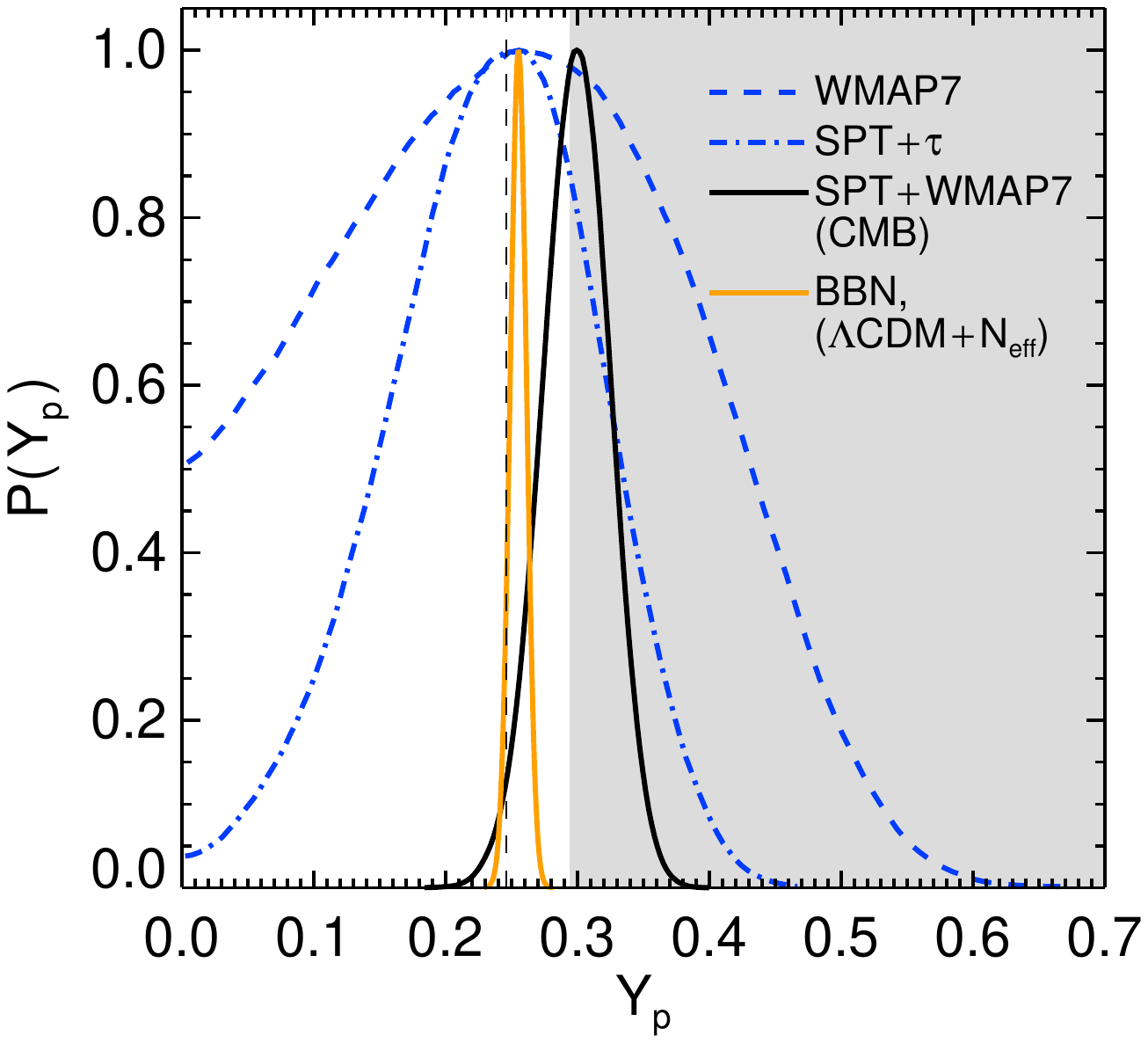}
\end{center}
\caption{The CMB data place independent constraints on the primordial helium abundance, \yp.
\textbf{\textit{Left panel:}}  The contours show the two-dimensional 68\% and
95\% confidence intervals for \yp{} and \ns{} for three datasets: SPT, \wseven, and
SPT+\wseven.
The BBN prediction for 3 neutrino species is marked by the horizontal black dashed line at $\yp=0.2478$.  
The change in the damping scale due to varying \yp{} can be mimicked by changing \ns{} for the SPT data. 
\wseven{} provides an independent measurement of \ns, greatly enhancing the SPT+\wseven{} measurement of the helium abundance. 
\textbf{\textit{Right panel:}} 
The marginalized one-dimensional posteriors for \yp{} in different model and data combinations. 
We show constraints from the SPT (blue dot-dashed line), \wseven{} (blue dashed line), and SPT+\wseven{} (black solid line).
The BBN prediction for 3 neutrino species is marked by the vertical black dashed line at $\yp=0.2478$.  
The orange solid line shows the marginally wider BBN prediction from the $\Lambda$CDM+$\neff$ MCMC chain. 
The grey area is ruled out by the 95\% upper limit from the protosolar helium abundance. 
The preference of the CMB for higher helium abundances can be reduced by introducing other extensions that affect small scale power, 
as is demonstrated in \S~\ref{sec:neff_yp}. 
}
\label{fig:yp}
\end{figure*}

When the universe cools to $T~\sim$ 0.1\,MeV, light nuclei begin to form, a
process known as big bang nucleosynthesis \citep[BBN,][]{schramm98, steigman07}.  
The primordial fraction of baryonic mass in $^4$He is denoted as \yp{}.  As mentioned in \S~\ref{subsec:neff}, BBN makes a precise prediction for the primordial helium abundance.  A useful analytic form
 is given by \citet{simha08}, valid for \neff{} near the standard model prediction:
\begin{equation}
\label{eq:yp}
\yp = 0.2485 + 0.0016\,[(273.9\,\omega_b-6)+100(S-1)], 
\end{equation}
where
\begin{equation}
S^2 = 1 + (7/43)(\neff-3.046).
\end{equation}
The $S^2$ factor accounts for any non-standard expansion rate prior
to and during BBN. 
We use the BBN prediction for \yp{} in nearly all MCMCs, unless we specifically state that \yp{} is free.

At fixed baryon density, increasing the helium fraction leads to increased damping. 
Helium has a higher binding energy and thus recombines earlier than
hydrogen; each neutral helium atom effectively absorbs four free electrons at the surface of last scattering.
This decreases the number of free electrons which increases the photon diffusion length, $r_d$, as seen in Equation~\ref{eqn:rd}. 
Mirroring \S~\ref{subsec:neff}, this single distance measure is degenerate with the angular diameter distance to last scattering. 
However, since the sound horizon scale is nearly independent of \yp{}, the ratio of $\theta_d/\theta_s$ breaks the degeneracy with $D_A$ and allows the CMB data to constrain \yp{} independently of BBN.

Precise experimental inferences of \yp{} have come from spectral observations of
extragalactic clouds of low-metallicity ionized gas \citep{izotov07,
peimbert07, izotov10, aver10, aver11, aver12}.  Historically, these results have shown substantial
variations between groups that greatly exceed the expected statistical
variations; for a more complete discussion of these
measurements, see e.g., \citet{aver10}.  Marginalizing over all possible trends of
the estimated \yp{} with metallicity increases the uncertainty
substantially: for example, \citet{aver12}  find $\yp = 0.2534
\pm0.0083$, $0.7\,\sigma$ above the BBN prediction for \yp{} assuming the three standard
neutrino species and the value of $\omega_b$ measured by \WMAP.

An upper limit  to \yp{} can be inferred from protosolar abundance
estimates resulting from helioseismology; \citet{serenelli10} find the protosolar helium abundance to be $Y_\mathrm{solar} = 0.278
\pm 0.008$.  This measurement relies upon well-studied gravitational and solar
evolution physics to back out the protosolar abundance from current measurements of the
helium fraction on the solar surface.  
It is difficult to imagine mechanisms by which the protosolar abundance is reduced below the primordial helium abundance. 
Therefore some measure of  caution is warranted in interpreting cosmological constraints with primordial helium abundances above
 the $2\,\sigma$ upper limit of $\yp < 0.294$ derived from these measurements.

 After combining \wseven{} with 
ACBAR and QUaD temperature power spectra, \citet{komatsu11} determine $\yp=0.326\pm0.075$.
\citet{dunkley11} obtain $\yp=0.313\pm0.044$ by combining the ACT and WMAP7 data.
Using the bandpowers from the first 790 $\deg^2$ SPT survey and WMAP7 data, K11 find 
$\yp=0.296\pm0.030$.  In this analysis,
we infer $\yp = 0.240\pm 0.079$ with the SPT bandpowers alone.
These results are shown in the right panel of Figure~\ref{fig:yp}.
The combined CMB data lead to:
\be
\yp = 0.300\pm 0.025.
\ee
As shown in the left panel of Figure~\ref{fig:yp}, the addition of the \wmap7{} data breaks a degeneracy present in the SPT data between \yp{} and \ns.  The fact that the resulting \yp{} is higher than the BBN-consistent value reflects the fact that, in the \lcdm{} model, where \yp{} is required to be BBN-consistent, SPT prefers a lower value of $n_s$ than does \wseven{}.
Since \yp{} has none of the late-time effects of neutrino species, adding BAO and \ho{} data hardly change the \yp{} constraint.
The constraint from CMB+BAO+\ho{} is $\yp = 0.305\pm 0.024$.  
However, slightly more than half the parameter space allowed by the CMB+BAO+\ho{} data is ruled out by the protosolar upper limit. 

The constraints presented above can also be compared to the BBN predicted value of $\yp = 0.24774\pm 0.00014$.
The $\sim$\,0.06\% uncertainty quoted here does not include the 0.2\%
theoretical uncertainty on \yp{} in the PArthENoPE code used in CAMB
\citep{pisanti08}.
The inferred \yp{} value for CMB+BAO+\ho{} is $2.4\,\sigma$ 
above the BBN prediction.  
We explore the relationship between the damping-tail inference of \yp{} and the BBN prediction in more detail in \S~\ref{sec:two}, where we consider an expanded model with both \yp{} and \neff{} free. 
It is worth repeating that the feature of the CMB data that drives these high values of \yp{} -- a trend of decreasing power at higher multipoles relative to \lcdm{} predictions -- can also be explained by other model extensions, such as \nrun{} or \neff{}, both discussed above.

\subsection{Early dark energy}
\label{subsec:ede}

The dark energy density could be considerable in the early universe if the equation of state parameter $w$ was much larger in the past, a situation commonly referred to as early dark energy. 
Early dark energy models lead to much larger signatures in the
CMB anisotropy than traditional dark energy models; see \citet{reichardt12a} for
a discussion of the effects of early dark energy. 
One reason these models are interesting is that they can be constructed with attractor solutions that
reduce the necessity to fine-tune the initial conditions \citep{wetterich88, ratra88}. 
For instance in ``tracer" models, the dark energy equation of state is required
to be equal to the background equation of state -- e.g., during radiation
domination, $w=1/3$.  The advantages are that the DE density can stay much closer to the
energy density of the dominant component
and that the DE evolution is independent of initial conditions. 

Instead of considering a specific model, we choose to constrain early dark energy in the
more general  tracer model parametrization by \cite{doran06}. 
This parametrization introduces two new parameters in addition to the $\Lambda$CDM set,
the dark energy equation of state at $z=0$, \oeqs, and the dark energy density
relative to the critical density at early times, \omegae, which is taken to be
constant at sufficiently high redshift ($z \gtrsim 10$). 
The dark energy density and equation of state are then given by
\beqa
\od{}(a) &=&  \frac{\odo{} - \omegae \left(1- a^{-3 \oeqs{}}\right) }{\odo{} +
\omo a^{3\oeqs}}  + \omegae \left (1- a^{-3 \oeqs}\right)\\
w(a) &=& -\frac{1}{3[1-\od(a)]} \frac{d\ln\od(a)}{d\ln a} + \frac{a_{eq}}{3(a + a_{eq})}. 
\eeqa

Here $a_{eq}$ is the scale factor at matter-radiation equality, 
and \odo{} ($\omo$) is the dark energy (matter) density relative to
critical density at $z=0$. 
The dark energy equation of state today is $w_0$. 
This parametrization assumes spatial flatness so that $\omo+ \odo = 1$.
Since we force $\od(a)$ to be constant at high redshift, the dark
energy equation of state mimics that of the dominant component at early times,
thus behaving like a tracer model.
Later, during matter domination at $z < 10$, the equation of state transitions towards
its current value, \oeqs, so it can account for cosmic acceleration.

To consistently describe the perturbations, we are motivated by quintessence models
to treat dark energy as a perfect fluid with a sound
speed, $c_s$, equal to the speed of light (see \citealt{hu98}). 
This choice, together with the parametrization for the background
evolution, completely specifies the behavior of dark energy. 
We require $\oeqs \ge -1$ and thus do not entertain the possibility of
``phantom crossing'' (see, e.g.,~\citealt{fanghulew08}). 
This restriction allows us to avoid pathologies
in perturbation evolution and to stay in the quintessence regime. 

Like all the other extensions, early dark energy changes the tilt
of the CMB power spectrum between small and large scales.  The
addition of SPT data therefore helps to further constrain the level of
early dark energy.  For a detailed physical explanation of how early dark energy is constrained by the CMB data, see \citet{reichardt12a}.

The small-scale CMB temperature
anisotropy power measurement from the SPT bandpowers improves the constraints on the
early dark energy density over WMAP7 alone by a factor of 3.5; the 95\% upper limit on \omegae{} is reduced from 0.052 for WMAP7-only to 0.013
for WMAP7+SPT. 
This is a 38\% improvement on the upper limit of $\omegae < 0.018$ reported for
WMAP7+K11 \citep{reichardt12a}. 
Adding low-redshift geometrical measurements does not help constrain early dark energy, although, these data have a dramatic effect on the quality of the
constraints on the late-time dark energy density and equation of state. 
The upper limit is essentially unchanged at  $\omegae < 0.014$ for WMAP7+SPT+BAO+SNe. 
The $\omegae < 0.013$ bound from WMAP+SPT is the best published constraint from the Cosmic Microwave Background.

\section{Two-parameter extensions}
\label{sec:two}

We will now consider two 2-parameter extensions to \LCDM: \neff+\yp{}
and \neff+\sumnu.  Of the many possible 2-parameter
combinations, we limit our discussion to these two extensions because
they are physically well-motivated.  The first case is an interesting
consistency test of BBN, while the second case is a natural space to
consider for sterile neutrinos.

\subsection{\neff{} and \yp}
\label{sec:neff_yp}

\begin{figure}
    \includegraphics[width=0.45\textwidth, trim=2.2cm 12.5cm 5.5cm
3cm]{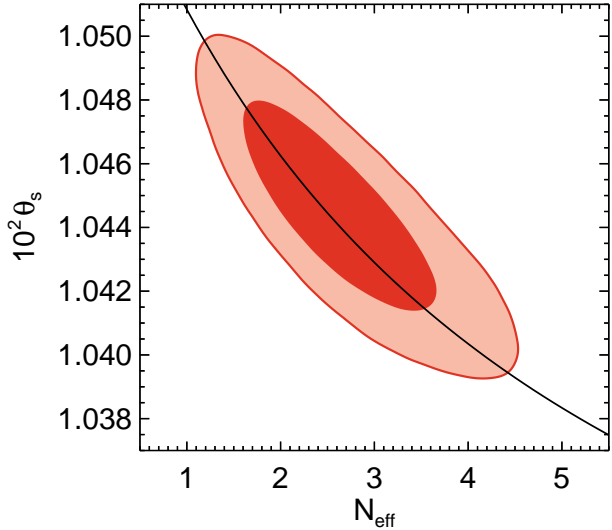}
\caption{
Changing \neff{} introduces phase shifts in the acoustic oscillations, however, constraints on these phase shifts are weakened by a degeneracy between \neff{} and $\theta_s$.
Here, we show the marginalized two-dimensional posteriors for
$\neff$ and $\theta_s$ from the CMB data for the $\Lambda$CDM+$\neff$+$\yp$ model.  
The black curve shows the correlation between the two parameters due to the phase shifts in the acoustic oscillations. 
This curve is defined using the \citet{bashinsky04} prescription for the phase shifts, normalized to the best-fit $\theta_s=1.0429\times 10^{-2}$ at $\neff=3.046$.  
}
\label{fig:neff_thetas}
\vskip 10 pt
\end{figure}

Here we examine simultaneous constraints on the effective number of 
neutrino species and the primordial abundance of $^4$He:
\lcdm+\neff+\yp.  This two-parameter extension to \LCDM{} is
moderately favored by the data with a $\Delta\chisq = 5.5$ (see
Table~\ref{tab:dchisq_ext}), though it does little to ease the tension
between datasets beyond the one-parameter \lcdm+\neff{} model.

We first discuss the physical mechanisms that lead to joint constraints on \neff{} and \yp{}.
As noted earlier in \S~\ref{subsec:neff}
and \ref{subsec:yp}, both the number of neutrinos and helium abundance
primarily impact the damping scale. 
However, the parameters are not fully degenerate due to several
additional effects of varying \neff{} as discussed in \citet{hou13a}.
Here we highlight in particular the role of shifts in the acoustic
peak locations induced by phase shifts in the acoustic oscillations \citep{bashinsky04}. 

Acoustic oscillations in a constant gravitational potential will have the form $\cos(k
r_s(\eta) + \varphi)$, where $k$ is the wavenumber and the phase
$\varphi = 0$ as a consequence of the initial conditions in inflation models. 
Changes in the gravitational potentials shift the phase away from
zero.  Unlike photons, neutrinos have a long free-streaming distance in the early plasma. 
They thus alter the evolution of the gravitational potential and
therefore the resulting phase shifts. For modes that entered the horizon in the
radiation-dominated era ($\ell > \ell_{eq} \simeq 434$),
\citet{bashinsky04} find that the change to the phase shift due to
neutrino free streaming is $\Delta \varphi = 0.19\pi \rho_\nu/(\rho_\nu+\rho_\gamma)$. 
Here $\rho_\gamma$ and $\rho_\nu$ are the energy densities of the photons and neutrinos respectively.
This phase shift changes the positions of the acoustic peaks in the CMB data. 

Constraints on the phase shifts are weakened by a partial
degeneracy between the angular size of the sound horizon
and a phase shift.  
This point is illustrated by  Figure~\ref{fig:neff_thetas} which shows the 68\% and 95\% confidence intervals  on \neff{} and \thetas{} from the CMB. 
Changing the phase shift by $\delta\varphi$ and the sound horizon by $\delta \theta_s$ moves the $m^{\rm th}$ acoustic peak by  $\delta \ell_m = (\delta \varphi - \ell_m
\delta \theta_s)/\theta_s$.  
We can thus hold fixed the position of any single acoustic peak by
varying $\theta_s$ to accommodate the phase shift.  
The black curve in the figure is the locus of points that does exactly
this, although not for a particular peak but for a particular $\ell$
value.   We chose $\ell = 1666$ because it is approximately where the signal-to-noise is maximized in the SPT bandpowers. 
The major axis of the constraint ellipse falls along the black line,
confirming that the calculated phase shifts explain the observed correlation between \neff{} and $\theta_s$. 
 The degeneracy between $\neff$ and $\theta_s$ is broken
(the contours close) in part because the compensating value of
$\theta_s$ is $\ell$-dependent; there is no one value of $\theta_s$
that preserves all the peak locations.  Other effects play a role in
the degeneracy breaking as well, including the early ISW effect and
effects due to the high values of the baryon fraction that occur at low $N_{\rm eff}$
\citep{hou13a}.

\begin{figure}
\begin{center}
    \includegraphics[width=0.48\textwidth, trim=1.5cm 15cm 1.2cm
1.8cm]{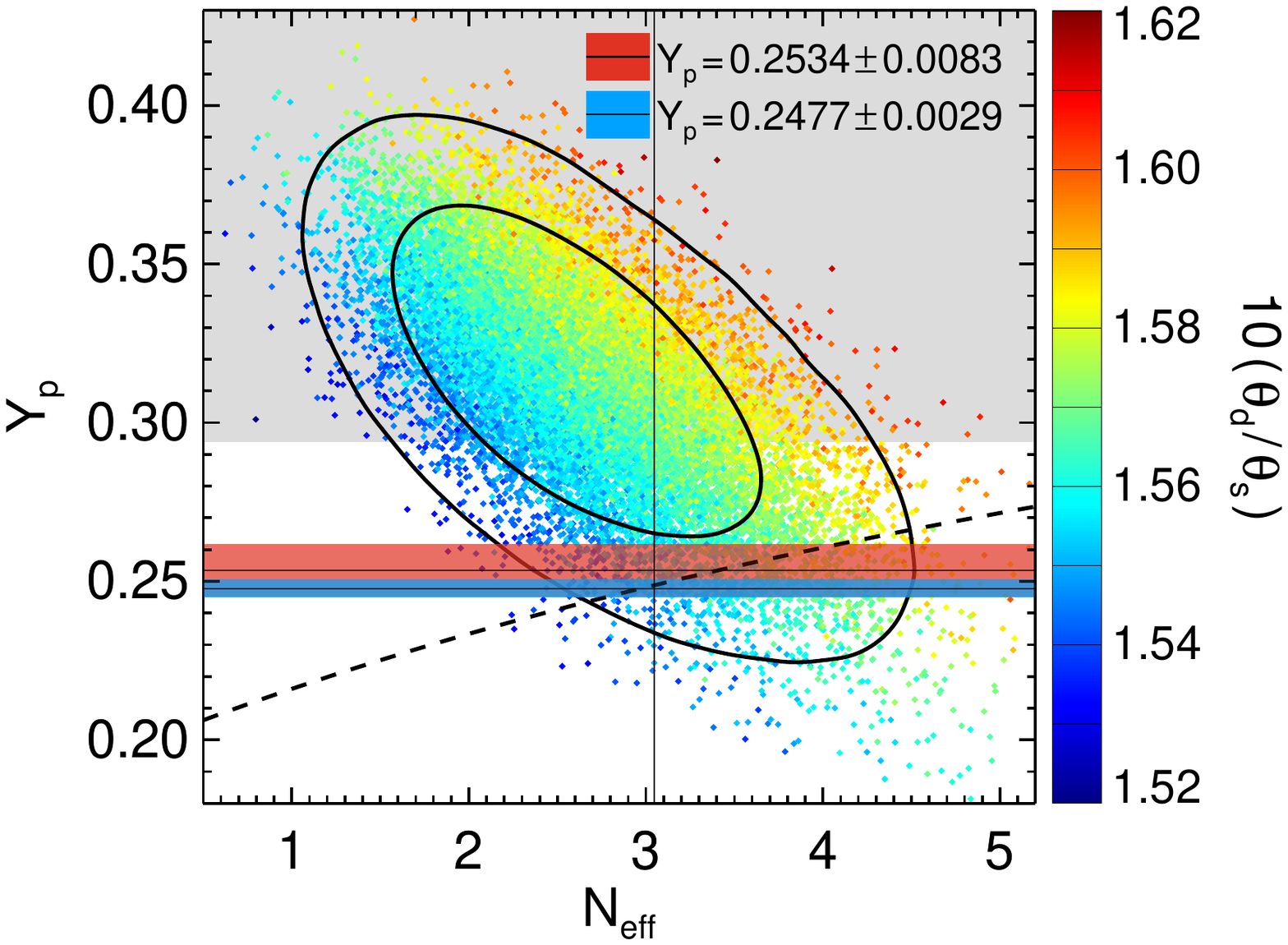}
\end{center}
\caption{
This plot contrasts the CMB-derived two-dimensional likelihood contours for \neff{} and \yp{}
with the predictions of BBN and other \yp{} measurements. 
The black solid contours are the 68\% and 95\% confidence intervals for
SPT+\wseven.   
The scattered points are samples from the Markov chain with the color encoding
the value of $\theta_d/\theta_s$.  
The dashed black curve marks the standard BBN relationship between \neff{}
and $Y_p$.  Similar to the right panel of Figure~\ref{fig:yp}, the gray area is ruled out by the 95\% upper limit 
of the measurement of solar initial helium abundance by \citet{serenelli10}. The $\yp$ measurement
from spectral observations of extragalactic low-metallicity ionized gas clouds by \citet{peimbert07} 
and \citet{aver12} are shown by the light blue and red bands respectively.
}
\label{fig:neffyp}
\end{figure}

The \neff-\yp{} contours from the CMB dataset are shown in
Figure~\ref{fig:neffyp}.  The contours form an ellipse, with the minor
axis (the better constrained direction) oriented in the direction of
varying $\theta_d/\theta_s$. 
Relative to \wseven+K11, the new SPT bandpowers have the most impact on the high \neff{} region. 
This improved constraint on \neff{} is due to improved constraints on the neutrino-induced phase shifts. 

The inferred values of \neff{} and \yp{} for the CMB data are:
\begin{eqnarray}
\yp & = & 0.314 \pm 0.033,\\
\neff & = &  2.60\pm0.67.\nonumber
\end{eqnarray}
When the BAO and \ho{} observations are added, the median values of these parameters 
shift slightly towards the standard model expectations, and the uncertainties tighten:
\begin{eqnarray}
\yp & = & 0.294 \pm 0.030,\\
\neff & = &  3.32\pm0.45.\nonumber
\end{eqnarray}

We compare these CMB-derived results with the predictions from BBN theory
(see Equation~\ref{eq:yp}) in Figure~\ref{fig:neffyp}.  
The dashed curve in Figure~\ref{fig:neffyp} denotes the BBN theory predictions.  
The data are consistent with BBN theory, although
they display a mild preference for higher-than-predicted values of \yp{}. 
 The region above the $2\,\sigma$ upper
limit from the initial solar helium abundance \citep{serenelli10} is greyed out. We also
plot the \citet{peimbert07} and \citet{aver12}
estimates of \yp{} based on spectroscopic observations of low-metallicity
extragalactic clouds (see \S~\ref{subsec:yp}) as horizontal bands.  The
CMB+BAO+\ho{} inference for the helium abundance is consistent with the
spectroscopic observations of \citet{aver12} at $1.3\,\sigma$.

\begin{figure*}[htbp]
\begin{center}
   \includegraphics[width=0.48\textwidth, trim=2.7cm 13cm 3.0cm
2.5cm]{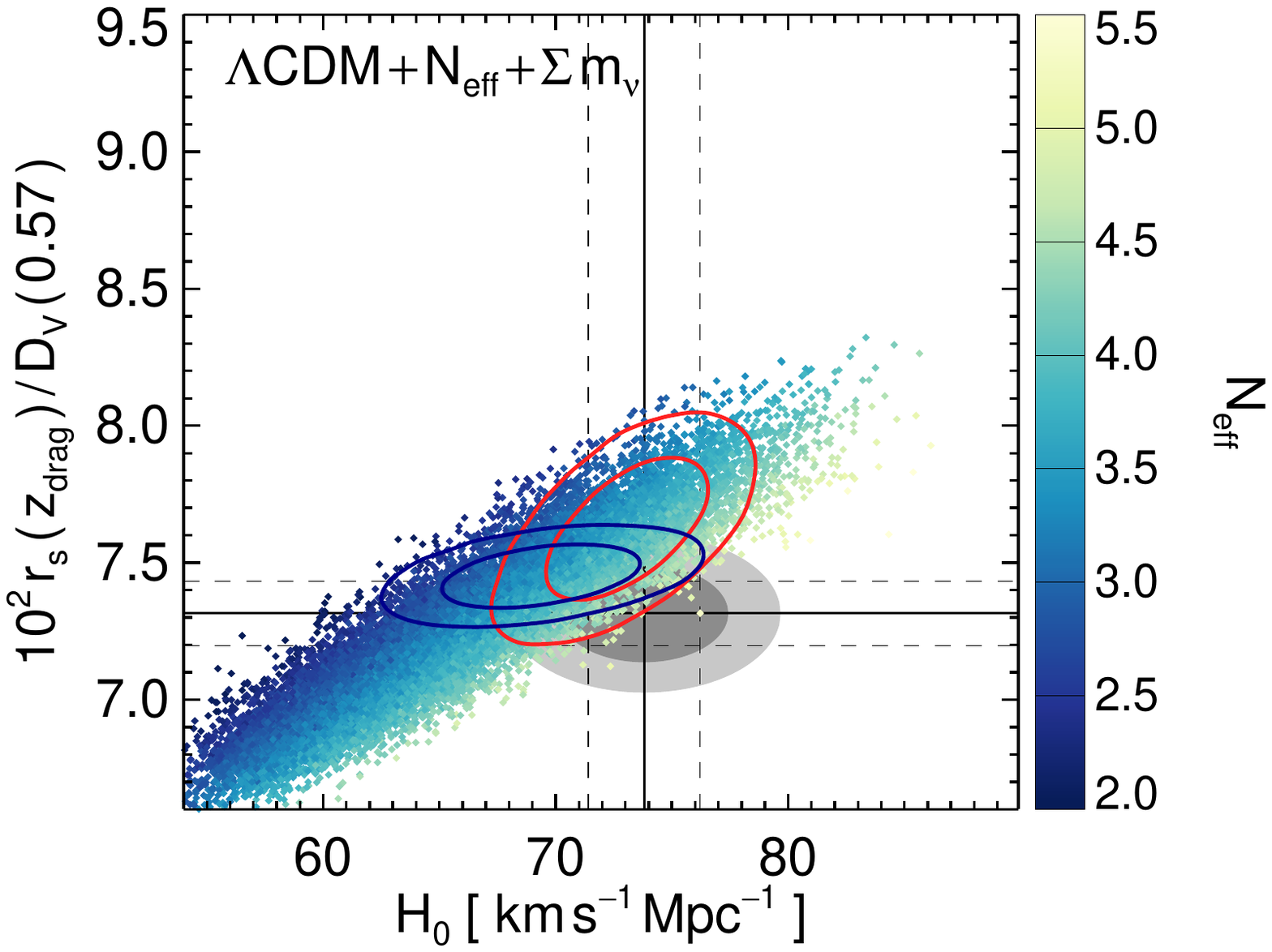}
   \includegraphics[width=0.48\textwidth, trim=2.3cm 13cm 3.4cm
2.5cm]{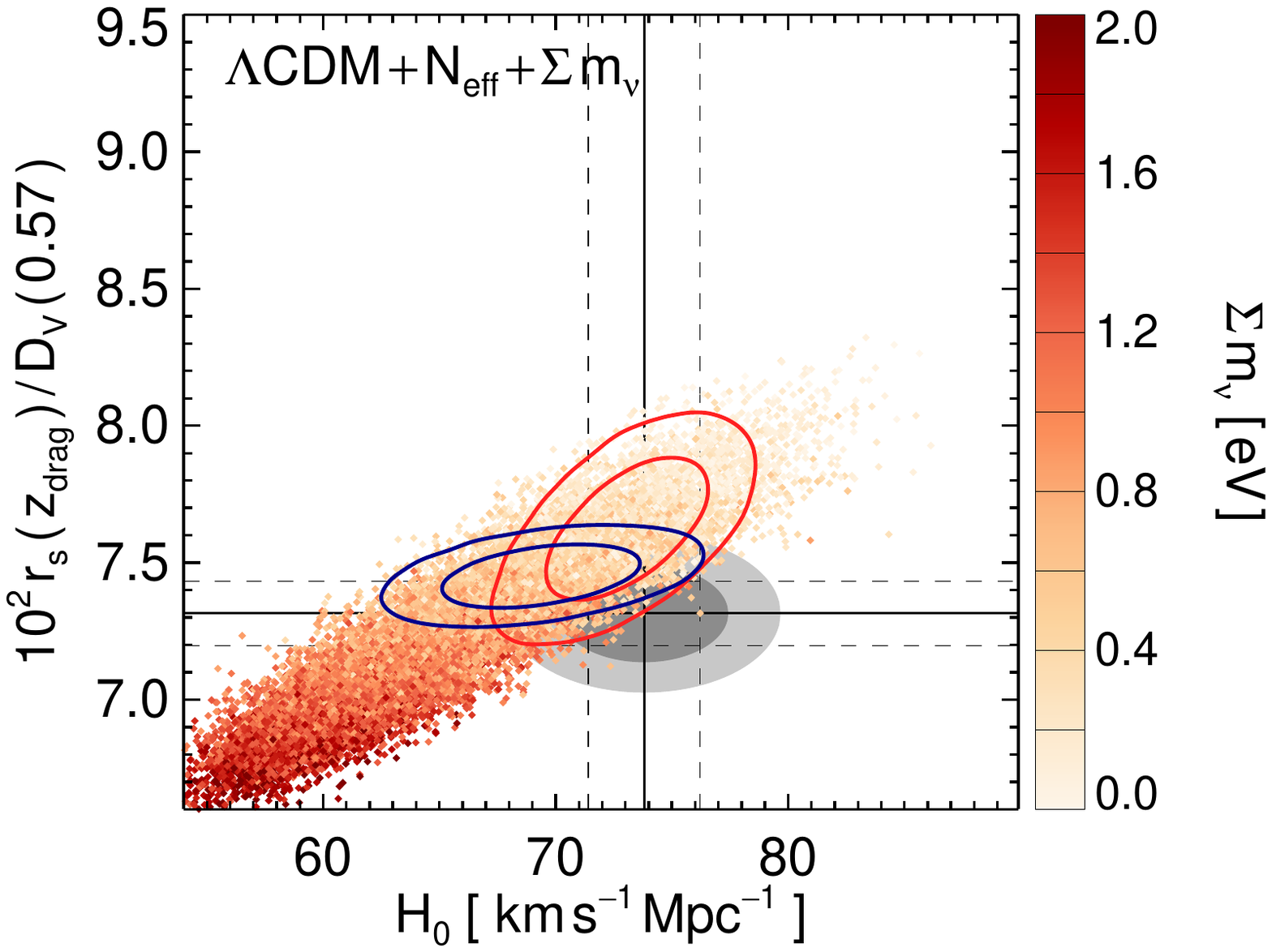}
\end{center}
\caption{The $H_0 - r_s/D_V$ plane for the two-parameter extension
$\Lambda$CDM+$\neff$+$\sumnu$ with the color of scattered points coding the
values of $\neff$ (left panel) and $\sumnu$ (right panel).  The colored
contours and the gray filled regions are as described in
Figure~\ref{fig:bao_vs_h}.}
\label{fig:neff_mnu_scatter}
\end{figure*}

In summary, the CMB data can constrain \yp{} and \neff{} simultaneously by measuring the damping scale and the location of the acoustic peaks.
The combination of CMB+BAO+\ho{} prefers a value of \neff{} which is consistent with three neutrinos species and a value of \yp{} slightly above the BBN prediction.

\subsection{\neff{} and \sumnu}
\label{sec:neff_mnu}

\begin{figure}
\begin{center}
    \includegraphics[width=0.48\textwidth, trim=2cm 13cm 2.5cm
1.8cm]{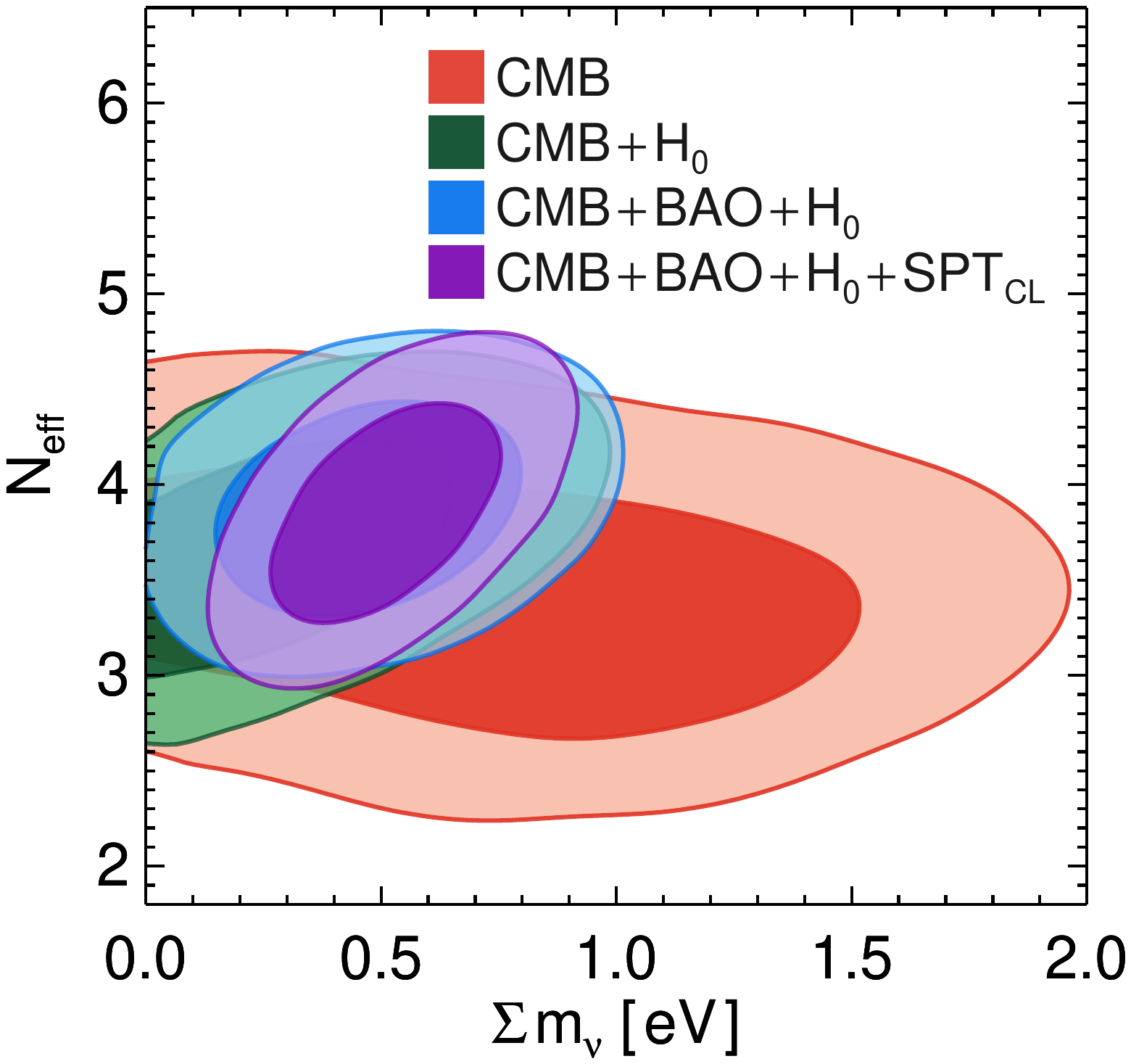}
\end{center}
\caption{This figure demonstrates the impact of each combination of datasets on the constraints on \sumnu{} and \neff. 
The shaded contours are the 68\% and 95\% confidence intervals for the following data combinations: SPT+\wseven{} (CMB; red), CMB+\ho{} (green), CMB+\ho+BAO (blue), CMB+\ho+BAO+\sptcl{} (purple). 
The combined data are in $>$\,2\,$\sigma$ tension with the \LCDM{} assumption of three massless neutrino species.}
\label{fig:mnu_neff}
\end{figure}

We finally focus on a two-parameter extension, \LCDM+\sumnu+\neff, in which both the mass and number of neutrino species are freed. 
This is a well-motivated, natural model expansion of the individual \sumnu{} and \neff{} 
extensions discussed earlier since (1) neutrino oscillation experiments have established that neutrinos are
massive \citep{robertson08} and (2) the event excess in the MiniBooNE search for 
$\bar\nu_\mu\rightarrow \bar \nu_e$ oscillations \citep{aguilar-arevalo10} and the
reactor antineutrino anomalies \citep{mention11} suggest one or more massive ($\sim$\,1\,eV) sterile neutrino species.
We will continue to assume equal masses in all neutrino species.  For treatments of (3+1) or (3+2) models in the literature, see e.g. \citet{kopp11} or \citet{gounti11}.

Cosmological observations can rule out large areas of the \neff-\sumnu{} plane. 
\citet{benson13} investigated constraints from \wseven, K11, BAO, \ho, and SPT galaxy clusters, and found $\neff = 3.91\pm0.42$ and $\sumnu=(0.34\pm0.17)$\,eV. 
A more recent study by  \citet{riemer-sorensen12b} used WMAP7+K11+WiggleZ+$H(z)$+SNLS3 to measure 
$\neff = 3.58^{+0.15}_{-0.16}\;(68\%\;{\rm CL})$ and $\sumnu<0.60\,{\rm eV}\;(95\%\;{\rm CL})$.  The tightness of the constraints on \neff{} from this study are driven by the inclusion of the $H(z)$ inference from measuring the age-redshift relation of passively evolving galaxies out to $z \sim 1.75$ \citep{moresco12}, which depends heavily on modeling stellar evolution.  The extended arm out to higher redshifts provides greater constraining power on the expansion history, placing constraints on \neff; conversely, this information has little impact on the neutrino mass constraint. 

As discussed in  \S~\ref{sec:extensions}, this two-parameter extension significantly improves the quality of fit to the CMB+BAO+\ho{} data  by $\chisq=7.9$,  equivalent to a $2.3\,\sigma$ Gaussian preference. 
This model also increases the consistency between the CMB, BAO, and \ho{} datasets, as  illustrated by Figure~\ref{fig:neff_mnu_scatter}. 
Both panels show  the $r_s/D_V(z=0.57)$--\ho{} plane and are identical except for the color-coding, which encodes \neff{} in the left panel and \sumnu{} in the right panel.
Like the \LCDM+\neff{} model shown in Figure~\ref{fig:bao_vs_h}, this 2-parameters extension allows the contours from the three datasets to overlap comfortably.

\neff{} and \sumnu{} are individually constrained by separate features in the CMB as described in \S~\ref{sec:mnu} and \S~\ref{subsec:neff} -- the early
ISW effect for $\sumnu$, the damping scale and the position of the acoustic peaks for $\neff$. 
We can use Figure~\ref{fig:neff_mnu_scatter} to understand phenomenologically the relative interplay between \neff, \sumnu, \ho, and the BAO features, given a CMB prior. 
In the left panel, \neff{} increases as \ho{} increases and $r_s/D_V$ decreases.  
The detailed direction means that the \ho{} measurement is much more constraining than BAO for \neff, similar to the constraint 
on individual $\neff$ discussed in \S~\ref{subsec:neff}.  
In the right panel, the BAO data are critical for the neutrino mass measurement, whereas
varying \ho{} at fixed $r_s/D_V$ (horizontal lines in the figure) has almost no effect on \sumnu{}.
This can also be seen by the significant tightening of the inferred mass uncertainty between the green and blue contours in Figure~\ref{fig:mnu_neff}.
We see in Figure~\ref{fig:mnu_neff} that there
is a positive correlation between $\neff$ and \sumnu{} when $H_0$ is one
of the datasets.  This correlation emerges because the only way to
increase $\neff$ in this model space while fixing $z_{\rm eq}$,
$\theta_s$ and $H_0$ is to increase $\sumnu$ (and decrease $\Omega_\Lambda$).  

The CMB constraints on \neff{} and \sumnu{} are nearly independent as
shown by the red contours of Figure~\ref{fig:mnu_neff}.  As
would be expected for independent parameters, the quality of the
constraints are nearly unchanged from the single-parameter extensions
discussed earlier.  The \neff{} constraint is $\neff = 3.40\pm 0.48$
with massive neutrinos compared to $\neff = 3.62\pm 0.48$ in the
earlier massless case.  In both the \LCDM{}+\sumnu{} and
\LCDM{}+\neff+\sumnu{} models, the CMB sets an upper limit on the
neutrino masses of $\sumnu < 1.6$\,eV (95\% CL).  This independence is
broken, and the constraints significantly tightened, once low-redshift
observations are added.  The green contours show the results of
including \ho, while the blue contours show the constraints from
CMB+BAO+\ho.  
For the combination of CMB+BAO+\ho{} we find:
\begin{eqnarray}
\sumnu &=& (0.48\pm 0.21)\,{\rm eV},\\
\neff & = &   3.89\pm 0.37.\nonumber
\end{eqnarray}
As should be expected from  \S~\ref{sec:mnu}, the mass constraint can be improved by adding a tracer of structure growth. 
For CMB+BAO+$H_0$+\sptcl{} (the purple contours, we find the tightest constraint of
\begin{eqnarray}
\sumnu &=& (0.51\pm 0.15)\,{\rm eV},\\
\neff & = &   3.86\pm 0.37.\nonumber
\end{eqnarray}
This is a greater than $3\,\sigma$ preference for $\sumnu > 0$ and a  $2.2\,\sigma$ preference for $\neff > 3.046$.

\section{Conclusions}
\label{sec:conclusions}

\begin{table*}
\begin{center}
\begin{threeparttable}
\caption{Hints for extensions to \lcdm}
\footnotesize
\begin{tabular}{| l | l | l | l |}
\hline
Model extension   & Preferred by the & Alleviates tension    & Statistical error on extension \\
                  & CMB data         & when combining   & parameter reduced by combining \\
            &      (see 1$^{\rm st}$ row of Table~\ref{tab:dchisq_ext}) & BAO and $H_0$ data & BAO and $H_0$ data \\
\hline
\sumnu{} & \marginal{}   & \no                    & \yes\\
&$\Delta \chi^2=2.4$, 1 dof  &  & $ \sigma_{{\rm CMB+BAO+}H_0}/\sigma_{\rm CMB}=0.37$ \\
         & Figure~\ref{fig:mnu_like} & Figure~\ref{fig:bao_vs_h} & Figure~\ref{fig:mnu_like} \\
\hline
\nrun{}  & \yes{}        & \no                    & \no         \\
& $\Delta \chi^2=4.9$, 1 dof & & $\sigma_{{\rm CMB+BAO+}H_0}/\sigma_{\rm CMB}=0.91$ \\
         & Figures~\ref{fig:ns_nrun}  and \ref{fig:1Dnrun}               & Figure~\ref{fig:bao_vs_h}              & Figure~\ref{fig:ns_nrun}  \\
\hline 
\neff{}  & \no{}         & \yes                   & \marginal \\
&$\Delta \chi^2=1.1$, 1 dof  & & $\sigma_{{\rm CMB+BAO+}H_0}/\sigma_{\rm CMB}=0.73$ \\
         & Figure~\ref{fig:neff}                 & Figure~\ref{fig:bao_vs_h}              & Figure~\ref{fig:neff} \\
\hline
\neff+\sumnu{} & \no{}    & \yes                   & \yes \\
&$\Delta \chi^2=2.6$, 2 dof & & $\sigma_{{\rm
    CMB+BAO+}H_0}/\sigma_{\rm CMB}=0.43$ \ $(\sumnu)$ \\
& & & $\sigma_{{\rm CMB+BAO+}H_0}/\sigma_{\rm CMB}=0.77$ \ ($\neff$) \\
               & Figure~\ref{fig:mnu_neff}             & Figure~\ref{fig:neff_mnu_scatter}      & Figure~\ref{fig:mnu_neff} \\
\hline
\end{tabular}
\label{tab:conclusion}
\begin{tablenotes}
\item This table summarizes the evidence for physics beyond the \lcdm{} model. 
We state our conclusions and relevant figures for each entry. 
The second column addresses whether the CMB alone shows a preference for the extension. The third column addresses whether 
the extension reduces the tension between the datasets. The $\Delta\chisq$s in Table~\ref{tab:dchisq_ext} quantify the preference and reduction in the tension.
The final column shows if the extension is favored because the statistical errors are significantly reduced by adding datasets.  
\end {tablenotes}
\end{threeparttable}
\end{center}
\vskip 5 pt
\end{table*}

The SPT bandpowers as presented in \citet{story13} are currently the best measurements of the CMB damping tail
from the third to ninth acoustic peaks $(650 \lesssim l \lesssim 3000)$.
The SPT bandpowers greatly improve measurements
of three CMB observables: the damping scale due to
photon diffusion, the locations of the acoustic peaks, and
the amplitude of the lensing potential. The combined CMB data show a slight, reddening, scale-dependent tilt unreproducible in the \lcdm{} model.
Therefore, we find that while the SPT bandpowers are well-fit by a \lcdm{} cosmology, there are tantalizing hints for extensions to this model.

With the aid of Table \ref{tab:conclusion}, we now summarize our most intriguing findings. 
The extension that the CMB data most prefer (at 2.4~$\sigma$) is the running of the spectral index, \nrun. 
This extension improves the fit to the CMB data, reducing the minimum
$\chi^2$ by $\Delta\chi^2 = 4.9$ from that achieved with \lcdm.   We
find, for the CMB alone, $\nrun =-0.024\pm 0.011$.  
Negative running of the spectral index reconciles the slightly redder spectral index preferred by the SPT data 
with the slightly bluer tilt preferred by the \wseven{} data; the same value of running is preferred by the \wseven{} data 
alone, although only at $\sim1\,\sigma$.
The combination of CMB, BAO, and \ho{} data prefers more negative running,  $\nrun =-0.028\pm 0.010$.  
Deviations from scale invariance of this magnitude are not expected in standard inflationary scenarios.

Increasing the primordial helium abundance increases the damping scale which has a similar impact as negative running on the 
CMB power spectrum. 
Freeing the helium abundance reduces the minimum $\chi^2$ of the combined CMB 
fit by $\Delta\chi^2 = 4.4$, resulting in $\yp = 0.300 \pm 0.025$. However, this value of $\yp$ exceeds the protosolar $95\%$ confidence upper 
limit of $\yp < 0.294$.  A summary of which extensions the CMB prefer can be found in Table~\ref{tab:conclusion}.

We also find that some extensions are more motivated than others with respect to how they reduce the mild tension between the CMB, 
\ho{}, and BAO datasets in the context of the \lcdm{} model. 
The ability of various extensions to reduce this tension is shown in 
the third column of Table 5. Of the extensions that we consider, the
effective number of neutrino species, \neff, is the only model
extension that significantly reduces the tension between the CMB, BAO,
and \ho{} datasets.  
It does so both on its own, and  when \sumnu{} is simultaneously free.   
The tension is reduced because \neff{} brings the model predictions for CMB and \ho{} into better agreement with 
data without much change in the prediction for the BAO observable $r_s/D_V$. 
Thus, this extension can bring the region of the CMB probability posterior that is more consistent with the measured $r_s/D_V$ 
toward simultaneous compatibility with the \ho{} measurement.
Combining the CMB with BAO and \ho{} data, we find $\neff = 3.71\pm0.35$.
Simultaneously freeing \sumnu{} leads to further improvement in the fit as higher neutrino masses bring the CMB into better consistency 
with $r_s/D_V$,  while increasing \neff{} leads to consistency with \ho.

The impact of combining low redshift measurements with CMB data is summarized for selected extensions in the last column of Table 5. 
We find that the CMB, CMB+BAO, and CMB+\sptcl{} all at least weakly prefer a non-zero neutrino mass (with only \ho{} preferring zero mass).
The combined  CMB+BAO+\ho+\sptcl{} dataset constrains $\sumnu\ = (0.31 \pm 0.11)$\,eV. 
The preference for massive neutrinos remains at $>2\,\sigma$ as  long as we keep at least one of the two most precise BAO measurements; however, it disappears if we do not include any BAO data. 

We find improved constraints on two additional model extensions, early dark energy and non-zero curvature. 
These extensions do not address the scale-dependent tilt present in the data or lead to a reduction of the tension between the CMB and low-redshift data.

In summary, we explore six theoretically motivated extensions to the 
\LCDM{} model, and find that the combination of the SPT data with other cosmological measurements yields a 2-3\,$\sigma$ 
preference for some of these extensions. 
Favored extensions are running of the primordial spectral index, a sum of neutrino masses of order $0.3$\,eV, 
or greater than 3 effective neutrino species. 
The evidence for any extension to the standard \lcdm{} model is currently weak
given we find a maximum preference of 3.0\,$\sigma$ after considering a number of extensions and data sets.
However, a significant detection of any of these extensions in future data sets such as {\it Planck} and high-resolution CMB polarization experiments would dramatically impact our understanding of particle physics and cosmology.

\begin{acknowledgements}
The SPT is supported by the National Science Foundation through grant ANT-0638937, with partial support provided by NSF grant PHY-1125897, the Kavli Foundation, and the Gordon and Betty Moore Foundation.
The McGill group acknowledges funding from the National Sciences and Engineering Research Council of Canada, Canada Research Chairs program, and the Canadian Institute for Advanced Research. 
R. Keisler acknowledges support from NASA Hubble Fellowship grant HF-51275.01,
B.A.\ Benson a KICP Fellowship,
M.\ Dobbs an Alfred P. Sloan Research Fellowship,
O.\ Zahn a BCCP fellowship.
This research used resources of the National Energy Research Scientific Computing Center (NERSC), which is supported by the Office of Science of the U.S. Department of Energy under Contract No. DE-AC02-05CH11231, and the resources of the University of Chicago Computing Cooperative (UC3), supported in part by the Open Science Grid, NSF grant NSF PHY 1148698.
We acknowledge the use of the Legacy Archive for Microwave Background Data Analysis (LAMBDA). 
Support for LAMBDA is provided by the NASA Office of Space Science.
\end{acknowledgements}

\appendix

\section{Impact of extragalactic foregrounds and other systematic effects}
\label{app:sys_effects}

One of the central findings of this paper is the preference in the data for lower power at small scales than would be predicted by the \lcdm{} model conditioned on the \wseven{} data. In this appendix, we test the robustness of this preference to the priors on the beam, calibration and extragalactic foregrounds. For simplicity, we only consider two typical extensions: running of the spectral index and changes to Silk damping (parameterized by the helium abundance $Y_p$).
We test how the marginalized constraints on \nrun and \yp{} change when the following systematic tests are performed:
\begin{itemize}
	\item Doubling the width of foreground priors,
	\item Removing the foreground prior (except for positive definiteness),
	\item Doubling the calibration uncertainty,
	\item Doubling the beam uncertainties.
\end{itemize}

\begin{figure*}
\begin{center}
    \includegraphics[width=0.48\textwidth, trim=2.2cm 13.5cm 5.8cm
2.5cm]{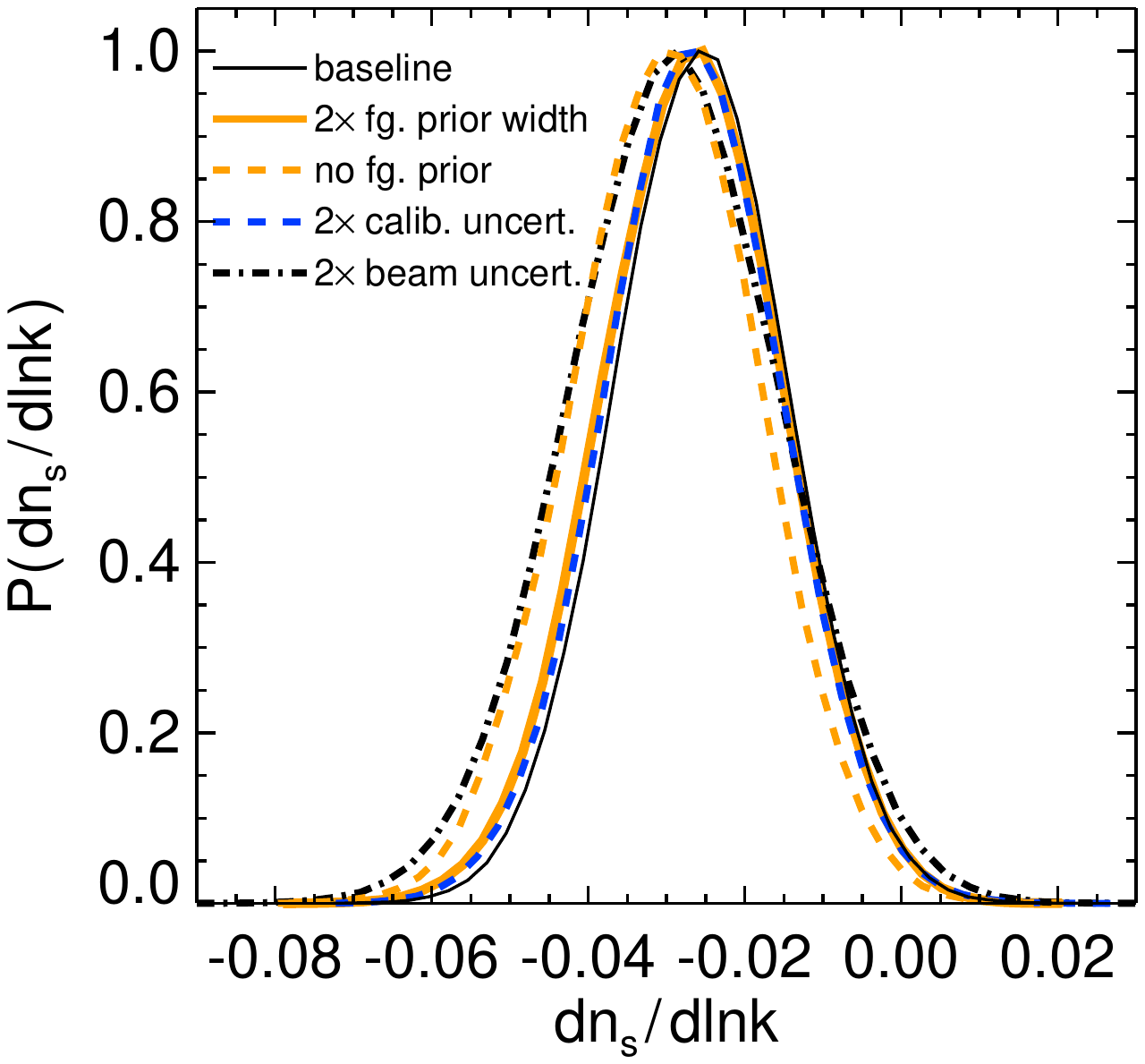}
    \includegraphics[width=0.48\textwidth, trim=2.2cm 13.5cm 5.8cm
2.5cm]{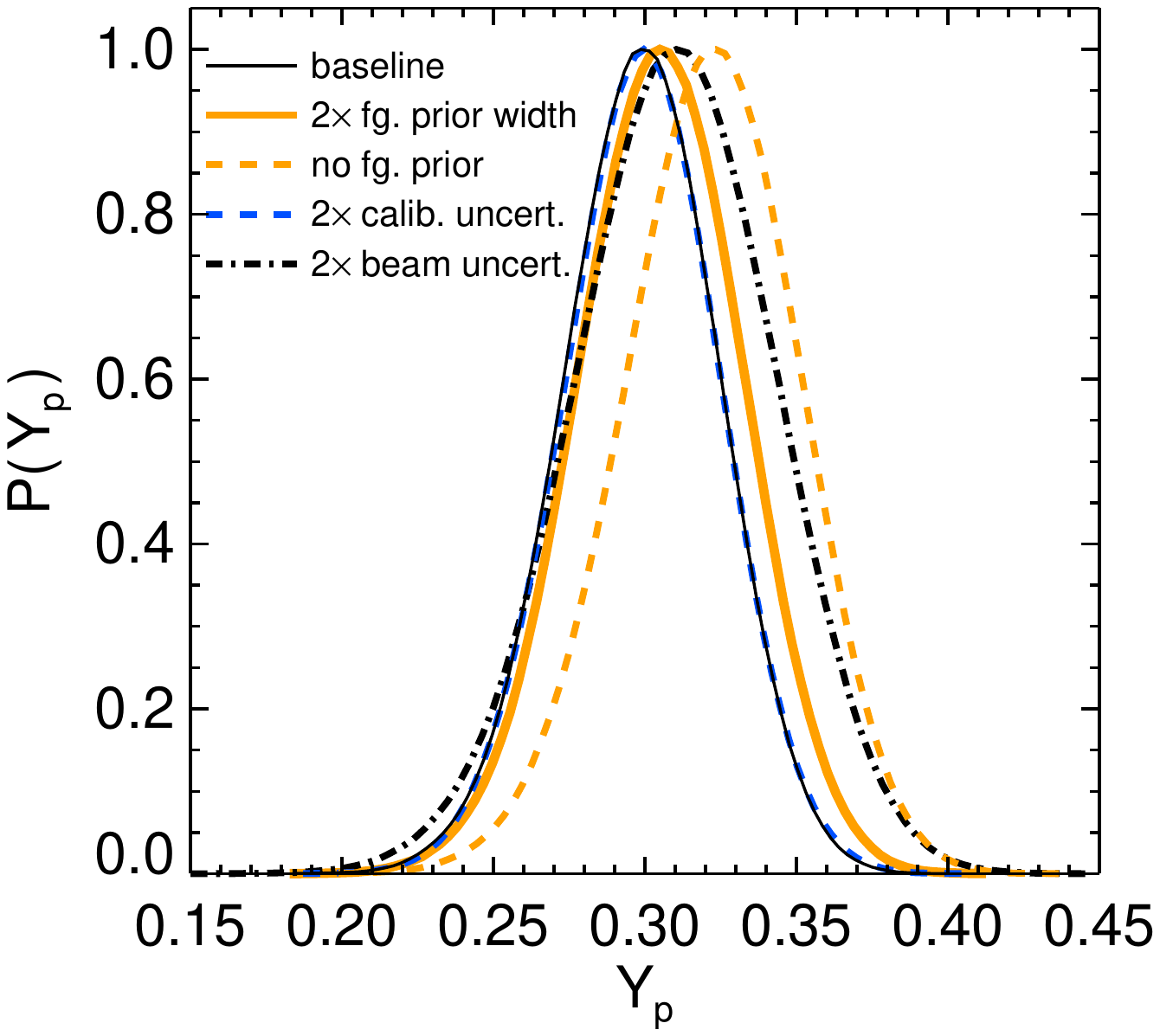}
\end{center}
\caption{The preference for negative running or more damping (in terms of higher $Y_p$) are robust to the extra Galactic foreground modeling and other systematic effects.  In both panel, 'fg.' refers to foreground, 'calib.' refers to calibration and 'uncert.' refers to uncertainty.}
\label{fig:app_sys}
\end{figure*}

\begin{table*}
\begin{center}
\begin{threeparttable}
\caption{The marginalized constraints on \nrun and $Y_p$ for different systematic effects}
\begin{tabular}{ c | c  c  c  c  c }
\hline
 Parameters & baseline & $2\times$ fg. prior width & no fg. prior & $2\times$ calib. uncert. & $2\times$ beam uncert. \\
\hline
\nrun & $-0.024 \pm 0.011$ & $-0.026 \pm 0.012$ & $-0.029 \pm 0.012$ & $-0.025 \pm 0.011$ & $-0.033\pm 0.014$ \\
$Y_p$ & $0.300 \pm 0.025$ & $0.306 \pm 0.027$ & $0.323 \pm 0.028$ & $0.300 \pm 0.025$ & $0.311 \pm 0.032$ \\
\hline
\end{tabular}
\label{tab:appendix_systematics}
\begin{tablenotes}
\item fg. - foreground
\item calib. - calibration
\item uncert. - uncertainty
\end{tablenotes}
\end{threeparttable}
\end{center}
\end{table*}
The marginalized constraints resulting from these tests are listed in Table~\ref{tab:appendix_systematics}, and the posterior distributions are illustrated in Figure~\ref{fig:app_sys}.  The preference for negative running or more damping is robust to the systematic effects considered here.  Doubling the calibration uncertainty has little impact on both parameters.  Loosening the foreground priors slightly moves the $\nrun$ ($Y_p$) median to lower (higher) values with the uncertainty slightly broadened, which makes the preference a bit stronger.  Doubling the beam uncertainties leads to stronger shifts and broadening of the constraints but the $2\sigma$ preference is almost unchanged.

In another test, we restrict the SPT bandpowers with $\ell_{\rm max} = 2200$ and $\ell_{\rm max} = 1500$.  Combined with \wseven{} data, we obtain $\nrun=-0.025 \pm 0.013$ and $\nrun = -0.033\pm 0.014$, respectively.  The $2\sigma$ preference is robust within the range in which the extragalactic foregrounds are not important.

In summary, we find the preference in the data for less power at small scales to be robust to the multipole range and the foreground, beam and calibration priors.

{\it Facilities:}
\facility{South Pole Telescope}

\bibliography{../../BIBTEX/spt}

\end{document}